\documentclass[twocolumn]{aastex631}
\usepackage{natbib}
\usepackage{graphicx}
\usepackage{mathrsfs}
\usepackage{amsmath}
\usepackage{hyperref}
\usepackage{enumitem}
\usepackage{rotating}[-90]
\usepackage{tablefootnote}
\makeatletter
\def\blfootnote{\xdef\@thefnmark{}\@footnotetext}
\makeatother

\begin{document}

\title{Astrometric Accelerations of Provisional Targets for the Habitable Worlds Observatory}

\author[0009-0004-3843-5285]{Katie E. Painter}
\affiliation{Department of Astronomy, The University of Texas at Austin, Austin, TX 78712, USA}
\author[0000-0003-2649-2288]{Brendan P. Bowler}
\affiliation{Department of Physics, University of California, Santa Barbara, Santa Barbara, CA 93106, USA}
\affiliation{Department of Astronomy, The University of Texas at Austin, Austin, TX 78712, USA}
\author[0000-0003-4557-414X]{Kyle Franson}
\affiliation{Department of Astronomy, The University of Texas at Austin, Austin, TX 78712, USA}
\author[0000-0002-7733-4522]{Juliette C. Becker}
\affiliation{Department of Astronomy, University of Wisconsin–Madison, 475 N. Charter Street, Madison, WI 53706, USA}
\author[0000-0002-0040-6815]{Jennifer A. Burt}
\affiliation{Jet Propulsion Laboratory, California Institute of Technology, 4800 Oak
Grove Drive, Pasadena, CA 91109, USA}

\blfootnote{Corresponding author: Katie E. Painter}

\blfootnote{\url{katie.teixeira@utexas.edu}}

\begin{abstract}
NASA's Habitable Worlds Observatory (HWO) will be the first space telescope capable of directly imaging Earth-like planets in the habitable zones of Sun-like stars to probe their atmospheres for signs of life. Now in its early stages of design, a list of the 164 most promising targets for HWO has been released to the community to carry out precursor science. Massive companions in these systems---stars, brown dwarfs, or giant planets---could preclude the existence of Earth-sized planets in the habitable zone by impacting their long-term dynamical stability. Here, we use astrometry from Hipparcos and Gaia EDR3 to identify stars in the HWO preliminary target list that exhibit astrometric accelerations and determine joint constraints on the expected mass and separation of these companions. We find that 54 HWO targets have significant astrometric accelerations, 37 of which are accounted for by known giant planets and stellar companions. Follow-up efforts are required to clarify the specific nature of the suspected companions around the remaining 17 accelerating stars. Stars without significant accelerations are used to rule out large regions of companion mass and separation down to planetary masses. We find that with Hipparcos and Gaia EDR3 we are $\sim$85$\%$ sensitive to 2~$M_\mathrm{Jup}$ planets between 4 and 10 AU. Future Gaia releases will provide sensitivity to sub-Jovian mass planets on Solar System scales for provisional HWO targets. Finally, using analytical estimates of dynamical stability, we find that 13 HWO targets have known stellar or planetary companions that are likely to disrupt habitable-zone planets.
\end{abstract}


\section{Introduction}
The Habitable Worlds Observatory (HWO) is a planned NASA flagship space telescope with the primary goal of searching for biosignatures on terrestrial planets in the habitable zones (HZs) of nearby Sun-like stars. It will be a large Infrared/Optical/Ultraviolet observatory capable of performing a wide range of astrophysics programs. Sensitivity to the characteristic reflected-light contrast of an Earth-like planet orbiting a Sun-like star at 1 AU will require HWO be equipped with high contrast imaging and spectroscopy capable of $10^{-10}$ contrast levels. HWO’s nominal mission objective, as outlined in the Astro2020 Decadal Survey \citep{Astro2020}, is to search for and characterize approximately 25 Earth-sized planets in the HZs of their host stars. This sample will offer the first quantitative assessment of the prevalence of biosignature gases in the atmospheres of nearby terrestrial planets in the HZ.

Given the technical challenges of identifying terrestrial planets in the HZs of nearby Sun-like stars with techniques such as radial velocities \citep{Ciardi_19, Crass_21, Newman_23}, HWO is expected to concurrently discover these planets and examine their atmospheres. This mission will more precisely constrain the occurrence rate of Earth-sized planets in the HZ, or $\eta_{\oplus}$, with current estimates of $\eta_{\oplus}$ ranging from 0.1 to 1.0 \citep{Burke_15, Bryson_21}. Before the mission begins, characterizing other members of these systems such as stellar binary companions, brown dwarfs, and giant planets will be critical to establish how likely it is for an Earth-sized planet to reside in the HZ. For example, a system harboring a giant planet in the HZ would be unlikely to also host an Earth analog, which would become dynamically unstable on short timescales. A planet as light as 10 $M_{\oplus}$ could still disrupt significant portions of the HZ \citep{Kopparapu_10}. A giant planet at farther distances could also impact the stability of an Earth analog, such as a Jupiter-sized planet at 4 AU \citep{Kopparapu_10, Agnew_19, Kane_19}. Similarly, a stellar companion could dynamically interact with an Earth-sized planet in the HZ at much larger orbital separations---up to 50 AU for a 0.5 $M_{\odot}$ star with $e=0.9$---and even preclude its formation in the protoplanetary disk \citep{David_03, Jaime_12, Kraus_16}. The dynamical stability of Earth analogs will therefore depend on the masses and orbits of other companions. Thus, constraining the existence of companions on Solar-System scales, including previously unknown giant planets, and measuring their orbital properties are key priorities for HWO precursor science.

Our focus in this study is on the list of 164 provisional targets that has been presented for HWO \citep{Mamajek_Stapelfeldt_23}. These include predominantly F, G, and K main sequence stars within 25 pc whose habitable zones are accessible to HWO's expected telescope architecture ($\sim$6-meter inscribed aperture) and coronagraph inner working angle. Considerations that went into the construction of the list included inner working angle, planet-star brightness ratio, planet apparent magnitude, distance, host star luminosity, and optical brightness. Stars with binary companions within 3$''$ were avoided due to the anticipated challenges of starlight suppression. 

Previous work has already begun to assemble what is known about these targets and their planetary systems in preparation for HWO. For example, \cite{Laliotis_23} presented an analysis of archival RV data of bright, southern Sun-like stars to identify new planets and map sensitivity to planet mass and orbital separation for individual systems. This sample was drawn from the EPRV Working Group's final report \citep{Crass_21}, which preceded \cite{Mamajek_Stapelfeldt_23}, but many stars appear on both lists. They found that for many stars, the existing archival RVs are not sensitive down to Saturn-mass planets at 1 AU, thus motivating future EPRV efforts. \cite{Harada_24} derived stellar properties for the 164 targets by modeling their spectral energy distributions, and compiled these along with abundances, photometry, and flare rates into an expanded catalog. They emphasized the need for more space-based UV and mid-IR measurements and abundance surveys. \cite{Tuchow_24} designed an input catalog including uniformly compiled and calculated stellar properties for about 13,000 stars which could be used to design a new list for a different telescope architecture or science goal. \cite{Kane_24} presented dynamical analysis for the 30 potential HWO targets from \cite{Mamajek_Stapelfeldt_23} known to host a planet, finding that for 11 systems, less than 50$\%$ of the HZ is dynamically viable for an Earth-sized planet. \cite{Harada_24b} analyzed RVs and stellar activity indicators from HIRES/Keck \citep{Vogt_94} and HARPS/ESO \citep{Mayor_03} for the 90 potential HWO targets which have at least 20 total RV epochs from these instruments. They identified 5 new RV signals and found the median HZ sensitivity in the sample to be about 66 $M_{\oplus}$.

Here, we crossmatch the list of 164 targets from \cite{Mamajek_Stapelfeldt_23} with the Hipparcos-Gaia Catalog of Accelerations (HGCA; \citealt{Brandt_18, Brandt_21}), which provides proper motion differences (long-term accelerations) between the Hipparcos \citep{Perryman_97} and Gaia \citep{Gaia_16} missions for over $10^5$ stars. 154 of the 164 targets from \cite{Mamajek_Stapelfeldt_23} have entries in the HGCA catalog. With the $\sim$25 year time baseline between Hipparcos and Gaia, these accelerations indicate the presence of giant planets, brown dwarfs, and stellar companions on wide orbits ($\sim$1--100 AU) in these systems. This long-baseline astrometry provides access to a region of parameter space that has been difficult to reach, complimenting radial velocities at close separations and direct imaging at wide separations. As an example of the utility of long-baseline accelerations, HGCA accelerations have recently been used to discover planets, brown dwarfs, white dwarfs, and stellar companions; measure their masses; and precisely constrain their orbits (e.g., \citealt{Brandt_et_al_21}, \citealt{Bowler_21}, \citealt{Franson_23b}, \citealt{deRosa_23}, \citealt{Mesa_23}, \citealt{Rickman_24}). Astrometric accelerations have also been used to measure the true masses and characterize the orbits of known RV planets (e.g., \citealt{Li_21}, \citealt{Venner_21}, \citealt{VanZandt_24}). In this work, we use HGCA accelerations, along with a semi-analytical modeling framework adapted from \cite{Kervella_19, Kervella_22}, to constrain the existence of new and previously known companions spanning three orders of magnitude in mass and separation in the HWO target systems.

This study is organized as follows. In Section \ref{sec:Methods}, we describe how the HGCA is used to extract long-term accelerations of the HWO preliminary target list. We also detail the framework to jointly constrain companion masses and separations, both for significant accelerations and in cases where no acceleration is detected. In Section \ref{sec:Results}, this framework is applied to HWO targets and the results are analyzed in conjunction with known companions. Individual targets and the implications of this work for HWO's mission are discussed in detail in Section \ref{sec:Discussion}, and Section \ref{sec:Conclusions} provides a summary of our findings.

\section{Astrometric Accelerations of Promising HWO Targets}
\label{sec:Methods}

The HGCA \citep{Brandt_18, Brandt_21} provides three separate, cross-calibrated proper motion measurements with uncertainties for over $10^5$ bright stars in the solar neighborhood: the Hipparcos proper motion at an average epoch of 1991.25, the Gaia EDR3 proper motion at an average epoch of 2015.50, and a long-term (``scaled positional difference'') proper motion calculated as the difference in sky positions between the two missions. The long-term proper motion ($\mu_{\rm{HG}}$) can be subtracted from the Gaia EDR3 proper motion ($\mu_{\rm{G}}$) to obtain a proper motion difference of the star, $\Delta\mu_{\rm{G}}$. Using the Gaia EDR3 proper motion instead of the Hipparcos proper motion produces a more precise proper motion difference. This corresponds to the time-averaged motion of the star around the barycenter of its system in the plane of the sky, and can be converted into a tangential velocity using the measured Gaia parallax. 

An astrometric acceleration (due to a change in the velocity vector) can also be calculated by incorporating the time-baseline between Hipparcos and Gaia. A non-zero acceleration correlates with the gravitational acceleration of the star induced by an unseen companion in the plane of the sky. From this information, the mass and separation of the companion can be jointly constrained, albeit with aliasing due to the long time baseline between Hipparcos and Gaia. Note that this analysis assumes that only one massive companion is accelerating the host star; more than one companion has the potential to bias the astrometric measurement and subsequent constraints on mass and separation. We calculate the magnitude of the acceleration in the plane of the sky ($a_{\alpha\delta}$) and its uncertainty for all 156 HWO targets cross-matched in the HGCA. Note that 4 of the 8 targets that are not listed in the HGCA are fainter binary companions of those with measurements in that catalog.\footnote{AK Leporis (HIP 27072 B), Xi Bootis B (HIP 72659 B), p Eridani B (HIP 7751 B), and 36 Ophiuchi B (HIP 84405 B) are companions of stars in the HGCA.} Alpha Centauri A and B (HIP 71683 and HIP 71681) are not included in the HGCA because they are too bright for Gaia. 70 Ophiuchi (HIP 88601) A and B are not included in the catalog, likely because they were spatially resolved in Gaia but unresolved in Hipparcos.

The proper motion differences from the HGCA are used to assess which HWO targets have a statistically significant acceleration \citep{Brandt_18, Brandt_21}. The HGCA provides $\chi^2$ values assuming a model of constant proper motion applied to the most precise proper motion measurements, which are typically $\mu_{\rm{G}}$ and $\mu_{\rm{HG}}$. This goodness of fit metric is calculated based on proper motion measurements in both R.A. and Declination, and thus has two degrees of freedom ($\nu$ = $N$ measurements -- $k$ model parameters, where $N$ = 4 and $k$ = 2). Using these $\chi^2$ measurements, we determine one-tail probabilities that $\chi^2$ at these values or greater would be expected by chance using the $\chi^2$ distribution with $\nu$=2. These probabilities ($P$) are then translated into Gaussian-equivalent significance levels as follows:
\begin{equation}
n_{\sigma} = \sqrt{2}\ \mathrm{erf}^{-1}(P)
\end{equation}

\begin{figure}
    \centering   
    \includegraphics[width=3.4in]{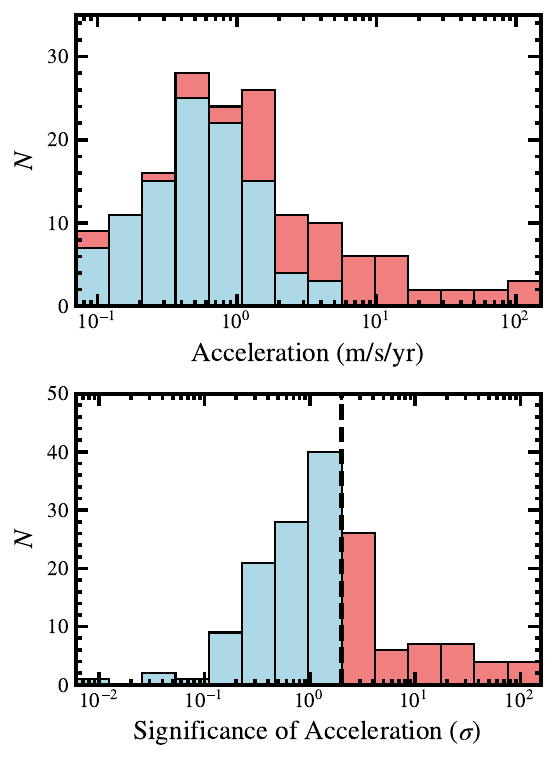}
    \caption{Top: HGCA acceleration measurements for HWO preliminary targets. Bottom: Significance of HGCA acceleration measurements for HWO targets, calculated from $\chi^2$ values for constant proper motion model fits (see Section \ref{sec:Methods} for details). We adopt a cutoff for significance at 2$\sigma$. The significant ($>2\sigma$) accelerations are shown in red in both the top and bottom panels.}
    \label{fig:Figure_1}
\end{figure}

We place our significance cutoff at 2$\sigma$ ($\chi^2$ = 6.16 with 2 degrees of freedom) to minimize the exclusion of real detections while mitigating false positives at lower significance levels. For instance, the orbital inclination and true mass of the $\approx$7~$M_\mathrm{Jup}$ planet HD 81040 b was determined from the $\approx$2.0$\sigma$ astrometric acceleration on its host star \citep{Li_21}, and a $\approx$16~$M_\mathrm{Jup}$ substellar companion to HIP 99770 was discovered with direct imaging based on a 2.2$\sigma$ acceleration \citep{Currie_23}. This 2$\sigma$ cutoff produces 54 HWO targets with statisically significant accelerations. There are 22 accelerations that have significance levels between 2$\sigma$ and 3$\sigma$. Among these systems with modest significance, given a false positive rate between 4.6\% (for a 2$\sigma$ acceleration) and 0.3\% (for 3$\sigma$), we expect between 0 and 1 of these detections to be a false positive. The remaining 32 systems have acceleration significance levels between 3 and 130$\sigma$, as shown in Figure \ref{fig:Figure_1}. The entire list of preliminary HWO targets cross-matched in the HGCA is displayed in Table \ref{tab:HWO_targets}, along with their corresponding astrometric accelerations.

\subsection{Translating Accelerations into Constraints on Mass and Separation}
\label{sec:mass-sep-constraints}
Given the mass and distance of each host star, it is possible to predict the possible range of masses and separations of companions consistent with the measured accelerations. For this study, we follow the semi-analytical formalism from \cite{Kervella_19, Kervella_22}. Inputs into this model include the mass of the host star $m_1$, its parallax from Gaia $\bar{\omega}_{\rm{G}}$, the proper motion measurements $\mu_{\rm{HG}}$ and $\mu_{\rm{G}}$ in R.A. and Declination, and the associated uncertainties of these measurements. We first calculate $\Delta\mu_{\rm{G}}$ from its orthogonal components in R.A. and Declination. Next, with the Gaia parallax, we determine the tangential velocity of the target in the plane of the sky, or $\Delta v_{\rm{T,G}}$. Then, the mass of a secondary companion $m_2$ can be inferred as a function of orbital radius $r$ \citep{Kervella_19}, which assumes $m_1 \gg m_2$:
\begin{equation}
    m_2(r) = \frac{1}{\gamma \ \eta \ \zeta} \ \sqrt{\frac{m_1}{G}} \ {\Delta v_{\rm{T,G}}} \sqrt{r}
    \label{eq:m2}
\end{equation}
There are three multiplicative correction factors in this function that relate to observational biases: $\gamma$, $\eta$, and $\zeta$. $\gamma$ accounts for the fact that a Gaia proper motion is not an instantaneous measurement, but is instead a time-averaged quantity
over a baseline, $\delta t_{\rm{G}}$, which in this case corresponds to EDR3 (34 months):
\begin{equation}
    \gamma(P) = \frac{P/\delta t_{\rm{G}}}{\sqrt{2}\pi}\sqrt{1-\mathrm{cos}\left(\frac{2\pi}{P/\delta t_{\rm{G}}}\right)}
    \label{eq:gamma}
\end{equation}
where $P$ is given by Kepler's third law and is a function of $m_1$ and $r$. Because $\Delta v_{\rm{T,G}}$ only measures velocity in the plane of the sky, $\eta$ acts to convert it into the magnitude of its 3D vector, including the effects of orbital phase, inclination, and eccentricity, and was found by \cite{Kervella_19} to be $0.87\substack{+0.12 \\ -0.32}$\ based on a Monte Carlo simulation of the orbital velocity vector. Finally, $\zeta$, the efficiency factor, accounts for the potential for $\mu_{\rm{HG}}$ to include part of the orbital velocity, as a function of period, which biases the measured $\Delta v_{\rm{T,G}}$. We extract this efficiency factor $\zeta$ as a function of normalized orbital period and its uncertainty $\sigma_{\zeta}$ from Figure 3 of \cite{Kervella_19} for this analysis.

\begin{figure*}
    \centering   
    \includegraphics[width=7.1in]{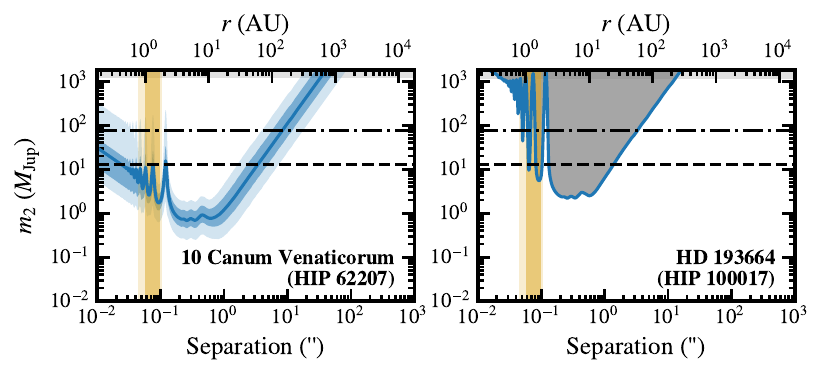}
    \caption{Examples of companion mass-separation predictions calculated using Hipparcos-Gaia astrometry from HGCA. For stars with significant accelerations, such as 10 Canum Venaticorum (left panel), 1 and 2$\sigma$ confidence intervals are computed to identify the joint constraints on companion mass and separation consistent with the measured acceleration of the host star. For non-accelerating stars, such as HD 193664 (right panel), we adopt a 3$\sigma$ upper limit above which companions can be ruled out with 99.7$\%$ confidence (gray region). Dashed lines correspond to boundaries between stars and brown dwarfs ($\approx$$75~M_\mathrm{Jup}$) and brown dwarfs and giant planets ($\approx$$13~M_\mathrm{Jup}$). Conservative and optimistic HZ limits are displayed as dark and light yellow rectangles, respectively; we detail the HZ limit calculations in Section \ref{sec:known-companions}.}
    \label{fig:Figure_2}
\end{figure*}

For each star in the HWO preliminary target list, we calculate $m_2$ as a function of $r$ in a Monte Carlo fashion for $10^6$ trials to account for uncertainties in 8 parameters: $m_1$, $\eta$, $\zeta(P)$, $\mu_{\delta,\rm{G}}$, $\mu_{\alpha,\rm{G}}$, $\mu_{\delta,\rm{HG}}$, $\mu_{\alpha,\rm{HG}}$, and $\bar{\omega}_{\rm{G}}$. For each trial, values are randomly drawn from normal distributions assuming no covariance among parameters. Host masses are adopted directly from \cite{Mamajek_Stapelfeldt_23}. Uncertainties were not reported for this parameter, so we assumed an error of 10$\%$, based on representative values from the literature (e.g., \citealt{Takeda_07}, \citealt{Bruntt_10}). Our results are not very sensitive to this assumption. For each accelerating star, we extract the mean $m_2$ as a function of $r$ and calculate the credible regions in the form of 1, 2, and 3$\sigma$ highest density intervals. An example of these constraints as a function of angular separation ($\rho$) and physical separation ($r$) for one star in the sample, 10 Canum Venaticorum, is shown in Figure \ref{fig:Figure_2}. 

For stars without significant accelerations ($<2\sigma$), we carry out the same Monte Carlo simulation to establish the region above which companions are ruled out with 99.7$\%$ (``$3\sigma$'') confidence as shown in the right-hand panel of Figure \ref{fig:Figure_2} for another HWO target, HD 193664. This immediately provides information about what types of companions can be excluded in these systems at characteristic separations of a few AU to tens of AU. We also calculate the corresponding 1, 2, 4, and $5\sigma$ upper limits for each non-accelerating star. We include these limits and the constraints for accelerating targets in Tables \ref{tab:accelerating_data} and \ref{tab:nonaccelerating_data} in Section \ref{sec:Appendix}.

\subsection{Identifying Known Companions Consistent with Accelerations}
\label{sec:known-companions}
With mass-separation constraints for all 156 HWO targets in the HGCA, known companions of these targets can be compared with these predictions to determine which, if any, are consistent with causing the measured acceleration. If a culprit of an acceleration is identified, either with RVs, imaging, or both, its orbit and dynamical mass can be measured when existing data is combined with the astrometric acceleration (e.g. \citealt{Brandt_19, Rickman_22, Matthews_24}). This makes it possible to assess the dynamical impact of a companion on the HZ and determine the inclination of companions previously detected by RVs, which in turn provides true masses for these companions (e.g., \citealt{Li_21}). 

We compile known companions to the HWO preliminary targets---planets, brown dwarfs, stars, and white dwarfs---from the literature (e.g., \citealt{Raghavan_10}, \citealt{Chini_14}, \citealt{Fuhrmann_14}, \citealt{Fuhrmann_17}), the NASA Exoplanet Archive, and the Washington Double Star Catalog (WDS) \citep{Mason_01}. Altogether, this amounts to 66 confirmed planets, 23 of which are giant planets above 0.3 $M_\mathrm{Jup}$; 59 stellar companions (3 of which are themselves close binary stars); 4 brown dwarf companions (one of which is a binary); and 4 white dwarf companions (one of which is a close binary). Note that 8 of the stars compiled here are companions to the HWO targets which were not included in the HGCA. For planets detected with radial velocities, we determine the probability distribution of true masses from the minimum mass based on an isotropic inclination distribution. 

Results are shown in Figure \ref{fig:Figure_8}, where we plot semi-major axis in both angular and physical units along with uncertainties associated with each companion. For imaged stellar, white dwarf, and brown dwarf companions, which may have historical measurements of relative astrometry dating back decades or, in some cases, even centuries, projected separations can change significantly from the earliest epochs to the most recent epochs. To test whether a companion is consistent with causing a measured acceleration, an epoch closest in time to the effective measurement of the Hipparcos-Gaia astrometric acceleration is needed.  For these systems we adopt a projected separation closest to the weighted average of the Gaia measurement epoch ($ t_{\rm{G}}$) and the Hipparcos-Gaia measurement epoch ($ t_{\rm{HG}}$): $1/4*t_{\rm{H}}+3/4*t_{\rm{G}}$ \citep{Brandt_19}. For all targets, the epoch of the measured acceleration is about 2009.8. Then, we convert projected separation into semi-major axis with corresponding uncertainties using the uniform-eccentricity conversion factor from \cite{Dupuy_11}.

We use mass estimates from the literature for 28 of the stellar or brown dwarf companions, and, for the remaining 40, determine masses based on evolutionary models or empirical relationships. For F, G, and K stars, we infer masses using the MESA Isochrones $\&$ Stellar Tracks (MIST) model grids \citep{Dotter_16, Choi_16, Paxton_11, Paxton_13, Paxton_15, Paxton_18}, along with the Gaia \textit{G} band magnitude, assuming a stellar age of 3 Gyr, which is the typical age of host stars in our sample in log-space. (The results are not strongly sensitive to the exact value except at the youngest ages $\lesssim 100$ Myr.). For M dwarfs, we use a mass versus $M_K$ empirical relationship from \cite{Mann_19}, which is calibrated with dynamical masses. For white dwarfs, we assign a mass of $0.6 M_\odot$ following \cite{Kepler_06}.

Finally, with companion separations and masses in hand, we assign each companion into one of two categories based on the degree to which they are responsible for the measured acceleration. If the companion falls within 2$\sigma$ of the joint mass and separation constraints predicted from an acceleration, we deem it as being ``consistent" with being the source of the observed acceleration. If it sits below the 2$\sigma$ lower-limit of the mass-separation curve, it is ``inconsistent'' because it is not massive enough to produce the acceleration. These evaluations are listed in Table \ref{tab:HWO_targets} and inherently assume that the source of the acceleration arises from a single companion. Of course, if multiple massive companions are present, this could alter the measured Hipparcos-Gaia acceleration and impact the predicted properties of the companion (e.g., HD 206893 B and c; \citealt{Grandjean_19}; \citealt{Hinkley_23}).

\begin{figure*}
    \centering   
    \includegraphics[width=7.1in]{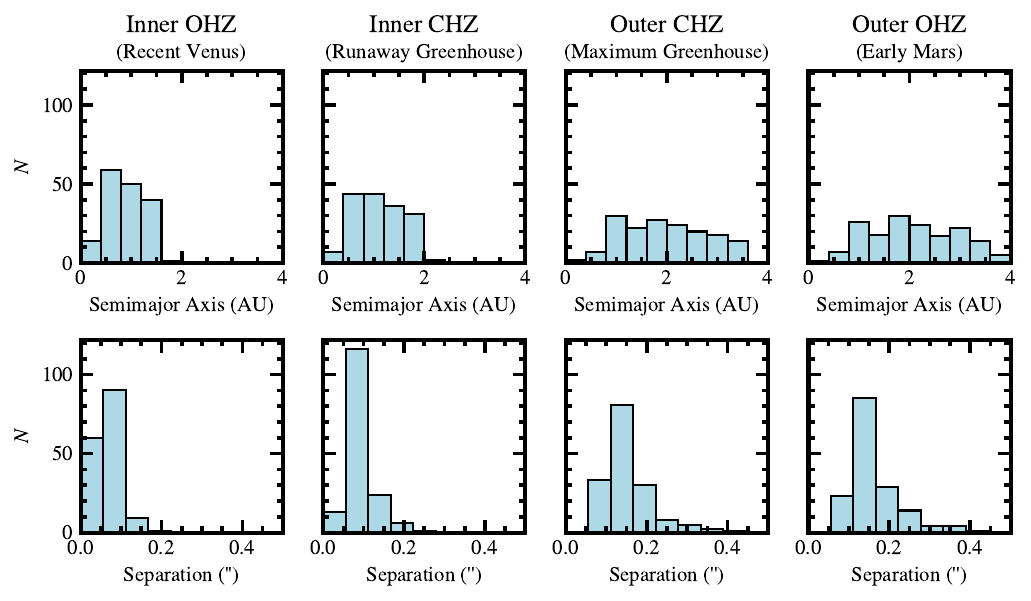}
    \caption{HZ limits in physical (top) and angular (bottom) separation for all HWO targets following the prescription from \cite{Kopparapu_14}. These display the separations that HWO will need to resolve. Optimistic bounds, defined by ``recent Venus'' and ``early Mars'' empirical constraints, are shown on the far left and right, respectively. Conservative bounds, defined by ``runaway greenhouse'' and ``maximum greenhouse'' calculations, are shown on the center left and right, respectively.}
    \label{fig:Figure_3}
\end{figure*}

In evaluating the suitability of targets for HWO, it is important to compare the region of stability based on known companions to the location of the HZ. There exist a broad range of definitions for the HZ, but it most commonly describes the region around a star where liquid water is possible on the surface of a geologically active rocky planet with a similar mass and atmospheric composition to Earth (e.g., \citealt{Kasting_93}, \citealt{Kopparapu13}, \citealt{Kaltenegger_17}). The ``conservative HZ'' calculated in \cite{Kopparapu13}, which spans 0.95-1.67 AU for a Sun-like star, was used in the construction of the precursor target list, scaled with each star's bolometric luminosity. Here, the HZ limits for each HWO target are calculated following the updated prescriptions from \cite{Kopparapu_14}. The conservative HZ (CHZ) is defined at the inner edge by the runaway greenhouse limit in which the incoming radiation causes water to evaporate, which heats the atmosphere in a positive feedback loop. The outer edge of the CHZ is defined by the ``maximum'' greenhouse limit in which the addition of $\rm{CO}_2$ to the atmosphere can no longer further heat the planet. The optimistic HZ (OHZ) is constructed based on empirical evidence of the past existence of liquid water on Venus and Mars. The inner edge is defined by the ``recent Venus'' limit at the inner edge, as there is evidence that Venus may have possessed liquid water as recently as $\sim$1 Gyr ago \citep{Solomon_91}. Then, the outer edge is defined by the ``early Mars'' limit based on evidence that Mars possessed liquid water $\sim$3.8 Gyr ago \citep{Carr_96}. These limits are computed based on the effective temperature and luminosity of the host star, which we draw directly from \cite{Mamajek_Stapelfeldt_23}, and they are listed in Table \ref{tab:HWO_targets}. 

Histograms of these boundaries for all of the HWO preliminary targets are shown in Figure \ref{fig:Figure_3}. Within the sample, inner limits for the optimistic case reach down as far as 0.12 AU or 0.04$''$ while conservative inner limits are as low as 0.15 AU or 0.05$''$. Outer HZ limits reach as high as 3.7 AU or 0.44$''$ for the optimistic case but only 3.5 AU or 0.42$''$ for the conservative case. The HZ ranges are much narrower for individual systems and depend on the mass of the host star. It will be important to compare these values against the technical limits of the proposed telescope architecture for HWO. Of course, the exact limits of the CHZ and OHZ can be altered by many factors, such as planetary atmospheric composition, albedo, water loss, and greenhouse effects.

\startlongtable
\begin{longrotatetable}
\begin{deluxetable*}{lcccccl}
\renewcommand\arraystretch{0.9}
\tabletypesize{\footnotesize}
\setlength{ \tabcolsep } {.2cm} 
\tablewidth{0pt}
\tablecolumns{11}
\tablecaption{HWO Provisional Targets with Hipparcos-Gaia Astrometric Accelerations\label{tab:HWO_targets}}
\tablehead{
 \colhead{Name} & \colhead{HIP ID} & \colhead{$a_{\alpha\delta}$ \tablenotemark{a}} &  \colhead{Significance} & \colhead{CHZ} & \colhead{OHZ}  & \colhead{Nature of $a_{\alpha\delta}$ \tablenotemark{b}}  \\
 &  & \colhead{(m/s/yr)} &  \colhead{($n_\sigma$)} & \colhead{(AU)} & \colhead{(AU)}  & \colhead{(CC = Consistent, IC = Inconsistent)}}
\startdata
HD 193664 & 100017 & 0.24$\pm$0.25 & 0.5 & (0.99, 1.74) & (0.78, 1.84) & -- \\
$\psi$ Capricorni & 102485 & 2.43$\pm$1.44 & 1.2 & (1.70, 2.94) & (1.34, 3.10) & -- \\
HD 199260 & 103389 & 1.45$\pm$0.78 & 1.4 & (1.29, 2.24) & (1.02, 2.37) & -- \\
61 Cygni A & 104214 & 6.43$\pm$0.33 & 17.6 & (0.38, 0.71) & (0.30, 0.75) & CC: 61 Cygni B \\
61 Cygni B & 104217 & 7.30$\pm$0.05 & 129.0 & (0.29, 0.54) & (0.23, 0.57) & CC: 61 Cygni A \\
AX Microscopii & 105090 & 0.08$\pm$0.07 & 0.6 & (0.30, 0.57) & (0.23, 0.60) & -- \\
$\gamma$ Pavonis & 105858 & 0.49$\pm$0.44 & 0.5 & (1.13, 1.98) & (0.89, 2.08) & -- \\
HN Pegasi A & 107350 & 0.24$\pm$0.33 & 0.3 & (1.00, 1.76) & (0.79, 1.86) & -- \\
HD 207129 & 107649 & 0.37$\pm$0.32 & 0.6 & (1.03, 1.82) & (0.82, 1.92) & -- \\
HD 14412 & 10798 & 0.27$\pm$0.18 & 1.0 & (0.65, 1.16) & (0.51, 1.22) & -- \\
$\varepsilon$ Indi A & 108870 & 2.71$\pm$0.17 & 16.8 & (0.47, 0.87) & (0.37, 0.92) & CC: $\varepsilon$ Indi Ab; IC: $\varepsilon$ Indi Ba/Bb \\
$\tau$ Piscis Austrini & 109422 & 0.95$\pm$0.79 & 0.6 & (1.57, 2.73) & (1.24, 2.88) & -- \\
HD 212330 A & 110649 & 83.17$\pm$0.72 & 130.3 & (1.58, 2.80) & (1.25, 2.95) & IC: HD 212330 B \\
$\upsilon$ Aquarii A & 111449 & 4.64$\pm$0.90 & 5.5 & (1.68, 2.92) & (1.33, 3.08) & CC: $\upsilon$ Aquarii B \\
$\xi$ Pegasi A & 112447 & 3.43$\pm$1.69 & 1.5 & (1.98, 3.45) & (1.56, 3.64) & -- \\
Fomalhaut B & 113283 & 0.23$\pm$0.12 & 1.8 & (0.45, 0.82) & (0.35, 0.87) & -- \\
GJ 887 & 114046 & 0.11$\pm$0.04 & 2.6 & (0.20, 0.38) & (0.16, 0.40) & IC: GJ 887 b; IC: GJ 887 c \\
HD 219134 & 114622 & 0.37$\pm$0.15 & 2.0 & (0.51, 0.94) & (0.41, 0.99) & CC: HD 219134 h, IC: HD 219134 f, IC: HD 219134 d \\
HD 219623 & 114924 & 0.91$\pm$0.48 & 1.7 & (1.33, 2.33) & (1.05, 2.46) & -- \\
HD 219482 & 114948 & 0.50$\pm$0.40 & 0.6 & (1.27, 2.21) & (1.00, 2.33) & -- \\
$\iota$ Piscium & 116771 & 1.39$\pm$1.49 & 0.5 & (1.70, 2.97) & (1.34, 3.13) & -- \\
$\iota$ Horologii & 12653 & 0.52$\pm$0.34 & 1.0 & (1.22, 2.14) & (0.97, 2.26) & -- \\
$\theta$ Persei A & 12777 & 0.86$\pm$1.03 & 0.4 & (1.38, 2.41) & (1.09, 2.55) & -- \\
$\tau^1$ Eridani & 12843 & 97.36$\pm$1.29 & 66.0 & (1.49, 2.59) & (1.17, 2.73) & -- \\
HD 17925 & 13402 & 0.37$\pm$0.20 & 1.2 & (0.62, 1.12) & (0.49, 1.18) & CC: $\tau^1$ Eridani B (Not imaged) \\
$\iota$ Persei & 14632 & 1.14$\pm$0.75 & 0.9 & (1.40, 2.47) & (1.11, 2.60) & -- \\
$\alpha$ Fornacis A & 14879 & 32.16$\pm$0.68 & 49.4 & (1.91, 3.34) & (1.51, 3.52) & CC: $\alpha$ Fornacis B \\
$\zeta^1$ Reticuli & 15330 & 0.16$\pm$0.39 & 0.1 & (0.85, 1.50) & (0.67, 1.59) & -- \\
$\zeta^2$ Reticuli & 15371 & 0.62$\pm$0.49 & 0.7 & (0.95, 1.68) & (0.75, 1.77) & -- \\
$\kappa^1$ Ceti & 15457 & 0.33$\pm$0.62 & 0.2 & (0.88, 1.56) & (0.70, 1.65) & -- \\
e Eridani & 15510 & 0.20$\pm$0.31 & 0.2 & (0.78, 1.40) & (0.62, 1.47) & -- \\
$\zeta$ Tucanae & 1599 & 1.14$\pm$0.73 & 1.1 & (1.06, 1.86) & (0.84, 1.96) & -- \\
$\kappa$ Reticuli A & 16245 & 3.39$\pm$1.51 & 1.8 & (1.99, 3.45) & (1.57, 3.64) & -- \\
$\varepsilon$ Eridani & 16537 & 1.12$\pm$0.22 & 5.5 & (0.57, 1.04) & (0.45, 1.09) & CC: $\varepsilon$ Eridani b \\
10 Tauri & 16852 & 1.96$\pm$1.14 & 1.2 & (1.64, 2.88) & (1.30, 3.04) & -- \\
$\delta$ Eridani & 17378 & 1.25$\pm$0.78 & 1.1 & (1.70, 3.07) & (1.34, 3.23) & -- \\
$\tau^6$ Eridani & 17651 & 2.67$\pm$1.32 & 2.3 & (2.03, 3.51) & (1.60, 3.71) & -- \\
HD 25457 & 18859 & 0.21$\pm$0.78 & 0.0 & (1.31, 2.29) & (1.04, 2.41) & -- \\
50 Persei & 19335 & 0.38$\pm$0.82 & 0.2 & (1.39, 2.42) & (1.10, 2.55) & -- \\
40 Eridani A & 19849 & 1.16$\pm$0.35 & 2.7 & (0.63, 1.14) & (0.50, 1.20) & CC: 40 Eridani B/C \\
$\beta$ Hydri & 2021 & 1.87$\pm$1.55 & 0.2 & (1.77, 3.12) & (1.40, 3.29) & -- \\
58 Eridani & 22263 & 0.18$\pm$0.29 & 0.2 & (0.93, 1.64) & (0.74, 1.73) & -- \\
$\pi^3$ Orionis & 22449 & 0.88$\pm$0.83 & 0.6 & (1.50, 2.61) & (1.19, 2.76) & -- \\
HD 32147 & 23311 & 0.07$\pm$0.11 & 0.2 & (0.54, 0.98) & (0.43, 1.04) & -- \\
$\zeta$ Doradus A & 23693 & 1.26$\pm$0.51 & 2.1 & (1.12, 1.97) & (0.89, 2.07) & IC: $\zeta$ Doradus B \\
104 Tauri & 23835 & 2.07$\pm$0.79 & 2.0 & (1.48, 2.62) & (1.17, 2.76) & -- \\
$\lambda$ Aurigae & 24813 & 0.31$\pm$0.82 & 0.2 & (1.25, 2.19) & (0.98, 2.31) & -- \\
HD 33564 & 25110 & 1.10$\pm$0.72 & 1.0 & (1.66, 2.90) & (1.31, 3.06) & -- \\
111 Tauri A & 25278 & 0.57$\pm$0.72 & 0.4 & (1.24, 2.17) & (0.98, 2.29) & -- \\
$\pi$ Mensae & 26394 & 4.34$\pm$0.56 & 7.5 & (1.16, 2.04) & (0.92, 2.15) & CC: $\pi$ Mensae b \\
HD 37394 & 26779 & 0.12$\pm$0.18 & 0.3 & (0.68, 1.22) & (0.54, 1.28) & -- \\
$\gamma$ Leporis & 27072 & 1.62$\pm$0.61 & 2.4 & (1.41, 2.46) & (1.11, 2.59) & CC: AK Leporis \\
HD 38858 & 27435 & 0.20$\pm$0.23 & 0.4 & (0.87, 1.53) & (0.68, 1.61) & -- \\
$\alpha$ Mensae A & 29271 & 0.38$\pm$0.39 & 0.6 & (0.89, 1.58) & (0.71, 1.67) & -- \\
71 Orionis & 29650 & 4.26$\pm$2.23 & 1.5 & (1.56, 2.70) & (1.23, 2.85) & -- \\
74 Orionis & 29800 & 1.52$\pm$0.99 & 1.1 & (1.57, 2.72) & (1.24, 2.87) & -- \\
54 Piscium A & 3093 & 0.43$\pm$0.37 & 1.1 & (0.72, 1.29) & (0.57, 1.36) & -- \\
HD 46588 & 32439 & 0.88$\pm$0.69 & 0.8 & (1.25, 2.19) & (0.99, 2.31) & -- \\
$\psi^5$ Aurigae & 32480 & 0.27$\pm$0.70 & 0.1 & (1.26, 2.21) & (1.00, 2.33) & -- \\
HD 50281 A & 32984 & 0.39$\pm$0.13 & 2.6 & (0.47, 0.86) & (0.37, 0.91) & CC: HD 50281 B \\
37 Geminorum & 33277 & 0.33$\pm$0.50 & 0.3 & (1.07, 1.88) & (0.85, 1.99) & -- \\
HD 53705 & 34065 & 1.74$\pm$0.60 & 2.4 & (1.16, 2.04) & (0.92, 2.16) & CC: HD 53706 \\
HD 55575 & 35136 & 0.70$\pm$0.35 & 1.8 & (1.14, 2.01) & (0.90, 2.12) & -- \\
HD 4391 & 3583 & 0.56$\pm$0.23 & 1.9 & (0.91, 1.60) & (0.72, 1.68) & -- \\
22 Lyncis & 36439 & 1.85$\pm$0.95 & 1.7 & (1.44, 2.50) & (1.13, 2.64) & -- \\
HD 4628 & 3765 & 0.21$\pm$0.29 & 0.4 & (0.54, 0.98) & (0.43, 1.04) & -- \\
$\eta$ Cassiopeiae A & 3821 & 13.61$\pm$0.44 & 25.5 & (1.06, 1.85) & (0.83, 1.96) & CC: $\eta$ Cassiopeiae B \\
212 Puppis A & 38423 & 1.74$\pm$0.67 & 2.1 & (1.45, 2.51) & (1.14, 2.65) & CC: 212 Puppis B \\
HD 65907 & 38908 & 0.78$\pm$0.41 & 1.7 & (1.07, 1.87) & (0.84, 1.97) & -- \\
$\phi^2$ Ceti & 3909 & 0.42$\pm$0.88 & 0.1 & (1.24, 2.16) & (0.98, 2.28) & -- \\
HD 69830 & 40693 & 0.33$\pm$0.24 & 1.0 & (0.76, 1.35) & (0.60, 1.42) & -- \\
$\chi$ Cancri & 40843 & 0.45$\pm$0.88 & 0.1 & (1.44, 2.51) & (1.13, 2.64) & -- \\
HD 5015 & 4151 & 3.16$\pm$0.97 & 2.7 & (1.73, 3.02) & (1.36, 3.19) & -- \\
HD 72673 & 41926 & 0.10$\pm$0.13 & 0.3 & (0.62, 1.11) & (0.49, 1.18) & -- \\
$\pi^1$ Ursae Majoris & 42438 & 0.24$\pm$0.37 & 0.3 & (0.93, 1.64) & (0.74, 1.73) & -- \\
HD 74576 & 42808 & 0.24$\pm$0.10 & 1.8 & (0.56, 1.01) & (0.44, 1.07) & -- \\
55 Cancri A & 43587 & 0.35$\pm$0.30 & 0.8 & (0.78, 1.39) & (0.62, 1.47) & -- \\
HD 76151 & 43726 & 0.57$\pm$0.31 & 1.8 & (0.94, 1.65) & (0.74, 1.74) & -- \\
HD 78366 & 44897 & 0.62$\pm$0.33 & 1.1 & (1.06, 1.85) & (0.83, 1.95) & -- \\
$\sigma^2$ Ursae Majoris A & 45038 & 9.18$\pm$0.85 & 11.5 & (1.86, 3.23) & (1.46, 3.41) & CC: $\sigma^2$ Ursae Majoris B \\
11 Leonis Minoris A & 47080 & 6.20$\pm$0.41 & 22.1 & (0.85, 1.52) & (0.67, 1.60) & CC: 11 Leonis Minoris B \\
HD 84117 & 47592 & 3.24$\pm$1.14 & 2.2 & (1.31, 2.28) & (1.03, 2.41) & -- \\
15 Leonis Minoris & 48113 & 1.80$\pm$0.58 & 2.5 & (1.58, 2.77) & (1.25, 2.93) & -- \\
20 Leonis Minoris A & 49081 & 0.53$\pm$0.67 & 0.3 & (1.12, 1.97) & (0.88, 2.08) & -- \\
HD 88230 & 49908 & 0.11$\pm$0.06 & 1.5 & (0.33, 0.62) & (0.26, 0.65) & -- \\
40 Leonis & 50564 & 3.80$\pm$1.74 & 3.1 & (1.88, 3.28) & (1.49, 3.46) & -- \\
I Carinae & 50954 & 3.54$\pm$1.36 & 2.2 & (2.02, 3.50) & (1.60, 3.69) & -- \\
36 Ursae Majoris A & 51459 & 0.86$\pm$0.49 & 1.4 & (1.18, 2.06) & (0.93, 2.17) & -- \\
HD 90089 A & 51502 & 99.87$\pm$11.15 & 10.9 & (1.62, 2.80) & (1.28, 2.96) & IC: HD 90089 B \\
HD 91324 & 51523 & 0.56$\pm$0.82 & 0.3 & (1.95, 3.41) & (1.54, 3.59) & -- \\
47 Ursae Majoris & 53721 & 1.08$\pm$0.82 & 0.9 & (1.19, 2.09) & (0.94, 2.20) & -- \\
Lalande 21185 & 54035 & 0.10$\pm$0.04 & 2.2 & (0.15, 0.28) & (0.12, 0.30) & CC: Lalande 21185 c, IC: Lalande 21185 b \\
HD 166 & 544 & 0.46$\pm$0.26 & 1.2 & (0.78, 1.38) & (0.61, 1.46) & -- \\
20 Crateris A & 56452 & 3.02$\pm$0.17 & 15.9 & (0.60, 1.07) & (0.47, 1.13) & CC: VB 04 \\
61 Ursae Majoris & 56997 & 0.18$\pm$0.59 & 0.0 & (0.75, 1.34) & (0.60, 1.41) & -- \\
HD 102365 A & 57443 & 1.27$\pm$0.36 & 3.1 & (0.88, 1.56) & (0.70, 1.65) & CC: HD 102365 B, IC: HD 102365 A b \\
$\beta$ Virginis & 57757 & 2.54$\pm$1.00 & 2.3 & (1.76, 3.08) & (1.39, 3.25) & -- \\
HD 103095 & 57939 & 0.11$\pm$0.10 & 0.7 & (0.46, 0.82) & (0.36, 0.87) & -- \\
$\nu$ Phoenicis & 5862 & 0.32$\pm$0.44 & 0.4 & (1.32, 2.31) & (1.04, 2.43) & -- \\
$\kappa$ Tucanae A & 5896 & 148.25$\pm$15.78 & 35.4 & (1.80, 3.14) & (1.42, 3.31) & IC: $\kappa$ Tucanae B \\
$\alpha$ Corvi A & 59199 & 7.40$\pm$1.38 & 5.5 & (1.87, 3.23) & (1.48, 3.41) & CC: $\alpha$ Corvi B \\
$\eta$ Corvi & 61174 & 2.61$\pm$1.56 & 0.9 & (1.92, 3.32) & (1.51, 3.50) & -- \\
$\beta$ Canum Venaticorum & 61317 & 0.83$\pm$0.56 & 1.0 & (1.01, 1.78) & (0.80, 1.88) & -- \\
10 Canum Venaticorum & 62207 & 0.51$\pm$0.19 & 2.4 & (0.99, 1.74) & (0.78, 1.83) & -- \\
$\beta$ Comae Berenices & 64394 & 1.14$\pm$0.53 & 1.6 & (1.09, 1.92) & (0.86, 2.02) & -- \\
HD 114613 & 64408 & 2.15$\pm$1.02 & 1.5 & (1.96, 3.47) & (1.55, 3.66) & -- \\
HD 114837 A & 64583 & 5.91$\pm$0.63 & 8.5 & (1.59, 2.78) & (1.26, 2.93) & CC: HD 114837 B \\
HD 115404 A & 64797 & 10.32$\pm$0.21 & 46.9 & (0.55, 1.00) & (0.43, 1.06) & CC: HD 115404 B, CC: HD 115404 A c \\
61 Virginis & 64924 & 1.31$\pm$0.61 & 1.4 & (0.88, 1.55) & (0.69, 1.64) & -- \\
HD 122064 & 68184 & 0.13$\pm$0.12 & 0.6 & (0.54, 0.98) & (0.43, 1.03) & -- \\
HD 125276 A & 69965 & 10.76$\pm$0.44 & 30.0 & (1.02, 1.79) & (0.81, 1.89) & CC: HD 125276 B \\
$\theta$ Bootis A & 70497 & 1.58$\pm$1.27 & 0.6 & (1.85, 3.23) & (1.46, 3.40) & -- \\
$\sigma$ Bootis & 71284 & 0.66$\pm$0.84 & 0.4 & (1.60, 2.77) & (1.26, 2.92) & -- \\
$\xi$ Bootis A & 72659 & 50.05$\pm$0.46 & 107.2 & (0.72, 1.28) & (0.57, 1.35) & CC: $\xi$ Bootis B \\
GJ 570 A & 73184 & 4.29$\pm$0.22 & 18.7 & (0.48, 0.87) & (0.38, 0.92) & CC: GJ 570 B/C \\
45 Bootis & 73996 & 0.66$\pm$0.94 & 0.4 & (1.65, 2.87) & (1.30, 3.02) & -- \\
$\upsilon$ Andromedae A & 7513 & 4.32$\pm$1.83 & 2.0 & (1.70, 2.97) & (1.34, 3.14) & CC: $\upsilon$ Andromedae d, IC: $\upsilon$ Andromedae D, IC: $\upsilon$ Andromedae c \\
$\nu^2$ Lupi & 75181 & 0.96$\pm$0.57 & 1.5 & (0.97, 1.71) & (0.77, 1.81) & -- \\
$\psi$ Serpentis A & 77052 & 24.04$\pm$0.43 & 70.4 & (0.87, 1.54) & (0.69, 1.63) & CC: $\psi$ Serpentis Ba/Bb \\
$\lambda$ Serpentis & 77257 & 1.06$\pm$1.79 & 0.6 & (1.34, 2.35) & (1.05, 2.48) & -- \\
HD 140901 A & 77358 & 1.36$\pm$0.31 & 3.9 & (0.87, 1.54) & (0.69, 1.62) & CC: HD 140901 B, CC: HD 140901 c \\
p Eridani A & 7751 & 15.21$\pm$0.48 & 32.7 & (0.58, 1.05) & (0.46, 1.11) & CC: p Eridani B \\
$\chi$ Herculis & 77760 & 0.76$\pm$0.70 & 0.6 & (1.68, 2.96) & (1.33, 3.12) & -- \\
$\gamma$ Serpentis & 78072 & 1.04$\pm$1.15 & 0.4 & (1.61, 2.80) & (1.27, 2.96) & -- \\
$\rho$ Coronae Borealis & 78459 & 0.48$\pm$0.41 & 0.7 & (1.28, 2.25) & (1.01, 2.37) & -- \\
18 Scorpii & 79672 & 0.78$\pm$0.51 & 1.0 & (0.99, 1.75) & (0.78, 1.85) & -- \\
$\rm{q}^1$ Eridani & 7978 & 0.84$\pm$0.34 & 2.0 & (1.16, 2.02) & (0.91, 2.13) & CC: $\rm{q}^1$ Eridani b \\
107 Piscium & 7981 & 0.86$\pm$0.46 & 1.3 & (0.66, 1.19) & (0.52, 1.25) & -- \\
HD 147513 & 80337 & 1.13$\pm$0.56 & 1.6 & (0.95, 1.66) & (0.75, 1.76) & -- \\
$\tau$ Ceti & 8102 & 0.28$\pm$0.23 & 0.9 & (0.69, 1.24) & (0.55, 1.30) & -- \\
12 Ophiuchi & 81300 & 0.37$\pm$0.24 & 1.1 & (0.67, 1.19) & (0.53, 1.26) & -- \\
HD 10780 & 8362 & 0.37$\pm$0.20 & 1.8 & (0.70, 1.25) & (0.55, 1.32) & -- \\
36 Ophiuchi A & 84405 & 1.21$\pm$0.24 & 5.0 & (0.56, 1.02) & (0.45, 1.07) & CC: 36 Ophiuchi B \\
36 Ophiuchi C & 84478 & 0.35$\pm$0.11 & 2.6 & (0.40, 0.74) & (0.32, 0.79) & IC: 36 Ophiuchi A/B \\
41 Arae A & 84720 & 12.15$\pm$0.41 & 28.7 & (0.66, 1.18) & (0.52, 1.25) & CC: 41 Arae Ba/Bb \\
w Herculis & 84862 & 0.45$\pm$0.41 & 0.6 & (1.08, 1.92) & (0.86, 2.02) & -- \\
$\xi$ Ophiuchi A & 84893 & 10.89$\pm$1.44 & 8.8 & (1.82, 3.16) & (1.44, 3.33) & CC: $\xi$ Ophiuchi B \\
HD 158633 & 85235 & 0.15$\pm$0.20 & 0.3 & (0.63, 1.12) & (0.50, 1.18) & -- \\
$\lambda$ Arae & 86486 & 1.27$\pm$1.19 & 0.5 & (1.93, 3.34) & (1.52, 3.53) & -- \\
58 Ophiuchi & 86736 & 1.04$\pm$1.14 & 0.4 & (1.50, 2.62) & (1.19, 2.76) & -- \\
$\mu$ Arae & 86796 & 0.91$\pm$0.57 & 1.0 & (1.31, 2.31) & (1.04, 2.44) & -- \\
HD 165185 & 88694 & 0.78$\pm$0.33 & 1.7 & (0.96, 1.70) & (0.76, 1.79) & -- \\
HD 166620 & 88972 & 0.17$\pm$0.12 & 0.9 & (0.60, 1.08) & (0.47, 1.14) & -- \\
$\iota$ Pavonis & 89042 & 21.82$\pm$1.56 & 13.7 & (1.24, 2.17) & (0.98, 2.29) & -- \\
36 Draconis A & 89348 & 5.12$\pm$1.57 & 2.8 & (1.86, 3.24) & (1.47, 3.41) & CC: 36 Draconis B \\
6 Ceti & 910 & 0.11$\pm$1.14 & 0.0 & (1.60, 2.80) & (1.27, 2.96) & -- \\
$\theta$ Sculptoris & 950 & 53.47$\pm$0.83 & 83.4 & (1.58, 2.75) & (1.25, 2.90) & -- \\
31 Aquilae & 95447 & 0.58$\pm$0.43 & 1.2 & (1.26, 2.23) & (0.99, 2.35) & -- \\
$\sigma$ Draconis & 96100 & 0.37$\pm$0.31 & 0.7 & (0.64, 1.15) & (0.51, 1.22) & -- \\
17 Cygni A & 97295 & 0.73$\pm$0.67 & 0.6 & (1.72, 2.98) & (1.35, 3.15) & -- \\
$o$ Aquilae A & 97675 & 1.69$\pm$0.68 & 2.0 & (1.57, 2.74) & (1.24, 2.89) & -- \\
GJ 777 A & 98767 & 1.16$\pm$0.28 & 3.4 & (1.04, 1.84) & (0.82, 1.95) & CC: GJ 777 Ab, IC: GJ 777 B \\
HD 189567 & 98959 & 0.11$\pm$0.19 & 0.2 & (0.97, 1.71) & (0.77, 1.81) & -- \\
$\delta$ Pavonis & 99240 & 0.51$\pm$0.27 & 1.3 & (1.08, 1.91) & (0.85, 2.01) & -- \\
HD 191408 A & 99461 & 0.75$\pm$0.25 & 2.6 & (0.53, 0.96) & (0.42, 1.01) & CC: HD 191408 B \\
HD 192310 & 99825 & 0.31$\pm$0.17 & 1.5 & (0.63, 1.13) & (0.50, 1.19) & -- \\
\enddata
\tablenotetext{a}{Magnitude of the HGCA astrometric acceleration in the plane of the sky.}
\tablenotetext{b}{In reference to causing the acceleration, CC=Consistent Companion and IC=Inconsistent Companion.}
\end{deluxetable*}
\end{longrotatetable}

\section{Results}
\label{sec:Results}

\begin{figure*}
    \centering   
    \includegraphics[width=6.8in]{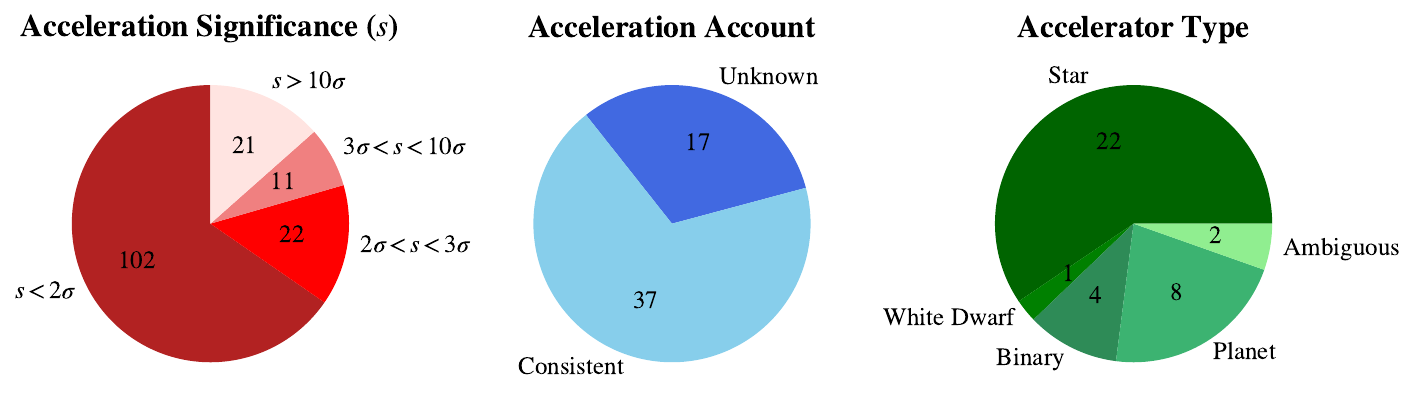}
    \caption{Left: Significance of acceleration ($s$) for all HWO targets in the HGCA. Middle: Of the 54 accelerating HWO targets, 37 have companions that are consistent with the acceleration and 17 have unknown accelerators. Right: Of the 37 accelerations that are consistent with a known source, 22 accelerations are caused by a stellar companion, 1 is caused by a white dwarf, 4 are caused by a binary, 8 are caused by a planet, and 2 have ambiguous accelerators.}
    \label{fig:Figure_4}
\end{figure*}

There are 54 HWO targets with statistically significant accelerations from the HGCA, or about 35\% of all of the HWO targets in the HGCA. Mass-separation predictions for these accelerating stars are displayed along with the locations of known companions in Figure \ref{fig:Figure_8}. Predictions of the companion properties causing the accelerations reach down to giant planet masses at orbital separations of a few to tens of AU in many cases, but are also consistent with stars and brown dwarfs at shorter periods and, similarly, at wider separations. Constraints on mass and separation for those HWO targets without significant accelerations are also displayed in Figure \ref{fig:Figure_8}. 

Here we provide a breakdown of the systems that have significant astrometric accelerations and previously known companions, although these known companions are not necessarily the origin of the acceleration because they may be too distant or too low in mass. 
\begin{itemize}
\item 44 of the accelerating targets have known companions.
\begin{itemize}
\item 36 accelerating stars have companions which are more massive than planets and have been imaged: 28 of these are apparently single (that is, they are in non-hierarchical configurations), 2 are single white dwarf companions, and 6 are companions which are themselves binaries. These include 3 M-dwarf binaries (GJ 570 B/C, $\psi$ Serpentis Ba/Bb, 41 Arae Ba/Bb), 1 K-dwarf binary (36 Ophiuchi A/B), 1 brown dwarf binary ($\varepsilon$ Indi Ba/Bb) and 1 white dwarf+M-dwarf binary (40 Eridani B/C). The total masses of binary companions to the accelerating stars (but not necessarily responsible for the acceleration) range from 0.021 to 1.7 $M_{\odot}$, and their separations range from 7 AU to 12000 AU. 
\item 2 accelerating targets are known spectroscopic binaries ($\tau^1$ Eridani and $\theta$ Sculptoris).
\item 12 accelerating targets have known planets, which were all discovered with radial velocities. They have minimum masses that range from 2.7 $M_{\oplus}$ to 10.3 $M_\mathrm{Jup}$ and semi-major axes that range from 0.038 AU to 28.4 AU.
\end{itemize}
\item 10 accelerating HWO targets have no confirmed companions, 2 of which are disputed binaries (104 Tauri and $\iota$ Pavonis).\footnote{7 of these have significance levels between 2 and 2.5$\sigma$, suggesting that they could be false positives.} The other 8 without previously confirmed companions are $\tau^6$ Eridani, HD 5015, HD 84117, 15 Leonis Minoris, 40 Leonis, I Carinae, $\beta$ Virginis, 10 Canum Venaticorum, $\iota$ Pavonis. These companions inferred here from the astrometric accelerations could have masses that fall anywhere from giant planets up to the stellar regime.
\end{itemize}

Among the 44 stars with known companions, 37 have companions which are classified as being ``consistent'' with the measured accelerations following the definitions in Section \ref{sec:known-companions}. 
\begin{itemize}
\item 21 targets have (apparently single) stellar companions causing the acceleration.
\item 1 target ($\tau^1$ Eridani) has a stellar companion inferred from radial velocities and it is consistent with the measured Hipparcos-Gaia astrometric acceleration, as detailed in Section \ref{sec:individual-targets}.
\item 1 target has a white dwarf causing the acceleration (20 Crateris A).
\item 4 have companions in hierarchical orbital configurations causing the acceleration (1 white-dwarf+M-dwarf binary---40 Eridani B/C---and 3 M-dwarf binaries---41 Arae Ba/Bb, GJ 570 B/C, and Psi Serpentis Ba/Bb). 
\item Another 8 targets have known planets causing the acceleration ($\epsilon$ Indi A, $\epsilon$ Eridani, $\rm{q}^1$ Eridani, $\pi$ Mensae, Lalande 21185, GJ 777 A, $\upsilon$ Andromedae A, and
HD 219134). 
\item Finally, for two cases (HD 140901 and HD 115404), the measured acceleration is consistent with both a known giant planet and a stellar companion. In these systems, the acceleration cannot be uniquely attributed to one or the other.
\end{itemize}
Of the 26 systems whose accelerations are caused by spatially resolved imaged single stars, close (unresolved) binary stars, or white dwarfs, the total masses of the companions range from 0.17 $M_{\odot}$ to 0.98 $M_{\odot}$, and their projected separations range from 29 AU to 936 AU. Note that there is one pair in the HWO target list that comprises two wide companions, each resolved as a separate entry in HGCA, and which are accelerating each other (61 Cygni A and B). The planets which are consistent with causing accelerations have minimum masses that range from 13.7 $M_{\oplus}$ (Lalande 21185 c) to 9.8 $M_\mathrm{Jup}$ ($\pi$ Mensae b), and semi-major axes that range from 2.0 AU to 28.4 AU.

There are 6 targets for which known companions are ``inconsistent'' with the measured acceleration. Thus, further work is needed to find and confirm companions for 17 unaccounted-for accelerations. The full breakdown of accelerating stars with and without consistent companions is displayed in Figure \ref{fig:Figure_4} and listed in Table \ref{tab:HWO_targets}. The significance breakdown of the 17 unaccounted-for accelerations is as follows: 11 are between 2-3$\sigma$, 1 is between 3-10$\sigma$, and 5 are greater than $10\sigma$. Thus, 65$\%$ of the unaccounted-for accelerations are detected at low significance levels between 2-3$\sigma$. By contrast, only 41$\%$ of the broader HWO precursor target list with 
HGCA accelerations have significance levels between 2-3$\sigma$. Therefore, the false positive rate may be somewhat higher among targets with unaccounted-for accelerations, but only by a modest amount; a difference of 24\% would correspond to an expectation value of 2-3 stars out of 11. We performed a common proper-motion companion search in Gaia DR3 out to 1000 AU for each of the 17 targets with unaccounted-for accelerations, limiting the parallax of potential companions to within 10\% of the target's parallax. We did not identify any previously unknown wide binary companions. This suggests that the companions causing these accelerations are likely close-in stellar companions, substellar companions, or giant planets.

In the Gaia catalog \citep{Gaia_Colloboration_23}, the Renormalised Unit Weight Error (RUWE) can be used to identify non-single stars based on whether the data is well-fit by a model of constant proper motion. A RUWE value greater than 1.4 signifies that the star may be non-single, and possibly has a close companion. A low RUWE value likely rules out close stellar companions. 42 of the HWO targets have elevated RUWE values, and 21 of these have known companions. Among the 17 unaccounted-for accelerations, 7 have RUWE above 1.4: $\kappa$ Tucanae A, HD 90089 A, $\iota$ Pavonis, $\beta$ Virginis, HD 212330 A, $\tau^6$ Eridani, and I Carinae. Thus, for these systems, the source of the acceleration is likely a close companion. The 17 unaccounted-for accelerating targets are listed in Table \ref{tab:ufa}, along with their RUWE values and the possible nature of the acceleration.

\startlongtable
\begin{deluxetable*}{lcccccll}
\renewcommand\arraystretch{0.9}
\tabletypesize{\footnotesize}
\setlength{ \tabcolsep } {.2cm} 
\tablewidth{0pt}
\tablecolumns{11}
\tablecaption{HWO Preliminary Targets with Astrometric Accelerations of Unknown Nature\label{tab:ufa}}
\tablehead{
 \colhead{Name} & \colhead{HIP ID} & \colhead{$a_{\alpha\delta}$\tablenotemark{a}} &  \colhead{Significance} & \colhead{SpT} & \colhead{Distance}  & \colhead{RUWE} & \colhead{Notes on Source of Acceleration\tablenotemark{b}}  \\
 &  & \colhead{(m/s/yr)} &  \colhead{($n_\sigma$)} & & \colhead{(pc)}  & }
\startdata
HD 212330 A & 110649 & 83.17$\pm$0.72 & 130.3 & G2IV-V & 20.3 & 2.4 & Close stellar companion \\
GJ 887 & 114046 & 0.11$\pm$0.04 & 2.6 & M1.0V & 3.3 & 0.9 & Possible giant planet \\
$\tau^6$ Eridani & 17651 & 2.67$\pm$1.32 & 2.3 & F5IV-V & 17.8 & 2.1 & Possible close giant planet \\
$\zeta$ Doradus A & 23693 & 1.26$\pm$0.51 & 2.1 & F9V Fe-0.5 & 11.7 & 0.9 & Possible giant planet \\
104 Tauri & 23835 & 2.07$\pm$0.79 & 2.0 & G1V & 15.9 & 0.9 & Possible giant planet \\
HD 5015 & 4151 & 3.16$\pm$0.97 & 2.7 & F8V & 18.8 & 1.1 & Possible giant planet \\
HD 84117 & 47592 & 3.24$\pm$1.14 & 2.2 & F9V & 15.0 & 0.9 & Possible giant planet \\
15 Leonis Minoris & 48113 & 1.80$\pm$0.58 & 2.5 & G0(V) & 18.8 & 0.8 & Possible giant planet \\
40 Leonis & 50564 & 3.80$\pm$1.74 & 3.1 & F6IV-V & 21.2 & 1.1 & Possible giant planet \\
I Carinae & 50954 & 3.54$\pm$1.36 & 2.2 & F3V & 16.2 & 2.0 & Possible close giant planet \\
HD 90089 A & 51502 & 99.87$\pm$11.15 & 10.9 & F4VkF2mF2 & 22.7 & 9.9 & Close stellar companion \\
$\beta$ Virginis & 57757 & 2.54$\pm$1.00 & 2.3 & F9V & 10.9 & 2.8 & Possible close giant planet \\
$\kappa$ Tucanae A & 5896 & 148.25$\pm$15.78 & 35.4 & F5V & 23.3 & 10.4 & Close stellar companion \\
10 Canum Venaticorum & 62207 & 0.51$\pm$0.19 & 2.4 & F9V Fe-0.3 & 17.6 & 0.8 & Possible giant planet \\
36 Ophiuchi C & 84478 & 0.35$\pm$0.11 & 2.6 & K5V(k) & 6.0 & 0.7 & Possible giant planet \\
$\iota$ Pavonis & 89042 & 21.82$\pm$1.56 & 13.7 & G0V & 17.8 & 4.8 & Possible close brown dwarf \\
$\theta$ Sculptoris & 950 & 53.47$\pm$0.83 & 83.4 & F5V & 21.7 & 1.2 & Stellar companion \\
\enddata
\tablenotetext{a}{Magnitude of the HGCA astrometric acceleration in the plane of the sky.}
\tablenotetext{b}{Here, the source of acceleration is attributed to a stellar companion if the mass-separation curve falls within the stellar regime. ``Possible'' giant planets or brown dwarfs correspond to mass-separation curves that reach below the mass limit for these objects, respectively.}
\end{deluxetable*}

\begin{figure*}
    \centering   
    \includegraphics[width=6.8in]{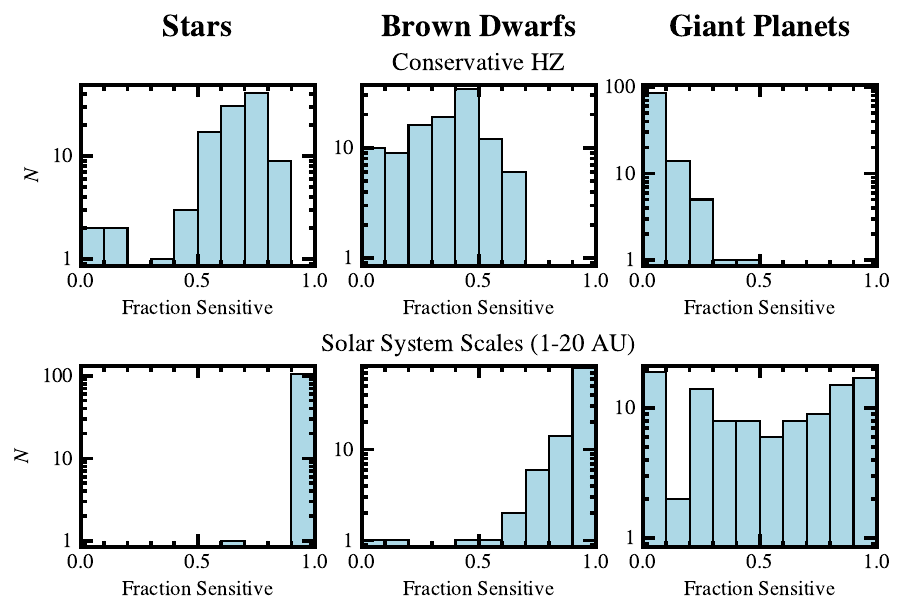}
    \caption{Fraction of parameter space (in companion mass and separation) that the HGCA method is sensitive to, to within at least 99.7$\%$ confidence. Each panel is calculated in different mass ranges---stellar ($m_2>75~M_\mathrm{Jup}$), brown dwarf ($13~M_\mathrm{Jup}<m_2<75~M_\mathrm{Jup}$), and giant planet ($5~M_\mathrm{Jup}<m_2<13~M_\mathrm{Jup}$)---and semi-major axis ranges---CHZ and 1--20 AU---for all non-accelerating targets. Hipparcos-Gaia accelerations are sensitive to most stellar companions orbiting within the CHZ with typical sensitivities of 50--90\%. For brown dwarfs, the sensitivity range spans 0--70\%, and for planets, it is typically less than 30\%. On solar system scales (1--20 AU), the HGCA method is almost 100\% sensitive to stellar companions typically spans 60--100\%. For planets, the sensitivity on solar system scales ranges from 0--100\%.}
    \label{fig:Figure_5}
\end{figure*}

Of the 156 HWO targets in the HGCA, there are 102 which do not have statistically significant accelerations. Figure \ref{fig:Figure_8} displays the region of companion mass and separation space where companions can be ruled out based on the non-detection of a long-term acceleration (within the gray shaded region) and, conversely, where companions could reside (outside the gray region). We use these constraints to evaluate the sensitivity of our method to companions at different masses and separations. Assuming the ``3$\sigma$'' upper limit is a strict threshold differentiating regions where a companion would and would not have been detected, we can estimate the fraction of parameter space for which Hipparcos-Gaia astrometric accelerations are sensitive in various mass-separation boxes. This is determined for each non-accelerating target in the stellar ($m_2>75~M_\mathrm{Jup}$), brown dwarf ($13~M_\mathrm{Jup}<m_2<75~M_\mathrm{Jup}$), and giant planet ($5~M_\mathrm{Jup}<m_2<13~M_\mathrm{Jup}$) regime over two different separation ranges: the conservative HZ and ``Solar-System scales'' (1-20 AU), totaling 6 boxes. Note that we use a narrow range of giant planet masses for this exercise, as the Hipparcos-Gaia accelerations are not typically sensitive to Jupiter- and Saturn-mass planets, at least with the current Gaia DR3 data release. The resulting fractions are displayed in histograms in Figure \ref{fig:Figure_5} and are averaged together in Table \ref{tab:sensitivity}. Hipparcos-Gaia accelerations are nearly completely sensitive to stellar masses across the 1-20 AU range (97\%) and about 66$\%$ sensitive in the CHZ, on average. For brown dwarf masses, they are almost 90$\%$ sensitive over 1--20 AU and about 36$\%$ sensitive in the CHZ. They are about 51$\%$ sensitive to the giant planet mass range over 1--20 AU, but only 5$\%$ sensitive to giant planets in the CHZ.

\begin{figure*}
    \centering   
    \includegraphics[width=6.8in]{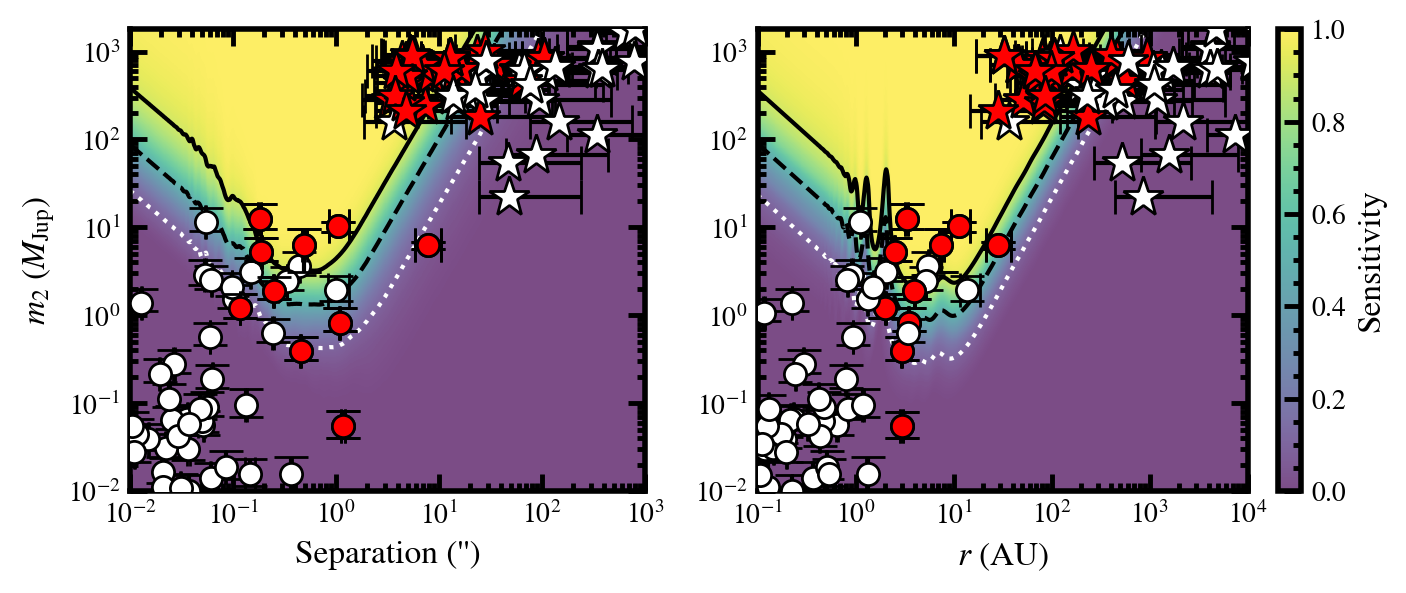}
    \caption{Average sensitivity of HGCA astrometric acceleration method to companions across mass and separation space, calculated from non-significant accelerations. 90$\%$, 50$\%$, and 10$\%$ contours are shown. All known companions of HWO targets in the HGCA are plotted. Red circles and stars represent planets and stars, respectively, that account for their host star's acceleration; white circles and stars represent planets and stars that do not account for the acceleration.}
    \label{fig:Figure_6}
\end{figure*}

A map of the typical sensitivity of astrometric accelerations over mass and separation can be generated by assembling constraints from the 102 non-detections in our sample. Rather than assuming a hard cutoff where companions are ruled out at the 3$\sigma$ level, we instead generate more nuanced sensitivity maps for each individual system by interpolating 1, 2, 3, 4 and 5$\sigma$ upper limits as a function of $m_2$ at each separation. These sensitivity maps are then calculated over a grid of $m_2$ and separation for all of the non-accelerating HWO targets; the average of the 102 grids is shown in the left panel of Figure \ref{fig:Figure_6}. This is repeated for physical separations ($r$) in the right panel of Figure \ref{fig:Figure_6}. The known companions of all HWO targets in the HGCA are plotted on these sensitivity maps, with red markers representing companions that account for their host star's acceleration and white markers representing those that do not. There are two broad clusters---the wide stellar companions and known close-in giant planets. The close binaries were intentionally removed from the sample by \cite{Mamajek_Stapelfeldt_23} from dynamical and practical considerations, and modern detection methods generally lack the combination of precision and baseline required to detect planets beyond a few AU. In Figure \ref{fig:Figure_6}, we also include the 90$\%$, 50$\%$, and 10$\%$ sensitivity contours for comparison. Note that the undetected (white) star that falls within the 90$\%$ contour is $\alpha$ Mensae B, and the undetected (white) planet that falls within the 90$\%$ contour is 55 Cancri A d.\footnote{In fact, both of these companions fall within the 97$\%$ confidence interval, as shown in Figure \ref{fig:Figure_8}.} For $\alpha$ Mensae B, this may be because of the poorly constrained semi-major axis; our statistical conversion from projected separation to true orbital separation allows for the stellar companion to fall outside the 3$\sigma$ astrometric sensitivity limits. The non-detection of 55 Cancri A d is less clear, though it seems to fall very close to the 3$\sigma$ sensitivity limit and may involve the statistical correction from $m_p\mathrm{sin}i$ (minimum mass) to $m_p$ (true mass) adopted for these plots. Figure \ref{fig:Figure_6} shows that the HGCA method is $\sim$85$\%$ sensitive to $2~M_\mathrm{Jup}$ planets between 4 and 10 AU. The sensitivity to a certain mass decreases at both shorter and longer semi-major axes. For instance, the method is only $\sim$17$\%$ sensitive to $2~M_\mathrm{Jup}$ companions at 1 AU and $\sim$40$\%$ sensitive to $2~M_\mathrm{Jup}$ companions at 20 AU.

\begin{deluxetable}{lrrr}
\renewcommand\arraystretch{0.9}
\tabletypesize{\footnotesize}
\setlength{ \tabcolsep } {.2cm} 
\tablewidth{0pt}
\tablecolumns{11}
\tablecaption{Average fraction of parameter space that HGCA is sensitive to with 99.7$\%$ confidence, calculated in 6 boxes of companion mass and separation.\label{tab:sensitivity}}
\tablehead{
 \colhead{Range} & \colhead{Stars} & \colhead{Brown Dwarfs} &  \colhead{Giant Planets}\\}
\startdata
CHZ & 0.66 & 0.36 & 0.05 \\
1--20 AU & 0.97 & 0.90 & 0.51 \\
\enddata
\end{deluxetable}

\section{Discussion}
\label{sec:Discussion}
In this study, we used astrometric accelerations from Hipparcos and Gaia to probe a previously unreached population of planets around our nearest Sun-like neighbors. One of the benefits of Hipparcos-Gaia accelerations is that they provide information both about companions that are accelerating their host stars and what is not there: non-detections can immediately be used to rule out companions in large regions of mass and separation. For the HWO stars, we found that approximately two-thirds of the targets do not have stellar companions or brown dwarfs between 4--10 AU, the characteristic semi-major axis range that the HGCA method accesses. There is of course a strong selection function avoiding binaries in this range \citep{Mamajek_Stapelfeldt_23}, but this fraction also aligns with previous estimates of the single-star fraction in the Galaxy \citep{Lada_06}. 

In addition to stellar and brown dwarf companions, the acceleration method is sensitive to some giant planets between 4--10 AU. Using the average sensitivity of 0.51 to giant planets ($5~M_\mathrm{Jup}<m_2<13~M_\mathrm{Jup}$) between 1--20 AU, we can estimate the occurrence rate of giant planets on wide orbits from this sample. 3 known planets ($\pi$ Mensae b, HD 140901 c, and HD 115404 A c) are recovered by accelerations in this region, and 6 unaccounted-for accelerations may be caused by planets in this regime. If the accelerations in these systems are indeed caused by giant planets, approximately 18 stars in the HWO sample would be expected to host a giant planet in this range based on the sensitivity of the method, corresponding to an occurrence rate of about $12\pm3\%$. This rough estimate is comparable to previous measurements of giant planet occurrence rates beyond the snow line. For instance, \cite{Nielsen_19} found the frequency of giant planets between 5--13 $M_\mathrm{Jup}$ around 1.5--5 $M_\odot$ stars at semi-major axes 10--100 AU to be $8.9\substack{+5.0 \\ -3.6}\%$. \cite{Fulton_21} determined the frequency of 30--6000 $M_\oplus$ planets around FGKM stars to be $14.1\substack{+2.0 \\ -1.8}\%$ between 2--8 AU and $8.9\substack{+3.0 \\ -2.4}\%$ between 8--32 AU. This suggests that the demographics of giant planets in the HWO preliminary target sample are broadly similar to those of Sun-like stars in the solar neighborhood ($\lesssim$100 pc).

\begin{figure*}
    \centering   
    \includegraphics[width=6.8in]{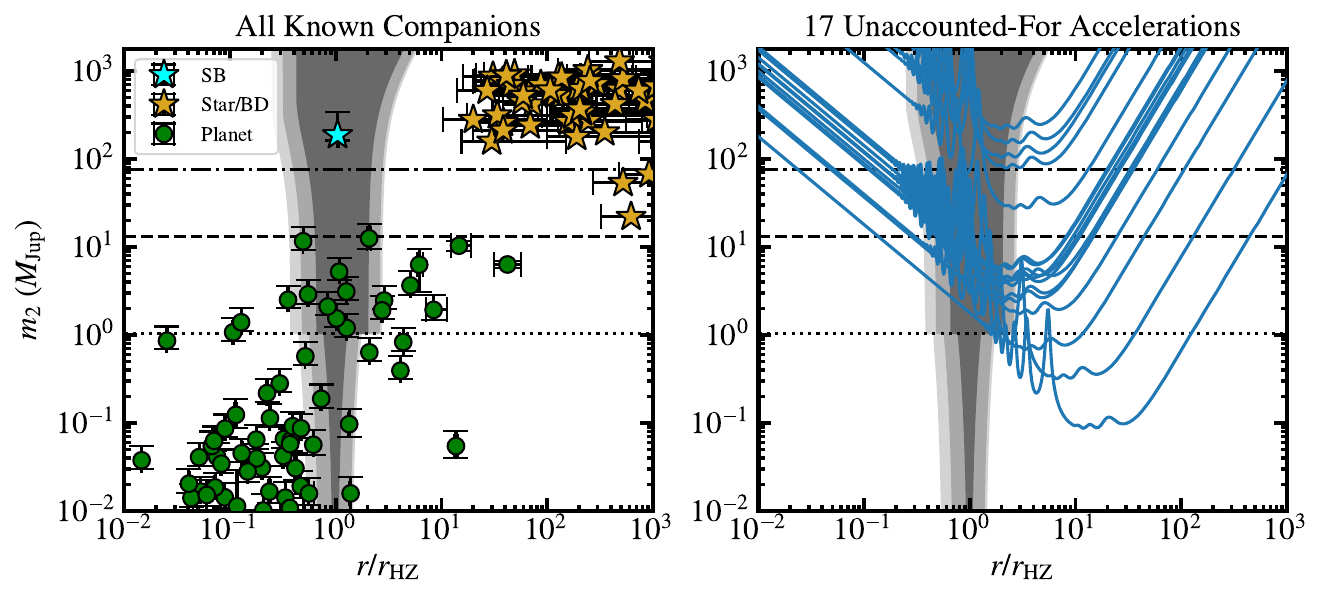}
    \caption{Left: Dynamical stability analysis of habitable zones. Known planets (green circles) and companions in the stellar/brown dwarf mass-range (yellow stars) are shown relative to the region of physical instability for an Earth-mass planet situated in the HZ (gray contours). The separations are normalized to the center of the habitable zone in each system. The stability limits are calculated per \cite{Quarles_18} and \cite{Quarles_20} above 1 $M_\mathrm{Jup}$ (dotted line) and using the Hill stability criterion below. The dashed-dotted line delineates the stellar-brown dwarf boundary and the dashed line delineates the brown dwarf-planet boundary. The dark gray shading represents the region where a second companion in the system would cause an Earth-mass planet located directly in the middle of HZ to be unstable. The medium gray shading represents the region where a companion would threaten an Earth-mass planet in some part of the CHZ. The light gray shading represents the region where a companion would threaten an Earth-mass planet in some portion of the OHZ. Note that the cyan star is attributed to a spectroscopic binary in the $\tau^1$ Eridani system detailed in Section \ref{sec:individual-targets}. Right: Constraints on systems with accelerations from previously unknown companions. The same instability contours are shown along with the mass-separation curves for the 17 unaccounted-for accelerations.}
    \label{fig:Figure_7}
\end{figure*}

This work has important implications for HWO and the potential to find small habitable-zone planets---and, by extension, signatures of life---orbiting these stars. We found that approximately one-third of these targets possess massive companions which are likely to impact the properties of other planets in each system, including Earth analogs, whether through formation or long-term dynamics. About half of these astrometric signals are attributed to known stellar companions with semi-major axes greater than $\sim$30 AU. (One acceleration is caused by a close-in stellar companion found by spectroscopic data.) Approximately one-sixth are attributed to known giant planets with semi-major axes spanning $\sim$2--30 AU. The remaining 17 unaccounted-for accelerations can be divided into groups based on their accelerating companions' likely masses. We find that 4 of these must be caused by a stellar companion from their mass-separation curves. One system likely has a previously unknown brown dwarf at a separation between 3--20 AU, but it could be a stellar companion with a smaller or larger orbit. The remaining 12 accelerations could be caused by planets with masses ranging from sub-Saturn to super-Jupiter. 

It is going to be important to understand how these unknown companions, along with all known companions, may impact the HZ of these systems, specifically the dynamical stability of an Earth-sized planet in the HZ. While detailed dynamical simulations for these systems are beyond the scope of this work, it is possible to estimate the general regions of dynamical instability in companion mass and separation based on analytical arguments. Here, we use stability limits found by \cite{Quarles_18} and \cite{Quarles_20} for binaries sampled down to a mass ratio of 0.001. For lower-mass planets, we employ the Hill stability criterion \citep{Marchal_82}, which describes the region in which the planet's gravity dominates over the star's gravity. Note that these relations assume circular orbits and coplanar planets, so this assessment will not capture non-zero orbital eccentricities or planetary orbits with mutual inclinations.

These stability limits are displayed in Figure \ref{fig:Figure_7} along with all known companions of the HWO targets in the HGCA and the mass-separation curves for the 17 unaccounted-for accelerations. It is immediately evident that no known binary companions are threatening to any region of the HZ, other than one spectroscopic binary companion ($\tau^1$ Eridani B) detailed in Section \ref{sec:individual-targets}. However, 15 known planets do fall within the region that would affect the stability of an Earth-mass planet in some portion of the HZ. The systems with planets which may be disruptive to Earth-mass planets in the HZ are $\iota$ Horologii (HD 17501), HD 33564, $\pi$ Mensae (HD 39091), HD 69830, 55 Cancri A (HD 75732 A), 47 Ursae Majoris (HD 95128), $\upsilon$ Andromedae (HD 9826 A), $q^1$ Eridani (HD 10647), HD 147513, $\tau$ Ceti (HD 10700), $\mu$ Arae (HD 160691), GJ 777 A (HD 190360), and HD 192310. 

\cite{Kane_24} recently presented the results of detailed dynamical simulations for the HWO preliminary targets with known planets. All of the systems in which we identify Earth-mass planets to be dynamically unstable in the HZ were similarly shown to have low survival rates in \cite{Kane_24}. However, they present an additional two systems with known planets (HD 20794 and HD 115617) for which less than 50$\%$ of the HZ is ``dynamically viable''. Thus, while our evaluations of dynamical stability do not capture the effects of eccentricity and planet interactions, they do encapsulate most of the systems that should be deemed low priority, or entirely avoided, for the HWO mission. 

The right panel of Figure \ref{fig:Figure_7} shows how the potential culprits of the 17 unaccounted-for accelerations could affect Earth-mass planets in the HZ. Because of the mass-separation degeneracy, these companions could exist outside or inside the region of instability. Many of these fall exclusively within the stellar mass regime and are likely to be previously unknown binary systems. Regardless, this representation highlights the urgency to recover and characterize these companions in preparation for HWO.

\subsection{Notes on Individual Targets}
\label{sec:individual-targets}
Here, we provide notes on individual accelerating targets and their companions. We specifically focus on those with known planets causing the acceleration, those with ambiguous acceleration culprits, those with multiple companions, known spectroscopic binaries, and those with plausible culprits for the acceleration. The following are ordered by increasing distance.

Lalande 21185 is a M2V star \citep{Kirkpatrick91} at a distance of 2.55 pc ($\bar{\omega}=392.7529\pm0.0321$ mas; \citealt{Gaia_22}) with $V=7.42$ mag \citep{Hauck_98}. It has two confirmed planets, Lalande 21185 b and c. Lalande 21185 b was revealed with Keck/HIRES RVs by \cite{Butler_17} and its most updated parameters are $M\mathrm{sin}i=0.00846\substack{+0.00060 \\ -0.00057} \ M_\mathrm{Jup}$, or $M\mathrm{sin}i=2.69\substack{+0.19 \\ -0.18} \ M_{\oplus}$, and $a=0.07879\substack{+0.00056 \\ -0.00055}$ AU \citep{Hurt_22}. Lalande 21185 c was confirmed by \cite{Rosenthal_21} and its most updated parameters are $M\mathrm{sin}i=0.0428\substack{+0.0076 \\ -0.0072} \ M_\mathrm{Jup}$ and $a=2.94\substack{+0.14 \\ -0.12}$ AU \citep{Hurt_22}. Lalande 21185 c accounts for the acceleration.

\textbf{\textit{$\varepsilon$ Eridani}} is a K2V star \citep{Keenan_89} at a distance of 3.22 pc ($\bar{\omega}=310.5773\pm0.1355$ mas; \citealt{Gaia_22}) with $V=3.72$ mag \citep{Hauck_98}. It has one known giant planet, $\varepsilon$ Eridani b, initially discovered through long-term precision RVs \citep{Hatzes_00}. Its most updated planet parameters are $M=0.66\substack{+0.12 \\ -0.09} \ M_\mathrm{Jup}$ and $a=3.53\substack{+0.03 \\ -0.04}$ AU, constrained through a combination of RVs, astrometry, and direct imaging \citep{Llop-Sayson_21, Mawet_19}. This planet accounts for the measured Hipparcos-Gaia acceleration.

\textbf{\textit{$\varepsilon$ Indi A}} is a K4V star \citep{Gray_06} at a distance of 3.64 pc ($\bar{\omega}=274.8431\pm0.0956$ mas; \citealt{Gaia_22}) with $V=4.67$ mag \citep{Hauck_98}. It has a binary companion, $\varepsilon$ Indi Ba/Bb, in a hierarchical orbital configuration at a projected separation of 403.11 arcsec ($\sim$1500 AU) \citep{Kirkpatrick_16}. $\varepsilon$ Indi Bab was originally discovered as a single brown dwarf companion \citep{Scholz_03, McCaughrean_04}, and soon after, it was resolved into two brown dwarf companions orbiting each other at a projected separation of 2.1 AU \citep{Volk_03}. $\varepsilon$ Indi Ba and Bb have spectral types of T1-1.5 and T6 and dynamical masses of $66.92\pm0.36 \ M_\mathrm{Jup}$ and $53.25\pm0.29 \ M_\mathrm{Jup}$, respectively \citep{King_10, Chen_22}. This distant companion to $\varepsilon$ Indi A does not account for its acceleration. $\varepsilon$ Indi A also possesses a planet $\varepsilon$ Indi Ab that was detected by \cite{Endl_02} and confirmed by \cite{Feng_19} with astrometry and RVs. This planet was recovered with direct imaging using the James Webb Space Telescope (JWST; \citealt{Matthews_24}) with Mid-InfraRed Instrument (MIRI). Its most updated parameters through the combination of direct imaging, astrometry, and RVs are $6.31\substack{+0.60 \\ -0.56} \ M_\mathrm{Jup}$ and $28.4\substack{+10 \\ -7.2}$ AU. This planet accounts for the measured astrometric acceleration.

\textbf{\textit{40 Eridani A}} is K0.5V star \citep{Gray_06} at a distance of 5.00 pc ($\bar{\omega}=199.608\pm0.1208$ mas; \citealt{Gaia_22}) with $V=4.42$ mag \citep{Hauck_98}. It has two stellar companions---40 Eridani B and C---which themselves form a binary pair ($a=6.930''$ \citep{Mason_17}). These were discovered by \cite{Herschel_1785} and orbit at a projected separation of 82.56$''$ ($\sim$ 410 AU) \citep{Mason_21} from the primary. 40 Eridani B is DA2.9 white dwarf \citep{Gianninas_11}. 40 Eridani C is M4.5V star \citep{Kirkpatrick91}. Their combined mass of 0.7766 $M_{\odot}$ \citep{Mason_17} accounts for the astrometric acceleration.

\textbf{\textit{GJ 570 A}} is a K4V star \citep{Keenan_89} at a distance of 5.89 pc ($\bar{\omega}=169.8843\pm0.0653$ mas; \citealt{Gaia_22}) with $V=5.72$ mag \citep{Hauck_98}. It has two companions---GJ 570 B and C---which form a double-lined spectroscopic binary system ($P= 308.884$ days \citep{Forveille_99}) and orbit at a projected separation of 25.68$''$ ($\sim$ 150 AU) \citep{Iverson_15} from the primary. Their joint spectral type is M1.5V \citep{Keenan_89} and their combined mass is 0.976 $M_{\odot}$ \citep{Forveille_99}. The T dwarf GJ 570 D was later discovered orbiting at 258.3$''$  from the GJ 570ABC system \citep{Burgasser_00}. GJ 570 B/C accounts for the acceleration here.

\textbf{\textit{36 Ophiuchi C}} is a K5V star \citep{Gray_06} at a distance of 5.95 pc ($\bar{\omega}=167.9617\pm0.0311$ mas; \citealt{Gaia_22}) with $V=6.30$ mag \citep{Hauck_98}. It has two companions which form a close binary, 36 Ophiuchi A and B, in a hierarchical configuration at 731.6$''$ ($\sim$ 4400 AU) \citep{Chaname_04} from 36 Ophiuchi C. 36 Ophiuchi A is a K2V star and 36 Ophiuchi B is a K1V star \citep{Torres_06}; they orbit each other at $\sim$5$''$ \citep{Anton_12}. Their combined mass is 1.7 $M_{\odot}$ \citep{Mamajek_Stapelfeldt_23}. However, this pair does not account for the acceleration.

\textbf{\textit{HD 219134}} is a K3V star \citep{Keenan_89} at a distance of 6.54 pc ($\bar{\omega}=152.864\pm0.0494$ mas; \citealt{Gaia_22}) with $V=5.54$ mag \citep{Hauck_98}. There is a complicated history of planet detections and refutations around this star. It has 5 confirmed planets. The four innermost planets, HD 219134 b, c, d, and e, were discovered by \cite{Motalebi_15} using HARPS RVs. Separately, 6 planets were fit to Keck-HIRES RV data by \cite{Vogt_15} which they named HD 219134 b, c, f, d, g, and h. The existence of HD 219134 f was challenged by \cite{Johnson_16}, attributing the signal to stellar activity, but was recovered again by \cite{Gillon_17} and \cite{Rosenthal_21}; however, neither recovered HD 219134 g. \cite{Rosenthal_21} reports minimum masses and semi-major axes of $M\mathrm{sin}i=0.0130\substack{+0.0010 \\ -0.0011} \ M_\mathrm{Jup}$ and $a=0.03838\pm0.00044$ AU for HD 219134 b, $M\mathrm{sin}i=0.0112\pm0.0014 \ M_\mathrm{Jup}$ and $a=0.06466\substack{+0.00074 \\ -0.00073}$ AU for HD 219134 c, $M\mathrm{sin}i=0.0243\substack{+0.0023 \\ -0.0022} \ M_\mathrm{Jup}$ and $a=0.1453\substack{+0.0017 \\ -0.0016}$ AU for HD 219134 f, $M\mathrm{sin}i=0.0516\substack{+0.0032 \\ -0.0030} \ M_\mathrm{Jup}$ and $a=0.2345\pm0.0027$ AU for HD 219134 d, and $M\mathrm{sin}i=0.308\pm0.014 \ M_\mathrm{Jup}$ and $a=2.968\pm0.037$ AU for HD 219134 h. HD 219134 h is consistent with the astrometric acceleration.

\textbf{\textit{41 Arae A}} is G9V star \citep{Corbally_84} at a distance of 8.79 pc ($\bar{\omega}=113.7513\pm0.0726$ mas; \citealt{Gaia_22}) with $V=5.47$ mag \citep{Hauck_98}. Originally, it was known to have one companion, 41 Arae B, which orbits at 10.2$''$ ($\sim$ 90 AU) \citep{Anton_12}. However, Gaia astrometry revealed that that 41 Arae B is actually 41 Arae Ba and Bb \citep{Reyle_22}, close companions which orbit another with a period of 87.91 days. Their joint spectral type is M0V \citep{Corbally_84} and we calculate their combined mass to be 0.58 $M_{\odot}$. This pair accounts for the acceleration.

\textbf{\textit{HD 115404 A}} is a K2.5V star \citep{Gray03} at a distance of 10.99 pc ($\bar{\omega}=91.0176\pm0.0236$ mas; \citealt{Gaia_22}) with $V=6.55$ mag \citep{Perryman_97}. It has a companion, HD 115404 B, with a spectral type of M0.5V \citep{Alonso-Floriano_15} and a projected separation of 7.59$''$ ($\sim$ 83 AU) \citep{Losse_10}. HD 115404 A also has two planets discovered by \cite{Feng_22} which reports $M\mathrm{sin}i=0.0970\substack{+0.0200 \\ -0.0220} \ M_\mathrm{Jup}$ and $a=0.088\substack{+0.003 \\ -0.004}$ AU for HD 115404 Ab and $M\mathrm{sin}i=10.319\substack{+1.473 \\ -1.209} \ M_\mathrm{Jup}$ and $a=11.364\substack{+3.301 \\ -1.905}$ AU for HD 115404 Ac. HD 115404 B and HD 115404 Ac are both consistent with the long-term astrometric acceleration.

\textbf{\textit{$\upsilon$ Andromedae A}} is a F8V star \citep{Gray_01} at a distance of 13.49 pc ($\bar{\omega}=74.12\pm0.19$ mas; \citealt{vanLeeuwen_07}) with $V=4.1$ mag \citep{Perryman_97}. It has a stellar companion with spectral type M4.5V, $\upsilon$ Andromedae B, discovered by \cite{Lowrance_02} at a projected separation of 55.38 arcsec ($\sim$750 AU; \citealt{Daley_11}). $\upsilon$ Andromedae A has three planets, $\upsilon$ Andromedae Ab, c, and d. $\upsilon$ Andromedae Ab was discovered by \cite{Butler_97} using RV data, and soon after $\upsilon$ Andromedae Ac and d were identified \citep{Butler_99}. \cite{Rosenthal_21} reports minimum masses and semi-major axes of $M\mathrm{sin}i=0.675\pm0.016 \ M_\mathrm{Jup}$ and $a=0.05914\substack{+0.00061 \\ -0.00063}$ AU for $\upsilon$ Andromedae Ab, $M\mathrm{sin}i=1.965\pm0.049 \ M_\mathrm{Jup}$ and $a=0.8265\substack{+0.0086 \\ -0.0088}$ AU for $\upsilon$ Andromedae Ac, and $M\mathrm{sin}i=4.1\pm0.1 \ M_\mathrm{Jup}$ and $a=2.517\substack{+0.026 \\ -0.027}$ AU for $\upsilon$ Andromedae Ad. $\upsilon$ Andromedae Ad is in good agreement with the observed astrometric acceleration.

\textbf{\textit{$\tau^1$ Eridani}} is a F7V star \citep{Abt_09} at a distance of 14.28 pc ($\bar{\omega}=70.0459\pm0.1599$ mas; \citealt{Gaia_22}) with $V=4.47$ mag \citep{Mermilliod_97}. It is a single-lined spectroscopic binary with a period of 958 days and a mass function $f$ equal to $1.91 \cdot 10^{-3} M_{\odot}$ \citep{Batten_78}. The equation for $f$ is given by:
\begin{equation}
f = \frac{(m_1\mathrm{sin}i)^3}{(m_1+m_2)^2}
\end{equation}
Since the mass of the primary is estimated by spectroscopy, the minimum mass of the secondary $m_2\mathrm{sin}i$ can by analytically solved for from $f$ using a cubic equation. In this case, $m_2\mathrm{sin}i= 0.1525 M_{\odot}$. However, drawing from an isotropic inclination distribution, in which $P(i) = \mathrm{sin}i$, produces true masses ranging from $\sim$ 0.15 to 
0.30 $M_{\odot}$. The semi-major axis of this unseen companion is estimated to be $\sim2.1$ AU. This makes it consistent with causing the measured acceleration. It also falls within the HZ of this system. This suggests that it is unlikely that an Earth analog would exist in this system, disfavoring this target for HWO.

\textbf{\textit{$\psi$ Serpentis A}} is a G5V star \citep{Gray_01} at a distance of 14.79 pc ($\bar{\omega}=67.6007\pm0.036$ mas; \citealt{Gaia_22}) with $V=5.87$ mag \citep{Hauck_98}. It has two companions, $\psi$ Serpentis Ba and Bb, which form a closer binary system spatially resolved by \cite{Rodriguez_15}. They orbit at a projected separation of 4.53$''$ ($\sim$ 67 AU) \citep{Hirsch_21}. Their joint spectral type is M3V and their combined mass is 0.50 $M_{\odot}$ \citep{Tokovinin_21}. This pair accounts for the astrometric acceleration.

\textbf{\textit{HD 140901 A}} is a G7IV-V star \citep{Gray_06} at a distance of 15.25 pc ($\bar{\omega}=65.5889\pm0.0342$ mas; \citealt{Gaia_22}) with $V=6.01$ mag \citep{Hauck_98}. Its distant companion, HD 140901 B, at a projected separation of 15$''$ ($\sim$ 230 AU) \citep{Zhao_11}, is a DA4.8 white dwarf \citep{Gianninas_11}. Two planets were discovered by \cite{Feng_22}, HD 140901 Ab and c. HD 140901 Ab was not recovered by \cite{Philipot_23} but HD 140901 Ac was. Its parameters are $M\mathrm{sin}i=1.8\pm0.5 \ M_\mathrm{Jup}$ and $a=11.8\substack{+4.1 \\ -2.5}$ AU. HD 140901 B and HD 140901 Ac are both consistent with the acceleration.

\textbf{\textit{GJ 777 A}} is a G7V star \citep{Gray_06} at a distance of 16.00 pc ($\bar{\omega}=62.4865\pm0.0354$ mas; \citealt{Gaia_22}) with $V=5.75$ mag \citep{Hauck_98}. It has a companion with spectral type M4.5V \citep{Alonso-Floriano_15}, GJ 777 B, at a separation of 178.05$''$ (about 3000 AU in projection) \citep{El-Badry_21}. It has two confirmed planets, GJ 777 Ac and b. GJ 777 Ab was discovered using RVs by \cite{Naef_03} and GJ 777 Ac was later found by \cite{Vogt_05}. GJ 777 Ac has $M\mathrm{sin}i=0.0675\pm0.0027 \ M_\mathrm{Jup}$ and $a=0.1294\pm0.0017$ AU and GJ 777 Ab has $M\mathrm{sin}i=1.492\pm0.043 \ M_\mathrm{Jup}$ and $a=3.955\substack{+0.051 \\ -0.053}$ AU \citep{Rosenthal_21}. GJ 777 Ab accounts for the astrometric acceleration.

\textbf{\textit{$\rm{q}^1$ Eridani}} is a F9V star \citep{Keenan_89} at a distance of 17.35 pc ($\bar{\omega}=57.6409\pm0.0453$ mas; \citealt{Gaia_22}) with $V=5.52$ mag \citep{Perryman_97}. It has one known giant planet, $\rm{q}^1$ Eridani b, which was discovered with RVs in \cite{Marmier_13}. Its most updated parameters, from that study, are $M\mathrm{sin}i=0.94\pm0.08 \ M_\mathrm{Jup}$ and $a=2.015\pm0.011$ AU. It accounts for the Hipparcos-Gaia acceleration.

\textbf{\textit{$\pi$ Mensae}} is a G0V star \citep{Gray_06} at a distance of 18.29 pc ($\bar{\omega}=54.6825\pm0.0354$ mas; \citealt{Gaia_22}) with $V=5.67$ mag \citep{Hauck_98}. It has at least two planets and possibly a third. $\pi$ Mensae b was found by \cite{Jones_02} through RVs, and its most updated parameters from a combination of astrometry, RVs, and imaging are $M=12.33\substack{+1.19 \\ -1.38} \ M_\mathrm{Jup}$ and $a=3.31\substack{+0.13 \\ -0.15}$ AU \citep{Feng_22}. $\pi$ Mensae c transits its host star and was discovered by the Transiting Exoplanet Survey Satellite (TESS; \citealt{Huang_18}; \citealt{Gandolfi_18}). Its most updated planet parameters are $M\mathrm{sin}i=3.5\pm0.3 \ M_\mathrm{Jup}$ and $a=0.069\pm0.003$ AU \citep{Feng_22}. \cite{Xuan_20} found that $\pi$ Mensae b and c have a non-zero mutual inclination. $\pi$ Mensae d was discovered by \cite{Hatzes_22} using RVs, was not recovered by \cite{Feng_22}, but then was detected by \cite{Laliotis_23}. Its period is $P=124.64\substack{+0.48 \\ -0.52}$ days and its minimum mass is $M\mathrm{sin}i=0.04210\pm0.00425 \ M_\mathrm{Jup}$. $\pi$ Mensae b accounts for the astrometric acceleration.

\textbf{\textit{$\theta$ Sculptoris}} is a F5V star \citep{Gray_06} at a distance of 21.72 pc ($\bar{\omega}=46.0425\pm0.0961$ mas; \citealt{Gaia_22}) with $V=5.24$ mag \citep{Hauck_98}. It is a single-lined spectroscopic binary with its orbital period only constrained to $179/n$ days and its radial velocity amplitude implies that it possesses a stellar-mass companion \citep{Fuhrmann_17}. From the constraint on orbital period, the maximum separation of this unseen companion is $\sim$1 AU or 0.05$''$ (interior to the HZ of the system). It likely accounts for the acceleration here. While Earth analogs may be present in this system, this target is unfavorable for HWO because starlight suppression will likely be too challenging with a close binary such as this \citep{Mamajek_Stapelfeldt_23}.

\textbf{\textit{$\kappa$ Tucanae A} }is a F5V star \citep{Gray_06} at a distance of 23.26 pc ($\bar{\omega}=42.9912\pm1.0581$ mas; \citealt{Gaia_22}) with $V=4.91$ mag \citep{Kharchenko_01}. It has a companion---$\kappa$ Tucanae B---at a separation of 4.60$''$ ($\sim$ 100 AU) \citep{Anton_19}. $\kappa$ Tucanae B is a K1V star \citep{Corbally_84} with an estimated mass of 0.88 $M_{\odot}$. It lies below the acceleration prediction curve, on the edge of the 4$\sigma$ contour, so it is uncertain whether this companion accounts for the acceleration.

\section{Conclusions}
\label{sec:Conclusions}
In this manuscript, we have presented the analysis of astrometric accelerations of the provisional HWO targets from the HGCA, which provides information about known and unknown stellar, brown dwarf, and giant planet companions. Our findings include:
\begin{itemize}
    \item 156 of the 164 HWO targets are in the HGCA, and 54 of these have significant astrometric accelerations ($>2\sigma$).
    \item 37 of these accelerations are accounted for by known stellar, white dwarf, brown dwarf, and planetary companions. Further study is needed to account for the other 17 accelerations.
    \item We provide constraints on companion mass and separation for HWO targets with and without detected accelerations, narrowing down the possible architectures of the systems and focusing future companion searches.
    \item  We show that the average sensitivity of the HGCA astrometric acceleration method to $2~M_\mathrm{Jup}$ companions between 4 and 10 AU is about 85$\%$. Due to the proximity and brightness of the HWO sample, astrometric accelerations can reach sub-Neptune-mass planets. For example, this method is sensitive to Lalande 21185 c, a planet with $M=13.7~M_\oplus$ at $a=2.94\substack{+0.14 \\ -0.12}$ \citep{Hurt_22}.
\end{itemize}

In addition to the analysis of astrometric accelerations, we carried out preliminary dynamical stability tests for all of the HWO targets with known planets. From this, we found that the following systems have companions which are likely disruptive to Earth-sized planets in the HZ: $\iota$ Horologii (HD 17501), HD 33564, $\pi$ Mensae (HD 39091), HD 69830, 55 Cancri A (HD 75732 A), 47 Ursae Majoris (HD 95128), $\upsilon$ Andromedae (HD 9826 A), $q^1$ Eridani (HD 10647), HD 147513, $\tau$ Ceti (HD 10700), $\mu$ Arae (HD 160691), GJ 777 A (HD 190360), and HD 192310. 

To learn more about the preliminary HWO targets and determine whether their HZs are likely to host Earth analogs, complimentary methods must be used to identify known planets and exclude regions of mass and separation. Precision RVs are most sensitive to shorter period planets, requiring long observational baselines to probe beyond the HZ. High-contrast imaging can probe the outer-most regions down to several AU for most systems. With a time baseline of 5.5 years, Gaia DR4 will fill in the gap between imaging and RVs, enabling acceleration measurements resulting from giant planets at semi-major axes of about 1--6 AU. Combining these constraints will allow a more comprehensive assessment of planets in the HWO systems. These methods will also constrain the orbital properties of known companions. Together, these datasets will be crucial in evaluating the probability that a terrestrial planet in the HZ could be dynamically stable in each of the HWO systems.

\section{Acknowledgements}
\label{sec:Acknowledgements}
K.E.P. and B.P.B. acknowledge support from a NASA
FINESST grant (80NSSC24K1551). B.P.B. acknowledges support from the National Science Foundation grant AST-1909209, NASA Exoplanet Research Program grant 20-XRP20$\_$2-0119, and the Alfred P. Sloan Foundation. This work was carried out in part at the Jet Propulsion Laboratory, California Institute of Technology, under contract 80NM00018D0004 with NASA. This research has made use of data from the NASA Exoplanet Archive, which is operated by the California Institute of Technology, under contract with the National Aeronautics and Space Administration under the Exoplanet Exploration Program \citep{NASA_Exo_Archive_PS}. This research has made use of the Washington Double Star Catalog maintained at the U.S. Naval Observatory. This research has made use of the SIMBAD database,
operated at CDS, Strasbourg, France. This research has made use of the Astrophysics Data System, funded by NASA under Cooperative Agreement 80NSSC21M00561.

\section{Appendix}
\label{sec:Appendix}

\begin{figure*}
  \resizebox{7.1in}{!}{\includegraphics{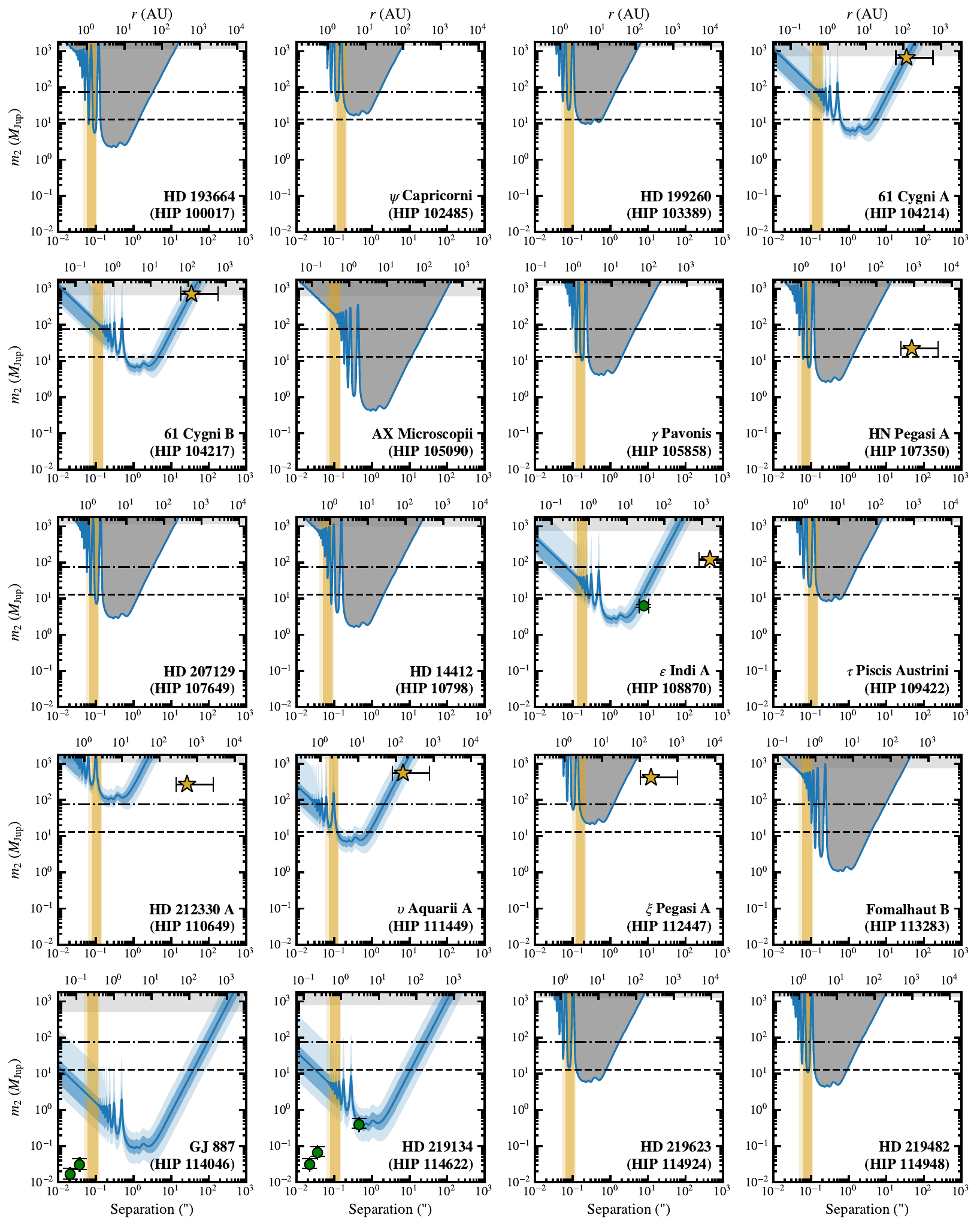}}
    \vskip -.1 in
 \caption{{Companion mass versus separation for all targets in the HGCA as shown in Figure \ref{fig:Figure_2}. Planets are plotted as green circles and stellar, white dwarf, brown dwarf companions as yellow stars. The light grey region at the top of each plot defines the mass range greater than the mass of the host star. 1$\sigma$ errors are shown for the masses of the planets and the separations of all the companions. A companion is consistent with the acceleration if these errors overlap with the 2$\sigma$ confidence interval derived from the astrometry. Note that some companions are outside of the bounds of these plots, and thus not shown, but they do not account for the acceleration in any of these cases.}
 \label{fig:Figure_8} }
 \end{figure*}

We include the astrometric constraints for each system in Figure \ref{fig:Figure_8}, Table \ref{tab:accelerating_data}, and Table \ref{tab:nonaccelerating_data}.

\newpage

\startlongtable
\begin{deluxetable*}{lcccccccc}
\renewcommand\arraystretch{0.9}
\tabletypesize{\footnotesize}
\setlength{ \tabcolsep } {.2cm} 
\tablewidth{0pt}
\tablecolumns{11}
\tablecaption{Joint Mass-Separation Constraints for Stars with Astrometric Accelerations\label{tab:accelerating_data}}
\tablehead{
 \colhead{Name} & \colhead{$\rho$} & \colhead{$-3\sigma$} &  \colhead{$-2\sigma$} & \colhead{$-1\sigma$} & \colhead{$\mu$}  & \colhead{$1\sigma$} & \colhead{$2\sigma$} & \colhead{$3\sigma$}  \\
 \colhead{} & \colhead{$('')$} & \colhead{($M_\odot$)} &  \colhead{($M_\odot$)} & \colhead{($M_\odot$)} & \colhead{($M_\odot$)}  & \colhead{($M_\odot$)} & \colhead{($M_\odot$)} & \colhead{($M_\odot$)}}
\startdata
HIP 104214 & 0.0100 & 0.26 & 0.32 & 0.41 & 0.68 & 0.68 & 0.68 & 0.68 \\
" & 0.0101 & 0.26 & 0.31 & 0.40 & 0.68 & 0.68 & 0.68 & 0.68 \\
" & 0.0102 & 0.25 & 0.31 & 0.40 & 0.68 & 0.68 & 0.68 & 0.68 \\
" & 0.0104 & 0.25 & 0.31 & 0.39 & 0.68 & 0.68 & 0.68 & 0.68 \\
" & 0.0105 & 0.24 & 0.30 & 0.39 & 0.68 & 0.68 & 0.68 & 0.68 \\
" & 0.0106 & 0.24 & 0.30 & 0.38 & 0.68 & 0.68 & 0.68 & 0.68 \\
" & 0.0107 & 0.24 & 0.30 & 0.38 & 0.68 & 0.68 & 0.68 & 0.68 \\
" & 0.0108 & 0.24 & 0.29 & 0.37 & 0.68 & 0.68 & 0.68 & 0.68 \\
" & 0.0110 & 0.23 & 0.29 & 0.37 & 0.68 & 0.68 & 0.68 & 0.68 \\
" & 0.0111 & 0.23 & 0.28 & 0.36 & 0.68 & 0.68 & 0.68 & 0.68 \\
" & 0.0112 & 0.23 & 0.28 & 0.36 & 0.68 & 0.68 & 0.68 & 0.68 \\
" & 0.0114 & 0.22 & 0.28 & 0.36 & 0.68 & 0.68 & 0.68 & 0.68 \\
" & 0.0115 & 0.23 & 0.28 & 0.35 & 0.68 & 0.68 & 0.68 & 0.68 \\
" & 0.0116 & 0.22 & 0.27 & 0.35 & 0.68 & 0.68 & 0.68 & 0.68 \\
" & 0.0118 & 0.22 & 0.27 & 0.34 & 0.68 & 0.68 & 0.68 & 0.68 \\
" & 0.0119 & 0.21 & 0.27 & 0.34 & 0.68 & 0.68 & 0.68 & 0.68 \\
" & 0.0120 & 0.21 & 0.26 & 0.34 & 0.68 & 0.68 & 0.68 & 0.68 \\
" & 0.0122 & 0.21 & 0.26 & 0.34 & 0.68 & 0.68 & 0.68 & 0.68 \\
" & 0.0123 & 0.21 & 0.26 & 0.33 & 0.68 & 0.68 & 0.68 & 0.68 \\
" & 0.0124 & 0.20 & 0.25 & 0.32 & 0.68 & 0.68 & 0.68 & 0.68 \\
" & 0.0126 & 0.20 & 0.25 & 0.32 & 0.68 & 0.68 & 0.68 & 0.68 \\
" & 0.0127 & 0.20 & 0.25 & 0.32 & 0.68 & 0.68 & 0.68 & 0.68 \\
" & 0.0129 & 0.20 & 0.25 & 0.31 & 0.68 & 0.68 & 0.68 & 0.68 \\
" & 0.0130 & 0.20 & 0.24 & 0.31 & 0.68 & 0.68 & 0.68 & 0.68 \\
" & 0.0132 & 0.20 & 0.24 & 0.31 & 0.68 & 0.68 & 0.68 & 0.68 \\
" & 0.0133 & 0.19 & 0.24 & 0.31 & 0.68 & 0.68 & 0.68 & 0.68 \\
" & 0.0135 & 0.19 & 0.24 & 0.30 & 0.68 & 0.68 & 0.68 & 0.68 \\
" & 0.0137 & 0.19 & 0.23 & 0.30 & 0.68 & 0.68 & 0.68 & 0.68 \\
" & 0.0138 & 0.18 & 0.23 & 0.29 & 0.68 & 0.68 & 0.68 & 0.68 \\
" & 0.0140 & 0.18 & 0.22 & 0.29 & 0.68 & 0.68 & 0.68 & 0.68 \\
" & 0.0141 & 0.18 & 0.22 & 0.29 & 0.68 & 0.68 & 0.68 & 0.68 \\
" & 0.0143 & 0.18 & 0.22 & 0.28 & 0.68 & 0.68 & 0.68 & 0.68 \\
" & 0.0145 & 0.18 & 0.22 & 0.28 & 0.68 & 0.68 & 0.68 & 0.68 \\
" & 0.0146 & 0.18 & 0.22 & 0.28 & 0.68 & 0.68 & 0.68 & 0.68 \\
" & 0.0148 & 0.18 & 0.21 & 0.27 & 0.68 & 0.68 & 0.68 & 0.68 \\
" & 0.0150 & 0.17 & 0.21 & 0.27 & 0.68 & 0.68 & 0.68 & 0.68 \\
" & 0.0151 & 0.17 & 0.21 & 0.27 & 0.68 & 0.68 & 0.68 & 0.68 \\
" & 0.0153 & 0.17 & 0.21 & 0.27 & 0.68 & 0.68 & 0.68 & 0.68 \\
" & 0.0155 & 0.17 & 0.20 & 0.26 & 0.68 & 0.68 & 0.68 & 0.68 \\
" & 0.0157 & 0.16 & 0.20 & 0.26 & 0.68 & 0.68 & 0.68 & 0.68 \\
... & ... & ... & ... & ... & ... & ... & ... & ... \\
\enddata
\tablenotetext{}{\textbf{Note:} This table is available in its entirety in machine-readable form.}
\end{deluxetable*}

\newpage

\startlongtable
\begin{deluxetable*}{lllllll}
\renewcommand\arraystretch{0.9}
\tabletypesize{\footnotesize}
\setlength{ \tabcolsep } {.2cm} 
\tablewidth{0pt}
\tablecolumns{11}
\tablecaption{Upper Limits for Stars without Astrometric Accelerations\label{tab:nonaccelerating_data}}
\tablehead{
 \colhead{Name} & \colhead{$\rho$} & \colhead{$1\sigma$} & \colhead{$2\sigma$} & \colhead{$3\sigma$} & \colhead{$4\sigma$} & \colhead{$5\sigma$}  \\
  \colhead{} & \colhead{$('')$} & \colhead{($M_\odot$)} & \colhead{($M_\odot$)} & \colhead{($M_\odot$)} & \colhead{($M_\odot$)} & \colhead{($M_\odot$)}}
\startdata
HIP 100017 & 0.0100 & 0.02 & 0.16 & 1.08 & 1.08 & 1.08 \\
" & 0.0101 & 0.02 & 0.15 & 1.08 & 1.08 & 1.08 \\
" & 0.0102 & 0.02 & 0.15 & 1.08 & 1.08 & 1.08 \\
" & 0.0104 & 0.02 & 0.15 & 1.08 & 1.08 & 1.08 \\
" & 0.0105 & 0.02 & 0.15 & 1.08 & 1.08 & 1.08 \\
" & 0.0106 & 0.02 & 0.15 & 1.08 & 1.08 & 1.08 \\
" & 0.0107 & 0.02 & 0.15 & 1.08 & 1.08 & 1.08 \\
" & 0.0108 & 0.02 & 0.14 & 1.08 & 1.08 & 1.08 \\
" & 0.0110 & 0.02 & 0.14 & 1.08 & 1.08 & 1.08 \\
" & 0.0111 & 0.02 & 0.14 & 1.08 & 1.08 & 1.08 \\
" & 0.0112 & 0.02 & 0.14 & 1.08 & 1.08 & 1.08 \\
" & 0.0114 & 0.02 & 0.14 & 1.08 & 1.08 & 1.08 \\
" & 0.0115 & 0.02 & 0.14 & 1.08 & 1.08 & 1.08 \\
" & 0.0116 & 0.02 & 0.13 & 1.08 & 1.08 & 1.08 \\
" & 0.0118 & 0.02 & 0.13 & 1.08 & 1.08 & 1.08 \\
" & 0.0119 & 0.02 & 0.13 & 1.08 & 1.08 & 1.08 \\
" & 0.0120 & 0.02 & 0.13 & 1.08 & 1.08 & 1.08 \\
" & 0.0122 & 0.02 & 0.13 & 1.08 & 1.08 & 1.08 \\
" & 0.0123 & 0.02 & 0.13 & 1.08 & 1.08 & 1.08 \\
" & 0.0124 & 0.02 & 0.13 & 1.08 & 1.08 & 1.08 \\
" & 0.0126 & 0.02 & 0.12 & 1.08 & 1.08 & 1.08 \\
" & 0.0127 & 0.02 & 0.12 & 1.08 & 1.08 & 1.08 \\
" & 0.0129 & 0.02 & 0.12 & 1.08 & 1.08 & 1.08 \\
" & 0.0130 & 0.02 & 0.12 & 1.08 & 1.08 & 1.08 \\
" & 0.0132 & 0.02 & 0.12 & 1.08 & 1.08 & 1.08 \\
" & 0.0133 & 0.02 & 0.12 & 1.08 & 1.08 & 1.08 \\
" & 0.0135 & 0.02 & 0.12 & 1.08 & 1.08 & 1.08 \\
" & 0.0137 & 0.02 & 0.11 & 1.08 & 1.08 & 1.08 \\
" & 0.0138 & 0.02 & 0.11 & 1.08 & 1.08 & 1.08 \\
" & 0.0140 & 0.02 & 0.11 & 1.08 & 1.08 & 1.08 \\
" & 0.0141 & 0.02 & 0.11 & 1.08 & 1.08 & 1.08 \\
" & 0.0143 & 0.02 & 0.11 & 1.08 & 1.08 & 1.08 \\
" & 0.0145 & 0.02 & 0.11 & 1.08 & 1.08 & 1.08 \\
" & 0.0146 & 0.02 & 0.11 & 1.08 & 1.08 & 1.08 \\
" & 0.0148 & 0.02 & 0.11 & 1.08 & 1.08 & 1.08 \\
" & 0.0150 & 0.02 & 0.10 & 1.08 & 1.08 & 1.08 \\
" & 0.0151 & 0.02 & 0.10 & 1.08 & 1.08 & 1.08 \\
" & 0.0153 & 0.02 & 0.10 & 1.08 & 1.08 & 1.08 \\
" & 0.0155 & 0.02 & 0.10 & 1.08 & 1.08 & 1.08 \\
" & 0.0157 & 0.02 & 0.10 & 1.08 & 1.08 & 1.08 \\
... & ... & ... & ... & ... & ... & ... \\
\enddata
\tablenotetext{}{\textbf{Note:} This table is available in its entirety in machine-readable form.}
\end{deluxetable*}

\renewcommand{\thefigure}{\arabic{figure} (Cont.)}
\addtocounter{figure}{-1}
\begin{figure*}
  \resizebox{7.1in}{!}{\includegraphics{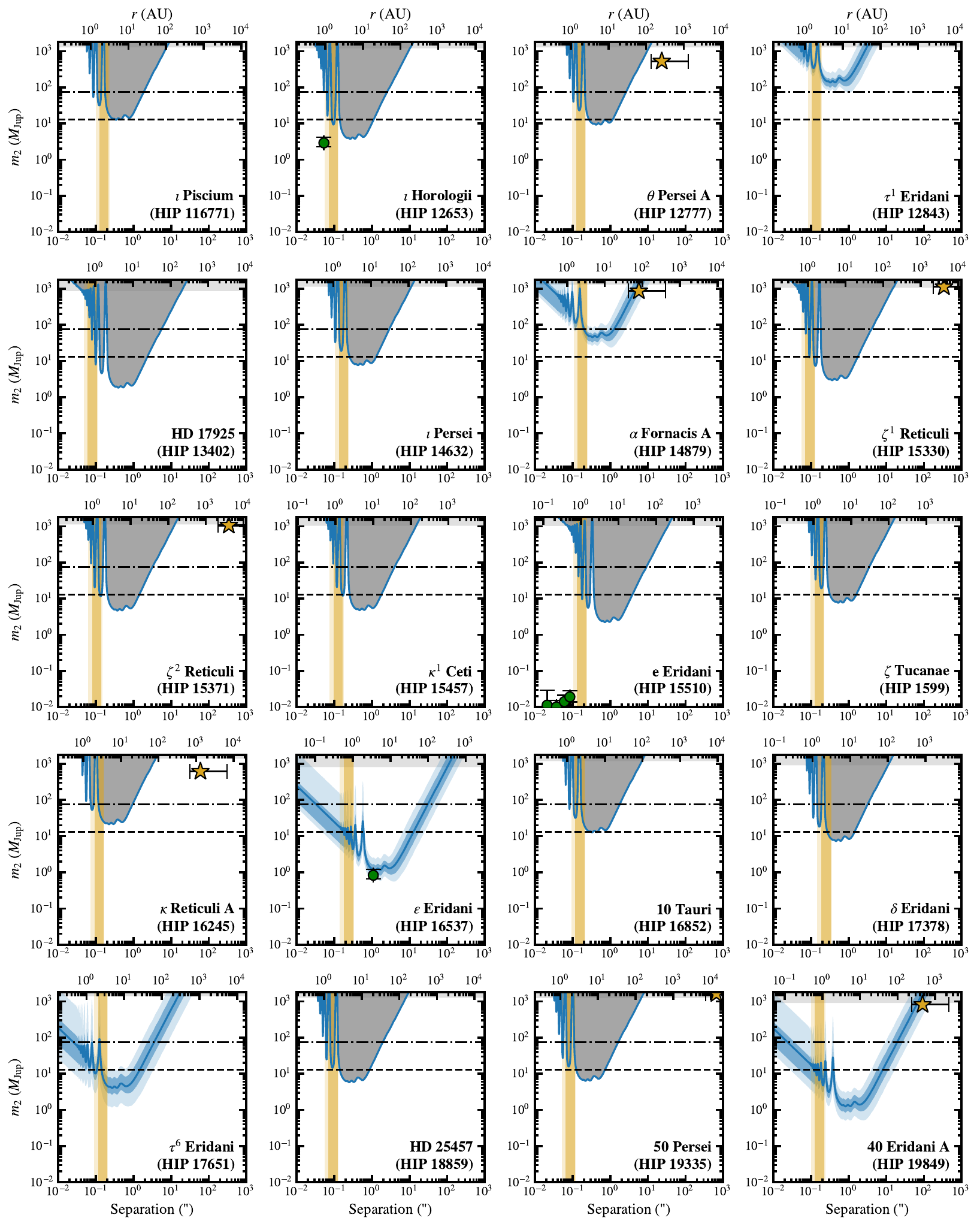}}
 \caption{}
 \end{figure*}
 \renewcommand{\thefigure}{\arabic{figure}}

\renewcommand{\thefigure}{\arabic{figure} (Cont.)}
\addtocounter{figure}{-1}
\begin{figure*}
  \resizebox{7.1in}{!}{\includegraphics{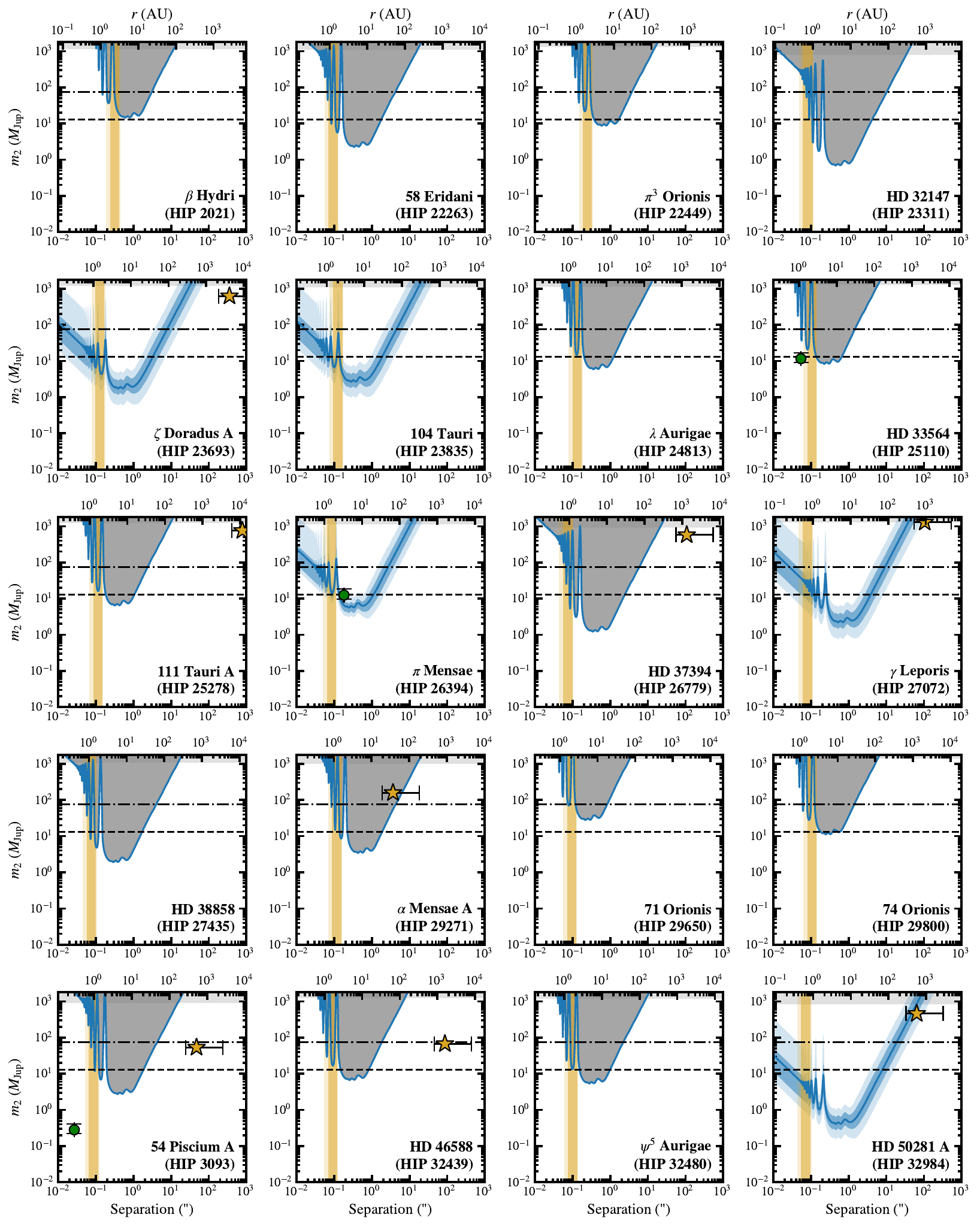}}
 \caption{}
 \end{figure*}
 \renewcommand{\thefigure}{\arabic{figure}}

\renewcommand{\thefigure}{\arabic{figure} (Cont.)}
\addtocounter{figure}{-1}
\begin{figure*}
  \resizebox{7.1in}{!}{\includegraphics{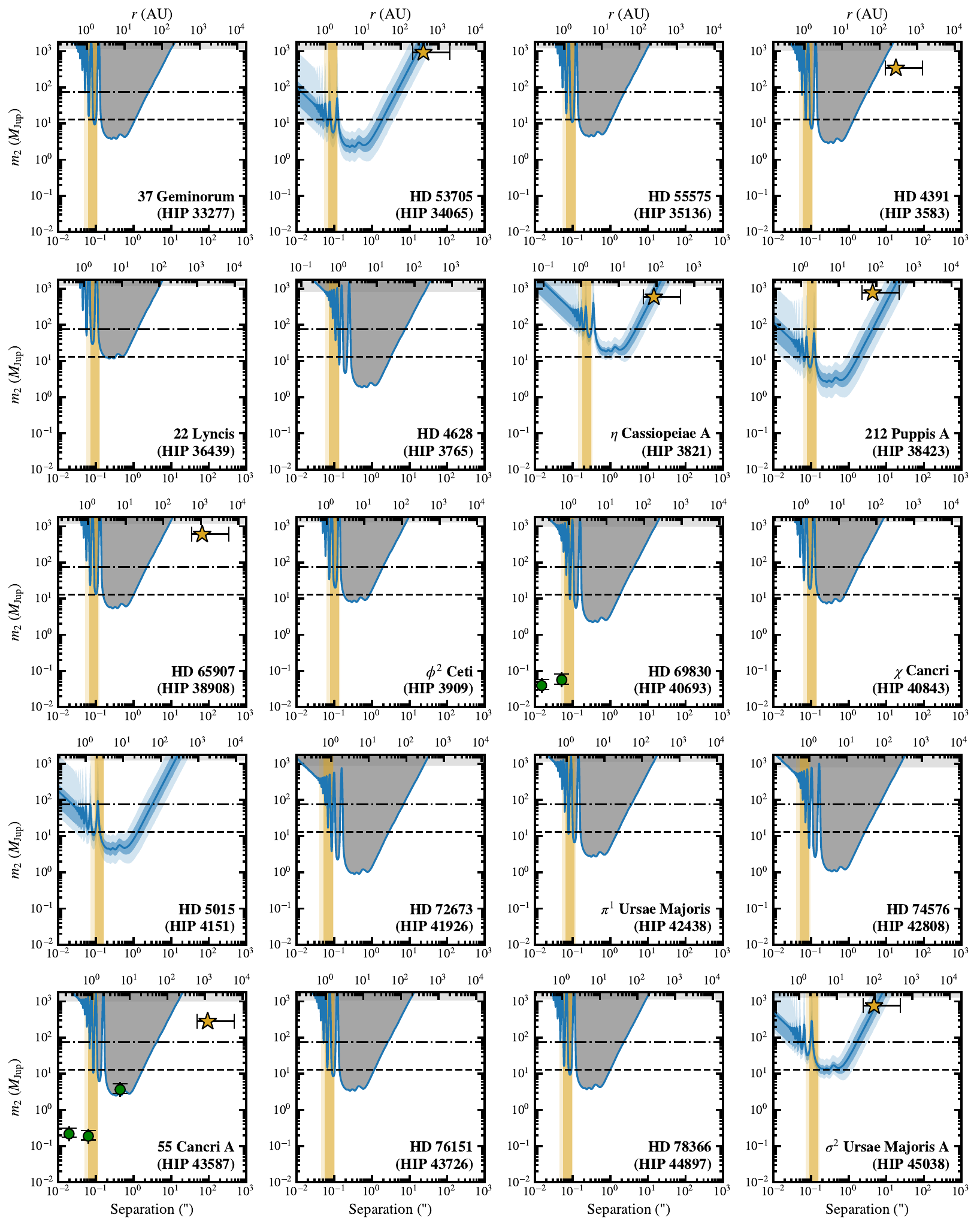}}
 \caption{}
 \end{figure*}
 \renewcommand{\thefigure}{\arabic{figure}}

\renewcommand{\thefigure}{\arabic{figure} (Cont.)}
\addtocounter{figure}{-1}
\begin{figure*}
  \resizebox{7.1in}{!}{\includegraphics{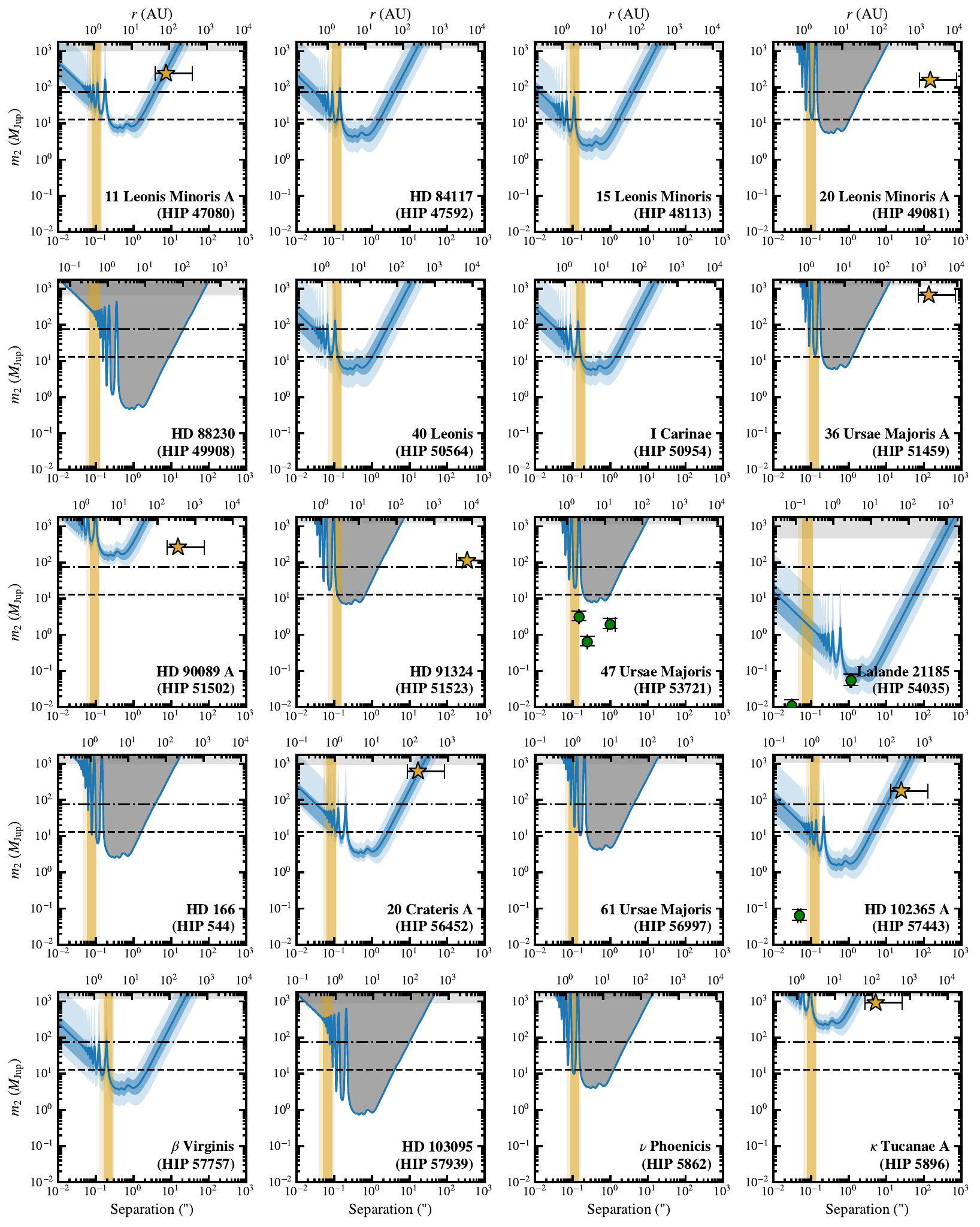}}
 \caption{}
 \end{figure*}
 \renewcommand{\thefigure}{\arabic{figure}}

\renewcommand{\thefigure}{\arabic{figure} (Cont.)}
\addtocounter{figure}{-1}
\begin{figure*}
  \resizebox{7.1in}{!}{\includegraphics{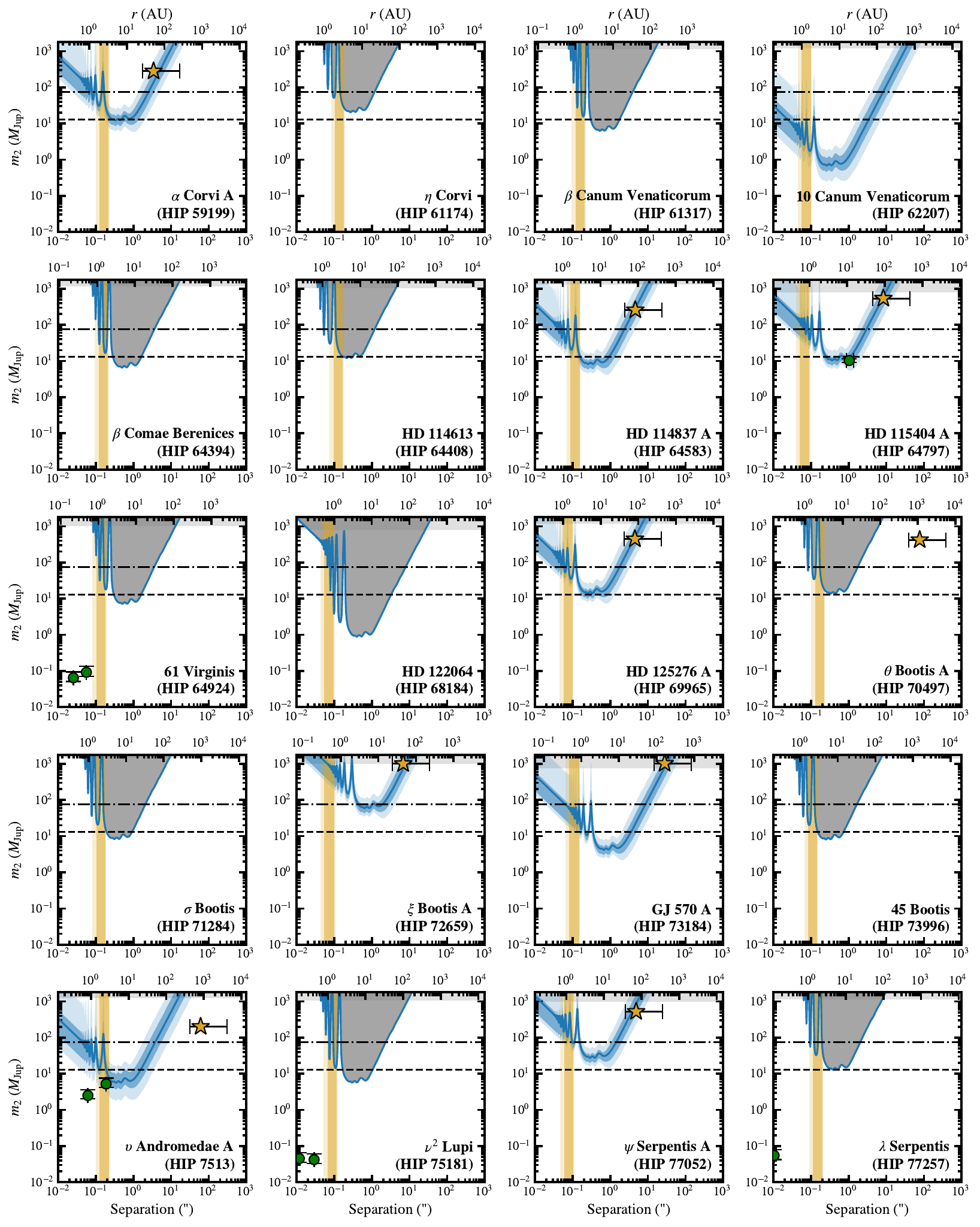}}
 \caption{}
 \end{figure*}
 \renewcommand{\thefigure}{\arabic{figure}}

\renewcommand{\thefigure}{\arabic{figure} (Cont.)}
\addtocounter{figure}{-1}
\begin{figure*}
  \resizebox{7.1in}{!}{\includegraphics{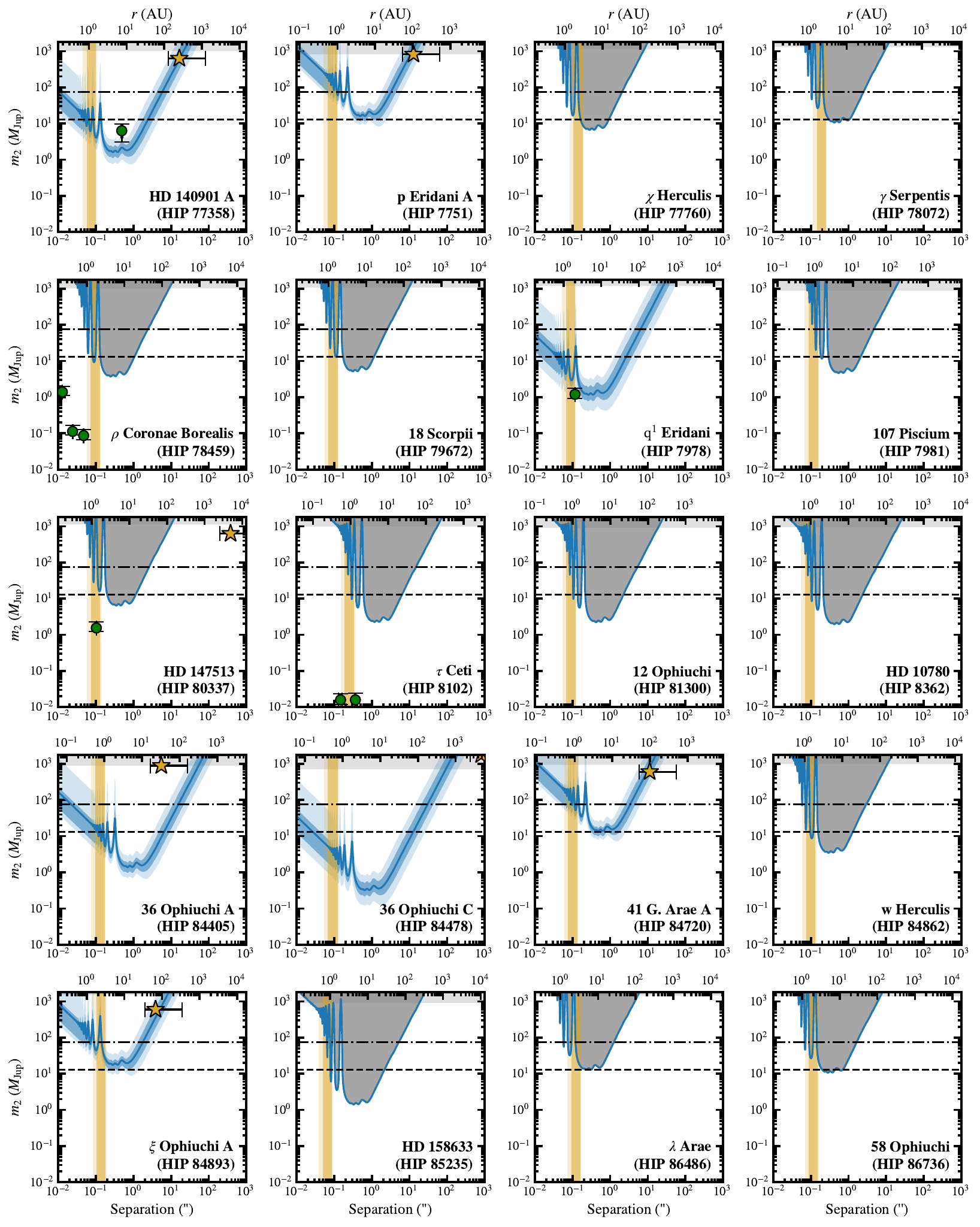}}
 \caption{}
 \end{figure*}
 \renewcommand{\thefigure}{\arabic{figure}}

\renewcommand{\thefigure}{\arabic{figure} (Cont.)}
\addtocounter{figure}{-1}
\begin{figure*}
  \resizebox{7.1in}{!}{\includegraphics{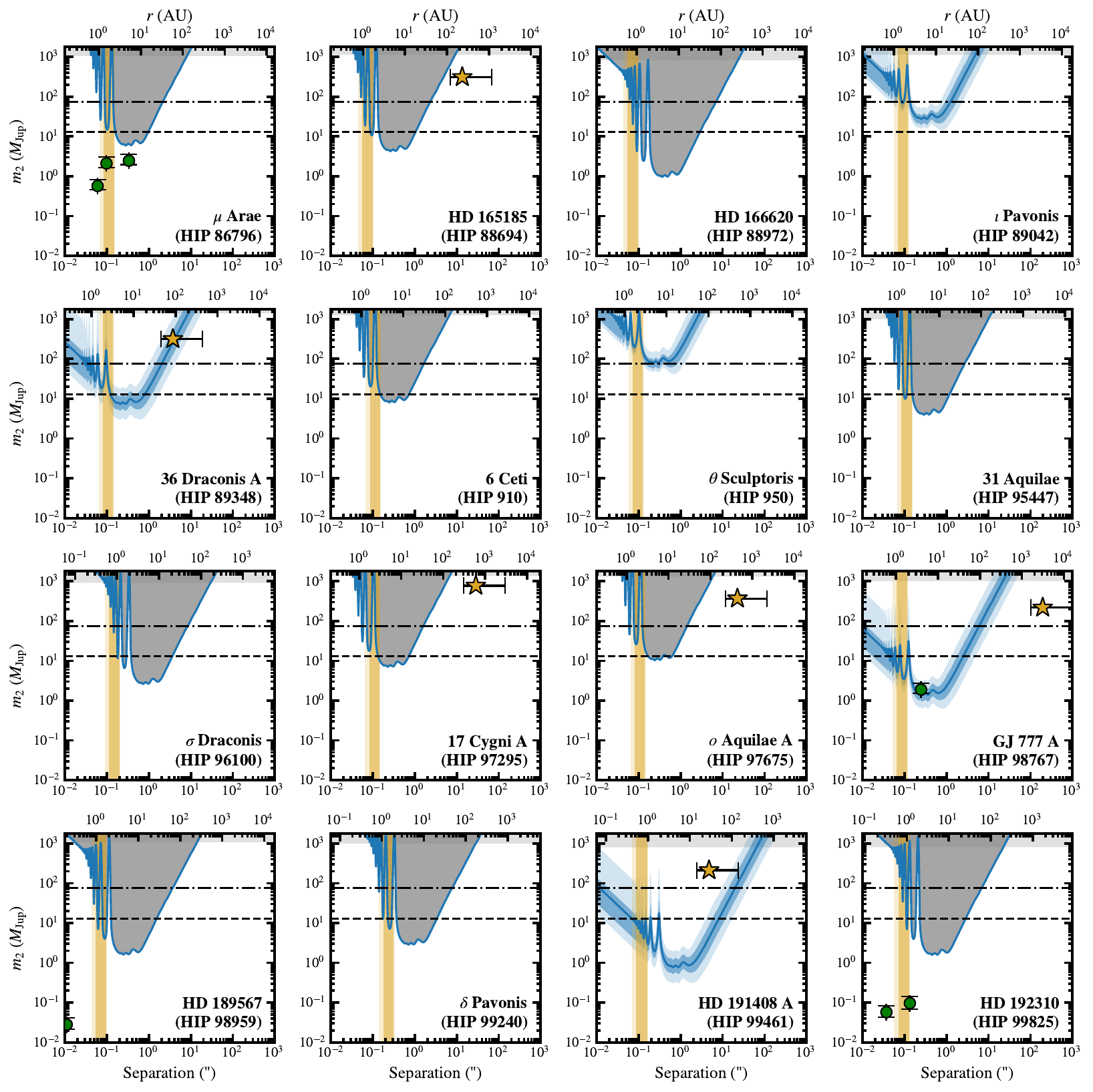}}
 \caption{}
 \end{figure*}
 \renewcommand{\thefigure}{\arabic{figure}}

\clearpage
\newpage

\bibliography{main.bib}

\begin{thebibliography}{}
\expandafter\ifx\csname natexlab\endcsname\relax\def\natexlab#1{#1}\fi
\providecommand{\url}[1]{\href{#1}{#1}}
\providecommand{\dodoi}[1]{doi:~\href{http://doi.org/#1}{\nolinkurl{#1}}}
\providecommand{\doeprint}[1]{\href{http://ascl.net/#1}{\nolinkurl{http://ascl.net/#1}}}
\providecommand{\doarXiv}[1]{\href{https://arxiv.org/abs/#1}{\nolinkurl{https://arxiv.org/abs/#1}}}

\bibitem[{{Abt}(2009)}]{Abt_09}
{Abt}, H.~A. 2009, \apjs, 180, 117, \dodoi{10.1088/0067-0049/180/1/117}

\bibitem[{{Agnew} {et~al.}(2019){Agnew}, {Maddison}, {Horner}, \&
  {Kane}}]{Agnew_19}
{Agnew}, M.~T., {Maddison}, S.~T., {Horner}, J., \& {Kane}, S.~R. 2019, \mnras,
  485, 4703, \dodoi{10.1093/mnras/stz345}

\bibitem[{{Alonso-Floriano} {et~al.}(2015){Alonso-Floriano}, {Morales},
  {Caballero}, {Montes}, {Klutsch}, {Mundt}, {Cort{\'e}s-Contreras}, {Ribas},
  {Reiners}, {Amado}, {Quirrenbach}, \& {Jeffers}}]{Alonso-Floriano_15}
{Alonso-Floriano}, F.~J., {Morales}, J.~C., {Caballero}, J.~A., {et~al.} 2015,
  \aap, 577, A128, \dodoi{10.1051/0004-6361/201525803}

\bibitem[{{Anton}(2012)}]{Anton_12}
{Anton}, R. 2012, Journal of Double Star Observations, 8, 15

\bibitem[{{Anton}(2019)}]{Anton_19}
---. 2019, Journal of Double Star Observations, 15, 336

\bibitem[{{Batten} {et~al.}(1978){Batten}, {Fletcher}, \& {Mann}}]{Batten_78}
{Batten}, A.~H., {Fletcher}, J.~M., \& {Mann}, P.~J. 1978, Publications of the
  Dominion Astrophysical Observatory Victoria, 15, 121

\bibitem[{{Bowler} {et~al.}(2021){Bowler}, {Endl}, {Cochran}, {MacQueen},
  {Crepp}, {Doppmann}, {Dulz}, {Brandt}, {Mirek Brandt}, {Li}, {Dupuy},
  {Franson}, {Kratter}, {Morley}, \& {Zhou}}]{Bowler_21}
{Bowler}, B.~P., {Endl}, M., {Cochran}, W.~D., {et~al.} 2021, \apjl, 913, L26,
  \dodoi{10.3847/2041-8213/abfec8}

\bibitem[{{Brandt}(2018)}]{Brandt_18}
{Brandt}, T.~D. 2018, \apjs, 239, 31, \dodoi{10.3847/1538-4365/aaec06}

\bibitem[{{Brandt}(2021)}]{Brandt_21}
---. 2021, \apjs, 254, 42, \dodoi{10.3847/1538-4365/abf93c}

\bibitem[{{Brandt} {et~al.}(2019){Brandt}, {Dupuy}, \& {Bowler}}]{Brandt_19}
{Brandt}, T.~D., {Dupuy}, T.~J., \& {Bowler}, B.~P. 2019, \aj, 158, 140,
  \dodoi{10.3847/1538-3881/ab04a8}

\bibitem[{{Brandt} {et~al.}(2021){Brandt}, {Dupuy}, {Li}, {Brandt}, {Zeng},
  {Michalik}, {Bardalez Gagliuffi}, \& {Raposo-Pulido}}]{Brandt_et_al_21}
{Brandt}, T.~D., {Dupuy}, T.~J., {Li}, Y., {et~al.} 2021, \aj, 162, 186,
  \dodoi{10.3847/1538-3881/ac042e}

\bibitem[{{Bruntt} {et~al.}(2010){Bruntt}, {Bedding}, {Quirion}, {Lo Curto},
  {Carrier}, {Smalley}, {Dall}, {Arentoft}, {Bazot}, \& {Butler}}]{Bruntt_10}
{Bruntt}, H., {Bedding}, T.~R., {Quirion}, P.~O., {et~al.} 2010, \mnras, 405,
  1907, \dodoi{10.1111/j.1365-2966.2010.16575.x}

\bibitem[{{Bryson} {et~al.}(2021){Bryson}, {Kunimoto}, {Kopparapu}, {Coughlin},
  {Borucki}, {Koch}, {Aguirre}, {Allen}, {Barentsen}, {Batalha}, {Berger},
  {Boss}, {Buchhave}, {Burke}, {Caldwell}, {Campbell}, {Catanzarite},
  {Chandrasekaran}, {Chaplin}, {Christiansen}, {Christensen-Dalsgaard},
  {Ciardi}, {Clarke}, {Cochran}, {Dotson}, {Doyle}, {Duarte}, {Dunham},
  {Dupree}, {Endl}, {Fanson}, {Ford}, {Fujieh}, {Gautier}, {Geary},
  {Gilliland}, {Girouard}, {Gould}, {Haas}, {Henze}, {Holman}, {Howard},
  {Howell}, {Huber}, {Hunter}, {Jenkins}, {Kjeldsen}, {Kolodziejczak},
  {Larson}, {Latham}, {Li}, {Mathur}, {Meibom}, {Middour}, {Morris}, {Morton},
  {Mullally}, {Mullally}, {Pletcher}, {Prsa}, {Quinn}, {Quintana}, {Ragozzine},
  {Ramirez}, {Sanderfer}, {Sasselov}, {Seader}, {Shabram}, {Shporer}, {Smith},
  {Steffen}, {Still}, {Torres}, {Troeltzsch}, {Twicken}, {Uddin}, {Van Cleve},
  {Voss}, {Weiss}, {Welsh}, {Wohler}, \& {Zamudio}}]{Bryson_21}
{Bryson}, S., {Kunimoto}, M., {Kopparapu}, R.~K., {et~al.} 2021, \aj, 161, 36,
  \dodoi{10.3847/1538-3881/abc418}

\bibitem[{{Burgasser} {et~al.}(2000){Burgasser}, {Kirkpatrick}, {Cutri},
  {McCallon}, {Kopan}, {Gizis}, {Liebert}, {Reid}, {Brown}, {Monet}, {Dahn},
  {Beichman}, \& {Skrutskie}}]{Burgasser_00}
{Burgasser}, A.~J., {Kirkpatrick}, J.~D., {Cutri}, R.~M., {et~al.} 2000, \apjl,
  531, L57, \dodoi{10.1086/312522}

\bibitem[{{Burke} {et~al.}(2015){Burke}, {Christiansen}, {Mullally}, {Seader},
  {Huber}, {Rowe}, {Coughlin}, {Thompson}, {Catanzarite}, {Clarke}, {Morton},
  {Caldwell}, {Bryson}, {Haas}, {Batalha}, {Jenkins}, {Tenenbaum}, {Twicken},
  {Li}, {Quintana}, {Barclay}, {Henze}, {Borucki}, {Howell}, \&
  {Still}}]{Burke_15}
{Burke}, C.~J., {Christiansen}, J.~L., {Mullally}, F., {et~al.} 2015, \apj,
  809, 8, \dodoi{10.1088/0004-637X/809/1/8}

\bibitem[{{Butler} {et~al.}(1999){Butler}, {Marcy}, {Fischer}, {Brown},
  {Contos}, {Korzennik}, {Nisenson}, \& {Noyes}}]{Butler_99}
{Butler}, R.~P., {Marcy}, G.~W., {Fischer}, D.~A., {et~al.} 1999, \apj, 526,
  916, \dodoi{10.1086/308035}

\bibitem[{{Butler} {et~al.}(1997){Butler}, {Marcy}, {Williams}, {Hauser}, \&
  {Shirts}}]{Butler_97}
{Butler}, R.~P., {Marcy}, G.~W., {Williams}, E., {Hauser}, H., \& {Shirts}, P.
  1997, \apjl, 474, L115, \dodoi{10.1086/310444}

\bibitem[{{Butler} {et~al.}(2017){Butler}, {Vogt}, {Laughlin}, {Burt},
  {Rivera}, {Tuomi}, {Teske}, {Arriagada}, {Diaz}, {Holden}, \&
  {Keiser}}]{Butler_17}
{Butler}, R.~P., {Vogt}, S.~S., {Laughlin}, G., {et~al.} 2017, \aj, 153, 208,
  \dodoi{10.3847/1538-3881/aa66ca}

\bibitem[{{Carr}(1996)}]{Carr_96}
{Carr}, M.~H. 1996, {Water on Mars}

\bibitem[{{Chanam{\'e}} \& {Gould}(2004)}]{Chaname_04}
{Chanam{\'e}}, J., \& {Gould}, A. 2004, \apj, 601, 289, \dodoi{10.1086/380442}

\bibitem[{{Chen} {et~al.}(2022){Chen}, {Li}, {Brandt}, {Dupuy}, {Cardoso}, \&
  {McCaughrean}}]{Chen_22}
{Chen}, M., {Li}, Y., {Brandt}, T.~D., {et~al.} 2022, \aj, 163, 288,
  \dodoi{10.3847/1538-3881/ac66d2}

\bibitem[{{Chini} {et~al.}(2014){Chini}, {Fuhrmann}, {Barr}, {Pozo},
  {Westhues}, \& {Hodapp}}]{Chini_14}
{Chini}, R., {Fuhrmann}, K., {Barr}, A., {et~al.} 2014, \mnras, 437, 879,
  \dodoi{10.1093/mnras/stt1953}

\bibitem[{{Choi} {et~al.}(2016){Choi}, {Dotter}, {Conroy}, {Cantiello},
  {Paxton}, \& {Johnson}}]{Choi_16}
{Choi}, J., {Dotter}, A., {Conroy}, C., {et~al.} 2016, \apj, 823, 102,
  \dodoi{10.3847/0004-637X/823/2/102}

\bibitem[{{Ciardi} {et~al.}(2019){Ciardi}, {Bean}, {Burt}, {Dragomir},
  {Gaidos}, {Johnson}, {Kempton}, {Konopacky}, {Meyer}, {Teske}, {Weiss}, \&
  {Zhou}}]{Ciardi_19}
{Ciardi}, D.~R., {Bean}, J., {Burt}, J., {et~al.} 2019, arXiv e-prints,
  arXiv:1903.05665, \dodoi{10.48550/arXiv.1903.05665}

\bibitem[{{Corbally}(1984)}]{Corbally_84}
{Corbally}, C.~J. 1984, \apjs, 55, 657, \dodoi{10.1086/190973}

\bibitem[{{Crass} {et~al.}(2021){Crass}, {Gaudi}, {Leifer}, {Beichman},
  {Bender}, {Blackwood}, {Burt}, {Callas}, {Cegla}, {Diddams}, {Dumusque},
  {Eastman}, {Ford}, {Fulton}, {Gibson}, {Halverson}, {Haywood}, {Hearty},
  {Howard}, {Latham}, {L{\"o}hner-B{\"o}ttcher}, {Mamajek}, {Mortier},
  {Newman}, {Plavchan}, {Quirrenbach}, {Reiners}, {Robertson}, {Roy}, {Schwab},
  {Seifahrt}, {Szentgyorgyi}, {Terrien}, {Teske}, {Thompson}, \&
  {Vasisht}}]{Crass_21}
{Crass}, J., {Gaudi}, B.~S., {Leifer}, S., {et~al.} 2021, arXiv e-prints,
  arXiv:2107.14291, \dodoi{10.48550/arXiv.2107.14291}

\bibitem[{{Currie} {et~al.}(2023){Currie}, {Brandt}, {Brandt}, {Lacy},
  {Burrows}, {Guyon}, {Tamura}, {Liu}, {Sagynbayeva}, {Tobin}, {Chilcote},
  {Groff}, {Marois}, {Thompson}, {Murphy}, {Kuzuhara}, {Lawson}, {Lozi}, {Deo},
  {Vievard}, {Skaf}, {Uyama}, {Jovanovic}, {Martinache}, {Kasdin}, {Kudo},
  {McElwain}, {Janson}, {Wisniewski}, {Hodapp}, {Nishikawa}, {He{\l}miniak},
  {Kwon}, \& {Hayashi}}]{Currie_23}
{Currie}, T., {Brandt}, G.~M., {Brandt}, T.~D., {et~al.} 2023, Science, 380,
  198, \dodoi{10.1126/science.abo6192}

\bibitem[{{Daley}(2011)}]{Daley_11}
{Daley}, J. 2011, Journal of Double Star Observations, 7, 104

\bibitem[{{David} {et~al.}(2003){David}, {Quintana}, {Fatuzzo}, \&
  {Adams}}]{David_03}
{David}, E.-M., {Quintana}, E.~V., {Fatuzzo}, M., \& {Adams}, F.~C. 2003,
  \pasp, 115, 825, \dodoi{10.1086/376395}

\bibitem[{{De Rosa} {et~al.}(2023){De Rosa}, {Nielsen}, {Wahhaj}, {Ruffio},
  {Kalas}, {Peck}, {Hirsch}, \& {Roberson}}]{deRosa_23}
{De Rosa}, R.~J., {Nielsen}, E.~L., {Wahhaj}, Z., {et~al.} 2023, \aap, 672,
  A94, \dodoi{10.1051/0004-6361/202345877}

\bibitem[{{Dotter}(2016)}]{Dotter_16}
{Dotter}, A. 2016, \apjs, 222, 8, \dodoi{10.3847/0067-0049/222/1/8}

\bibitem[{{Dupuy} \& {Liu}(2011)}]{Dupuy_11}
{Dupuy}, T.~J., \& {Liu}, M.~C. 2011, \apj, 733, 122,
  \dodoi{10.1088/0004-637X/733/2/122}

\bibitem[{{El-Badry} {et~al.}(2021){El-Badry}, {Rix}, \&
  {Heintz}}]{El-Badry_21}
{El-Badry}, K., {Rix}, H.-W., \& {Heintz}, T.~M. 2021, \mnras, 506, 2269,
  \dodoi{10.1093/mnras/stab323}

\bibitem[{{Endl} {et~al.}(2002){Endl}, {K{\"u}rster}, {Els}, {Hatzes},
  {Cochran}, {Dennerl}, \& {D{\"o}bereiner}}]{Endl_02}
{Endl}, M., {K{\"u}rster}, M., {Els}, S., {et~al.} 2002, \aap, 392, 671,
  \dodoi{10.1051/0004-6361:20020937}

\bibitem[{{Feng} {et~al.}(2019){Feng}, {Anglada-Escud{\'e}}, {Tuomi}, {Jones},
  {Chanam{\'e}}, {Butler}, \& {Janson}}]{Feng_19}
{Feng}, F., {Anglada-Escud{\'e}}, G., {Tuomi}, M., {et~al.} 2019, \mnras, 490,
  5002, \dodoi{10.1093/mnras/stz2912}

\bibitem[{{Feng} {et~al.}(2022){Feng}, {Butler}, {Vogt}, {Clement}, {Tinney},
  {Cui}, {Aizawa}, {Jones}, {Bailey}, {Burt}, {Carter}, {Crane}, {Flammini
  Dotti}, {Holden}, {Ma}, {Ogihara}, {Oppenheimer}, {O'Toole}, {Shectman},
  {Wittenmyer}, {Wang}, {Wright}, \& {Xuan}}]{Feng_22}
{Feng}, F., {Butler}, R.~P., {Vogt}, S.~S., {et~al.} 2022, \apjs, 262, 21,
  \dodoi{10.3847/1538-4365/ac7e57}

\bibitem[{{Forveille} {et~al.}(1999){Forveille}, {Beuzit}, {Delfosse},
  {Segransan}, {Beck}, {Mayor}, {Perrier}, {Tokovinin}, \&
  {Udry}}]{Forveille_99}
{Forveille}, T., {Beuzit}, J.-L., {Delfosse}, X., {et~al.} 1999, \aap, 351,
  619, \dodoi{10.48550/arXiv.astro-ph/9909342}

\bibitem[{{Franson} {et~al.}(2023){Franson}, {Bowler}, {Zhou}, {Pearce},
  {Bardalez Gagliuffi}, {Biddle}, {Brandt}, {Crepp}, {Dupuy}, {Faherty},
  {Jensen-Clem}, {Morgan}, {Sanghi}, {Theissen}, {Tran}, \&
  {Wolf}}]{Franson_23b}
{Franson}, K., {Bowler}, B.~P., {Zhou}, Y., {et~al.} 2023, \apjl, 950, L19,
  \dodoi{10.3847/2041-8213/acd6f6}

\bibitem[{{Fuhrmann} \& {Chini}(2015)}]{Fuhrmann_14}
{Fuhrmann}, K., \& {Chini}, R. 2015, \apj, 809, 107,
  \dodoi{10.1088/0004-637X/809/1/107}

\bibitem[{{Fuhrmann} {et~al.}(2017){Fuhrmann}, {Chini}, {Kaderhandt}, \&
  {Chen}}]{Fuhrmann_17}
{Fuhrmann}, K., {Chini}, R., {Kaderhandt}, L., \& {Chen}, Z. 2017, \apj, 836,
  139, \dodoi{10.3847/1538-4357/836/1/139}

\bibitem[{{Fulton} {et~al.}(2021){Fulton}, {Rosenthal}, {Hirsch}, {Isaacson},
  {Howard}, {Dedrick}, {Sherstyuk}, {Blunt}, {Petigura}, {Knutson}, {Behmard},
  {Chontos}, {Crepp}, {Crossfield}, {Dalba}, {Fischer}, {Henry}, {Kane},
  {Kosiarek}, {Marcy}, {Rubenzahl}, {Weiss}, \& {Wright}}]{Fulton_21}
{Fulton}, B.~J., {Rosenthal}, L.~J., {Hirsch}, L.~A., {et~al.} 2021, \apjs,
  255, 14, \dodoi{10.3847/1538-4365/abfcc1}

\bibitem[{{Gaia Collaboration}(2022)}]{Gaia_22}
{Gaia Collaboration}. 2022, {VizieR Online Data Catalog: Gaia DR3 Part 1. Main
  source (Gaia Collaboration, 2022)}, VizieR On-line Data Catalog: I/355.
  Originally published in: Astron. Astrophys., in prep. (2022),
  \dodoi{10.26093/cds/vizier.1355}

\bibitem[{{Gaia Collaboration} {et~al.}(2016){Gaia Collaboration}, {Prusti},
  {de Bruijne}, {Brown}, {Vallenari}, {Babusiaux}, {Bailer-Jones}, {Bastian},
  {Biermann}, {Evans}, {Eyer}, {Jansen}, {Jordi}, {Klioner}, {Lammers},
  {Lindegren}, {Luri}, {Mignard}, {Milligan}, {Panem}, {Poinsignon},
  {Pourbaix}, {Randich}, {Sarri}, {Sartoretti}, {Siddiqui}, {Soubiran},
  {Valette}, {van Leeuwen}, {Walton}, {Aerts}, {Arenou}, {Cropper}, {Drimmel},
  {H{\o}g}, {Katz}, {Lattanzi}, {O'Mullane}, {Grebel}, {Holland}, {Huc},
  {Passot}, {Bramante}, {Cacciari}, {Casta{\~n}eda}, {Chaoul}, {Cheek}, {De
  Angeli}, {Fabricius}, {Guerra}, {Hern{\'a}ndez}, {Jean-Antoine-Piccolo},
  {Masana}, {Messineo}, {Mowlavi}, {Nienartowicz}, {Ord{\'o}{\~n}ez-Blanco},
  {Panuzzo}, {Portell}, {Richards}, {Riello}, {Seabroke}, {Tanga},
  {Th{\'e}venin}, {Torra}, {Els}, {Gracia-Abril}, {Comoretto},
  {Garcia-Reinaldos}, {Lock}, {Mercier}, {Altmann}, {Andrae}, {Astraatmadja},
  {Bellas-Velidis}, {Benson}, {Berthier}, {Blomme}, {Busso}, {Carry},
  {Cellino}, {Clementini}, {Cowell}, {Creevey}, {Cuypers}, {Davidson}, {De
  Ridder}, {de Torres}, {Delchambre}, {Dell'Oro}, {Ducourant}, {Fr{\'e}mat},
  {Garc{\'\i}a-Torres}, {Gosset}, {Halbwachs}, {Hambly}, {Harrison}, {Hauser},
  {Hestroffer}, {Hodgkin}, {Huckle}, {Hutton}, {Jasniewicz}, {Jordan},
  {Kontizas}, {Korn}, {Lanzafame}, {Manteiga}, {Moitinho}, {Muinonen},
  {Osinde}, {Pancino}, {Pauwels}, {Petit}, {Recio-Blanco}, {Robin}, {Sarro},
  {Siopis}, {Smith}, {Smith}, {Sozzetti}, {Thuillot}, {van Reeven}, {Viala},
  {Abbas}, {Abreu Aramburu}, {Accart}, {Aguado}, {Allan}, {Allasia},
  {Altavilla}, {{\'A}lvarez}, {Alves}, {Anderson}, {Andrei}, {Anglada Varela},
  {Antiche}, {Antoja}, {Ant{\'o}n}, {Arcay}, {Atzei}, {Ayache}, {Bach},
  {Baker}, {Balaguer-N{\'u}{\~n}ez}, {Barache}, {Barata}, {Barbier}, {Barblan},
  {Baroni}, {Barrado y Navascu{\'e}s}, {Barros}, {Barstow}, {Becciani},
  {Bellazzini}, {Bellei}, {Bello Garc{\'\i}a}, {Belokurov}, {Bendjoya},
  {Berihuete}, {Bianchi}, {Bienaym{\'e}}, {Billebaud}, {Blagorodnova},
  {Blanco-Cuaresma}, {Boch}, {Bombrun}, {Borrachero}, {Bouquillon}, {Bourda},
  {Bouy}, {Bragaglia}, {Breddels}, {Brouillet}, {Br{\"u}semeister},
  {Bucciarelli}, {Budnik}, {Burgess}, {Burgon}, {Burlacu}, {Busonero}, {Buzzi},
  {Caffau}, {Cambras}, {Campbell}, {Cancelliere}, {Cantat-Gaudin}, {Carlucci},
  {Carrasco}, {Castellani}, {Charlot}, {Charnas}, {Charvet}, {Chassat},
  {Chiavassa}, {Clotet}, {Cocozza}, {Collins}, {Collins}, {Costigan}, {Crifo},
  {Cross}, {Crosta}, {Crowley}, {Dafonte}, {Damerdji}, {Dapergolas}, {David},
  {David}, {De Cat}, {de Felice}, {de Laverny}, {De Luise}, {De March}, {de
  Martino}, {de Souza}, {Debosscher}, {del Pozo}, {Delbo}, {Delgado},
  {Delgado}, {di Marco}, {Di Matteo}, {Diakite}, {Distefano}, {Dolding}, {Dos
  Anjos}, {Drazinos}, {Dur{\'a}n}, {Dzigan}, {Ecale}, {Edvardsson}, {Enke},
  {Erdmann}, {Escolar}, {Espina}, {Evans}, {Eynard Bontemps}, {Fabre},
  {Fabrizio}, {Faigler}, {Falc{\~a}o}, {Farr{\`a}s Casas}, {Faye}, {Federici},
  {Fedorets}, {Fern{\'a}ndez-Hern{\'a}ndez}, {Fernique}, {Fienga}, {Figueras},
  {Filippi}, {Findeisen}, {Fonti}, {Fouesneau}, {Fraile}, {Fraser}, {Fuchs},
  {Furnell}, {Gai}, {Galleti}, {Galluccio}, {Garabato}, {Garc{\'\i}a-Sedano},
  {Gar{\'e}}, {Garofalo}, {Garralda}, {Gavras}, {Gerssen}, {Geyer}, {Gilmore},
  {Girona}, {Giuffrida}, {Gomes}, {Gonz{\'a}lez-Marcos},
  {Gonz{\'a}lez-N{\'u}{\~n}ez}, {Gonz{\'a}lez-Vidal}, {Granvik}, {Guerrier},
  {Guillout}, {Guiraud}, {G{\'u}rpide}, {Guti{\'e}rrez-S{\'a}nchez}, {Guy},
  {Haigron}, {Hatzidimitriou}, {Haywood}, {Heiter}, {Helmi}, {Hobbs},
  {Hofmann}, {Holl}, {Holland}, {Hunt}, {Hypki}, {Icardi}, {Irwin}, {Jevardat
  de Fombelle}, {Jofr{\'e}}, {Jonker}, {Jorissen}, {Julbe}, {Karampelas},
  {Kochoska}, {Kohley}, {Kolenberg}, {Kontizas}, {Koposov}, {Kordopatis},
  {Koubsky}, {Kowalczyk}, {Krone-Martins}, {Kudryashova}, {Kull}, {Bachchan},
  {Lacoste-Seris}, {Lanza}, {Lavigne}, {Le Poncin-Lafitte}, {Lebreton},
  {Lebzelter}, {Leccia}, {Leclerc}, {Lecoeur-Taibi}, {Lemaitre}, {Lenhardt},
  {Leroux}, {Liao}, {Licata}, {Lindstr{\o}m}, {Lister}, {Livanou}, {Lobel},
  {L{\"o}ffler}, {L{\'o}pez}, {Lopez-Lozano}, {Lorenz}, {Loureiro},
  {MacDonald}, {Magalh{\~a}es Fernandes}, {Managau}, {Mann}, {Mantelet},
  {Marchal}, {Marchant}, {Marconi}, {Marie}, {Marinoni}, {Marrese},
  {Marschalk{\'o}}, {Marshall}, {Mart{\'\i}n-Fleitas}, {Martino}, {Mary},
  {Matijevi{\v{c}}}, {Mazeh}, {McMillan}, {Messina}, {Mestre}, {Michalik},
  {Millar}, {Miranda}, {Molina}, {Molinaro}, {Molinaro}, {Moln{\'a}r},
  {Moniez}, {Montegriffo}, {Monteiro}, {Mor}, {Mora}, {Morbidelli}, {Morel},
  {Morgenthaler}, {Morley}, {Morris}, {Mulone}, {Muraveva}, {Musella},
  {Narbonne}, {Nelemans}, {Nicastro}, {Noval}, {Ord{\'e}novic},
  {Ordieres-Mer{\'e}}, {Osborne}, {Pagani}, {Pagano}, {Pailler}, {Palacin},
  {Palaversa}, {Parsons}, {Paulsen}, {Pecoraro}, {Pedrosa}, {Pentik{\"a}inen},
  {Pereira}, {Pichon}, {Piersimoni}, {Pineau}, {Plachy}, {Plum}, {Poujoulet},
  {Pr{\v{s}}a}, {Pulone}, {Ragaini}, {Rago}, {Rambaux}, {Ramos-Lerate},
  {Ranalli}, {Rauw}, {Read}, {Regibo}, {Renk}, {Reyl{\'e}}, {Ribeiro},
  {Rimoldini}, {Ripepi}, {Riva}, {Rixon}, {Roelens}, {Romero-G{\'o}mez},
  {Rowell}, {Royer}, {Rudolph}, {Ruiz-Dern}, {Sadowski}, {Sagrist{\`a}
  Sell{\'e}s}, {Sahlmann}, {Salgado}, {Salguero}, {Sarasso}, {Savietto},
  {Schnorhk}, {Schultheis}, {Sciacca}, {Segol}, {Segovia}, {Segransan},
  {Serpell}, {Shih}, {Smareglia}, {Smart}, {Smith}, {Solano}, {Solitro},
  {Sordo}, {Soria Nieto}, {Souchay}, {Spagna}, {Spoto}, {Stampa}, {Steele},
  {Steidelm{\"u}ller}, {Stephenson}, {Stoev}, {Suess}, {S{\"u}veges}, {Surdej},
  {Szabados}, {Szegedi-Elek}, {Tapiador}, {Taris}, {Tauran}, {Taylor},
  {Teixeira}, {Terrett}, {Tingley}, {Trager}, {Turon}, {Ulla}, {Utrilla},
  {Valentini}, {van Elteren}, {Van Hemelryck}, {van Leeuwen}, {Varadi},
  {Vecchiato}, {Veljanoski}, {Via}, {Vicente}, {Vogt}, {Voss}, {Votruba},
  {Voutsinas}, {Walmsley}, {Weiler}, {Weingrill}, {Werner}, {Wevers},
  {Whitehead}, {Wyrzykowski}, {Yoldas}, {{\v{Z}}erjal}, {Zucker}, {Zurbach},
  {Zwitter}, {Alecu}, {Allen}, {Allende Prieto}, {Amorim},
  {Anglada-Escud{\'e}}, {Arsenijevic}, {Azaz}, {Balm}, {Beck}, {Bernstein},
  {Bigot}, {Bijaoui}, {Blasco}, {Bonfigli}, {Bono}, {Boudreault}, {Bressan},
  {Brown}, {Brunet}, {Bunclark}, {Buonanno}, {Butkevich}, {Carret}, {Carrion},
  {Chemin}, {Ch{\'e}reau}, {Corcione}, {Darmigny}, {de Boer}, {de Teodoro}, {de
  Zeeuw}, {Delle Luche}, {Domingues}, {Dubath}, {Fodor}, {Fr{\'e}zouls},
  {Fries}, {Fustes}, {Fyfe}, {Gallardo}, {Gallegos}, {Gardiol}, {Gebran},
  {Gomboc}, {G{\'o}mez}, {Grux}, {Gueguen}, {Heyrovsky}, {Hoar}, {Iannicola},
  {Isasi Parache}, {Janotto}, {Joliet}, {Jonckheere}, {Keil}, {Kim},
  {Klagyivik}, {Klar}, {Knude}, {Kochukhov}, {Kolka}, {Kos}, {Kutka}, {Lainey},
  {LeBouquin}, {Liu}, {Loreggia}, {Makarov}, {Marseille}, {Martayan},
  {Martinez-Rubi}, {Massart}, {Meynadier}, {Mignot}, {Munari}, {Nguyen},
  {Nordlander}, {Ocvirk}, {O'Flaherty}, {Olias Sanz}, {Ortiz}, {Osorio},
  {Oszkiewicz}, {Ouzounis}, {Palmer}, {Park}, {Pasquato}, {Peltzer}, {Peralta},
  {P{\'e}turaud}, {Pieniluoma}, {Pigozzi}, {Poels}, {Prat}, {Prod'homme},
  {Raison}, {Rebordao}, {Risquez}, {Rocca-Volmerange}, {Rosen}, {Ruiz-Fuertes},
  {Russo}, {Sembay}, {Serraller Vizcaino}, {Short}, {Siebert}, {Silva},
  {Sinachopoulos}, {Slezak}, {Soffel}, {Sosnowska}, {Strai{\v{z}}ys}, {ter
  Linden}, {Terrell}, {Theil}, {Tiede}, {Troisi}, {Tsalmantza}, {Tur},
  {Vaccari}, {Vachier}, {Valles}, {Van Hamme}, {Veltz}, {Virtanen}, {Wallut},
  {Wichmann}, {Wilkinson}, {Ziaeepour}, \& {Zschocke}}]{Gaia_16}
{Gaia Collaboration}, {Prusti}, T., {de Bruijne}, J.~H.~J., {et~al.} 2016,
  \aap, 595, A1, \dodoi{10.1051/0004-6361/201629272}

\bibitem[{{Gaia Collaboration} {et~al.}(2023){Gaia Collaboration}, {Vallenari},
  {Brown}, {Prusti}, {de Bruijne}, {Arenou}, {Babusiaux}, {Biermann},
  {Creevey}, {Ducourant}, {Evans}, {Eyer}, {Guerra}, {Hutton}, {Jordi},
  {Klioner}, {Lammers}, {Lindegren}, {Luri}, {Mignard}, {Panem}, {Pourbaix},
  {Randich}, {Sartoretti}, {Soubiran}, {Tanga}, {Walton}, {Bailer-Jones},
  {Bastian}, {Drimmel}, {Jansen}, {Katz}, {Lattanzi}, {van Leeuwen}, {Bakker},
  {Cacciari}, {Casta{\~n}eda}, {De Angeli}, {Fabricius}, {Fouesneau},
  {Fr{\'e}mat}, {Galluccio}, {Guerrier}, {Heiter}, {Masana}, {Messineo},
  {Mowlavi}, {Nicolas}, {Nienartowicz}, {Pailler}, {Panuzzo}, {Riclet}, {Roux},
  {Seabroke}, {Sordo}, {Th{\'e}venin}, {Gracia-Abril}, {Portell}, {Teyssier},
  {Altmann}, {Andrae}, {Audard}, {Bellas-Velidis}, {Benson}, {Berthier},
  {Blomme}, {Burgess}, {Busonero}, {Busso}, {C{\'a}novas}, {Carry}, {Cellino},
  {Cheek}, {Clementini}, {Damerdji}, {Davidson}, {de Teodoro}, {Nu{\~n}ez
  Campos}, {Delchambre}, {Dell'Oro}, {Esquej}, {Fern{\'a}ndez-Hern{\'a}ndez},
  {Fraile}, {Garabato}, {Garc{\'\i}a-Lario}, {Gosset}, {Haigron}, {Halbwachs},
  {Hambly}, {Harrison}, {Hern{\'a}ndez}, {Hestroffer}, {Hodgkin}, {Holl},
  {Jan{\ss}en}, {Jevardat de Fombelle}, {Jordan}, {Krone-Martins}, {Lanzafame},
  {L{\"o}ffler}, {Marchal}, {Marrese}, {Moitinho}, {Muinonen}, {Osborne},
  {Pancino}, {Pauwels}, {Recio-Blanco}, {Reyl{\'e}}, {Riello}, {Rimoldini},
  {Roegiers}, {Rybizki}, {Sarro}, {Siopis}, {Smith}, {Sozzetti}, {Utrilla},
  {van Leeuwen}, {Abbas}, {{\'A}brah{\'a}m}, {Abreu Aramburu}, {Aerts},
  {Aguado}, {Ajaj}, {Aldea-Montero}, {Altavilla}, {{\'A}lvarez}, {Alves},
  {Anders}, {Anderson}, {Anglada Varela}, {Antoja}, {Baines}, {Baker},
  {Balaguer-N{\'u}{\~n}ez}, {Balbinot}, {Balog}, {Barache}, {Barbato},
  {Barros}, {Barstow}, {Bartolom{\'e}}, {Bassilana}, {Bauchet}, {Becciani},
  {Bellazzini}, {Berihuete}, {Bernet}, {Bertone}, {Bianchi}, {Binnenfeld},
  {Blanco-Cuaresma}, {Blazere}, {Boch}, {Bombrun}, {Bossini}, {Bouquillon},
  {Bragaglia}, {Bramante}, {Breedt}, {Bressan}, {Brouillet}, {Brugaletta},
  {Bucciarelli}, {Burlacu}, {Butkevich}, {Buzzi}, {Caffau}, {Cancelliere},
  {Cantat-Gaudin}, {Carballo}, {Carlucci}, {Carnerero}, {Carrasco},
  {Casamiquela}, {Castellani}, {Castro-Ginard}, {Chaoul}, {Charlot}, {Chemin},
  {Chiaramida}, {Chiavassa}, {Chornay}, {Comoretto}, {Contursi}, {Cooper},
  {Cornez}, {Cowell}, {Crifo}, {Cropper}, {Crosta}, {Crowley}, {Dafonte},
  {Dapergolas}, {David}, {David}, {de Laverny}, {De Luise}, {De March}, {De
  Ridder}, {de Souza}, {de Torres}, {del Peloso}, {del Pozo}, {Delbo},
  {Delgado}, {Delisle}, {Demouchy}, {Dharmawardena}, {Di Matteo}, {Diakite},
  {Diener}, {Distefano}, {Dolding}, {Edvardsson}, {Enke}, {Fabre}, {Fabrizio},
  {Faigler}, {Fedorets}, {Fernique}, {Fienga}, {Figueras}, {Fournier},
  {Fouron}, {Fragkoudi}, {Gai}, {Garcia-Gutierrez}, {Garcia-Reinaldos},
  {Garc{\'\i}a-Torres}, {Garofalo}, {Gavel}, {Gavras}, {Gerlach}, {Geyer},
  {Giacobbe}, {Gilmore}, {Girona}, {Giuffrida}, {Gomel}, {Gomez},
  {Gonz{\'a}lez-N{\'u}{\~n}ez}, {Gonz{\'a}lez-Santamar{\'\i}a},
  {Gonz{\'a}lez-Vidal}, {Granvik}, {Guillout}, {Guiraud},
  {Guti{\'e}rrez-S{\'a}nchez}, {Guy}, {Hatzidimitriou}, {Hauser}, {Haywood},
  {Helmer}, {Helmi}, {Sarmiento}, {Hidalgo}, {Hilger}, {H{\l}adczuk}, {Hobbs},
  {Holland}, {Huckle}, {Jardine}, {Jasniewicz}, {Jean-Antoine Piccolo},
  {Jim{\'e}nez-Arranz}, {Jorissen}, {Juaristi Campillo}, {Julbe}, {Karbevska},
  {Kervella}, {Khanna}, {Kontizas}, {Kordopatis}, {Korn}, {K{\'o}sp{\'a}l},
  {Kostrzewa-Rutkowska}, {Kruszy{\'n}ska}, {Kun}, {Laizeau}, {Lambert},
  {Lanza}, {Lasne}, {Le Campion}, {Lebreton}, {Lebzelter}, {Leccia}, {Leclerc},
  {Lecoeur-Taibi}, {Liao}, {Licata}, {Lindstr{\o}m}, {Lister}, {Livanou},
  {Lobel}, {Lorca}, {Loup}, {Madrero Pardo}, {Magdaleno Romeo}, {Managau},
  {Mann}, {Manteiga}, {Marchant}, {Marconi}, {Marcos}, {Marcos Santos},
  {Mar{\'\i}n Pina}, {Marinoni}, {Marocco}, {Marshall}, {Martin Polo},
  {Mart{\'\i}n-Fleitas}, {Marton}, {Mary}, {Masip}, {Massari},
  {Mastrobuono-Battisti}, {Mazeh}, {McMillan}, {Messina}, {Michalik}, {Millar},
  {Mints}, {Molina}, {Molinaro}, {Moln{\'a}r}, {Monari}, {Mongui{\'o}},
  {Montegriffo}, {Montero}, {Mor}, {Mora}, {Morbidelli}, {Morel}, {Morris},
  {Muraveva}, {Murphy}, {Musella}, {Nagy}, {Noval}, {Oca{\~n}a}, {Ogden},
  {Ordenovic}, {Osinde}, {Pagani}, {Pagano}, {Palaversa}, {Palicio},
  {Pallas-Quintela}, {Panahi}, {Payne-Wardenaar}, {Pe{\~n}alosa Esteller},
  {Penttil{\"a}}, {Pichon}, {Piersimoni}, {Pineau}, {Plachy}, {Plum}, {Poggio},
  {Pr{\v{s}}a}, {Pulone}, {Racero}, {Ragaini}, {Rainer}, {Raiteri}, {Rambaux},
  {Ramos}, {Ramos-Lerate}, {Re Fiorentin}, {Regibo}, {Richards}, {Rios Diaz},
  {Ripepi}, {Riva}, {Rix}, {Rixon}, {Robichon}, {Robin}, {Robin}, {Roelens},
  {Rogues}, {Rohrbasser}, {Romero-G{\'o}mez}, {Rowell}, {Royer}, {Ruz Mieres},
  {Rybicki}, {Sadowski}, {S{\'a}ez N{\'u}{\~n}ez}, {Sagrist{\`a} Sell{\'e}s},
  {Sahlmann}, {Salguero}, {Samaras}, {Sanchez Gimenez}, {Sanna},
  {Santove{\~n}a}, {Sarasso}, {Schultheis}, {Sciacca}, {Segol}, {Segovia},
  {S{\'e}gransan}, {Semeux}, {Shahaf}, {Siddiqui}, {Siebert}, {Siltala},
  {Silvelo}, {Slezak}, {Slezak}, {Smart}, {Snaith}, {Solano}, {Solitro},
  {Souami}, {Souchay}, {Spagna}, {Spina}, {Spoto}, {Steele},
  {Steidelm{\"u}ller}, {Stephenson}, {S{\"u}veges}, {Surdej}, {Szabados},
  {Szegedi-Elek}, {Taris}, {Taylor}, {Teixeira}, {Tolomei}, {Tonello}, {Torra},
  {Torra}, {Torralba Elipe}, {Trabucchi}, {Tsounis}, {Turon}, {Ulla}, {Unger},
  {Vaillant}, {van Dillen}, {van Reeven}, {Vanel}, {Vecchiato}, {Viala},
  {Vicente}, {Voutsinas}, {Weiler}, {Wevers}, {Wyrzykowski}, {Yoldas}, {Yvard},
  {Zhao}, {Zorec}, {Zucker}, \& {Zwitter}}]{Gaia_Colloboration_23}
{Gaia Collaboration}, {Vallenari}, A., {Brown}, A.~G.~A., {et~al.} 2023, \aap,
  674, A1, \dodoi{10.1051/0004-6361/202243940}

\bibitem[{{Gandolfi} {et~al.}(2018){Gandolfi}, {Barrag{\'a}n}, {Livingston},
  {Fridlund}, {Justesen}, {Redfield}, {Fossati}, {Mathur}, {Grziwa}, {Cabrera},
  {Garc{\'\i}a}, {Persson}, {Van Eylen}, {Hatzes}, {Hidalgo}, {Albrecht},
  {Bugnet}, {Cochran}, {Csizmadia}, {Deeg}, {Eigm{\"u}ller}, {Endl}, {Erikson},
  {Esposito}, {Guenther}, {Korth}, {Luque}, {Monta{\~n}es Rodr{\'\i}guez},
  {Nespral}, {Nowak}, {P{\"a}tzold}, \& {Prieto-Arranz}}]{Gandolfi_18}
{Gandolfi}, D., {Barrag{\'a}n}, O., {Livingston}, J.~H., {et~al.} 2018, \aap,
  619, L10, \dodoi{10.1051/0004-6361/201834289}

\bibitem[{{Gianninas} {et~al.}(2011){Gianninas}, {Bergeron}, \&
  {Ruiz}}]{Gianninas_11}
{Gianninas}, A., {Bergeron}, P., \& {Ruiz}, M.~T. 2011, \apj, 743, 138,
  \dodoi{10.1088/0004-637X/743/2/138}

\bibitem[{{Gillon} {et~al.}(2017){Gillon}, {Demory}, {Van Grootel}, {Motalebi},
  {Lovis}, {Collier Cameron}, {Charbonneau}, {Latham}, {Molinari}, {Pepe},
  {S{\'e}gransan}, {Sasselov}, {Udry}, {Mayor}, {Micela}, {Piotto}, \&
  {Sozzetti}}]{Gillon_17}
{Gillon}, M., {Demory}, B.-O., {Van Grootel}, V., {et~al.} 2017, Nature
  Astronomy, 1, 0056, \dodoi{10.1038/s41550-017-0056}

\bibitem[{{Grandjean} {et~al.}(2019){Grandjean}, {Lagrange}, {Beust}, {Rodet},
  {Milli}, {Rubini}, {Babusiaux}, {Meunier}, {Delorme}, {Aigrain}, {Zicher},
  {Bonnefoy}, {Biller}, {Baudino}, {Bonavita}, {Boccaletti}, {Cheetham},
  {Girard}, {Hagelberg}, {Janson}, {Lannier}, {Lazzoni}, {Ligi}, {Maire},
  {Mesa}, {Perrot}, {Rouan}, \& {Zurlo}}]{Grandjean_19}
{Grandjean}, A., {Lagrange}, A.~M., {Beust}, H., {et~al.} 2019, \aap, 627, L9,
  \dodoi{10.1051/0004-6361/201935044}

\bibitem[{{Gray} {et~al.}(2006){Gray}, {Corbally}, {Garrison}, {McFadden},
  {Bubar}, {McGahee}, {O'Donoghue}, \& {Knox}}]{Gray_06}
{Gray}, R.~O., {Corbally}, C.~J., {Garrison}, R.~F., {et~al.} 2006, \aj, 132,
  161, \dodoi{10.1086/504637}

\bibitem[{{Gray} {et~al.}(2003){Gray}, {Corbally}, {Garrison}, {McFadden}, \&
  {Robinson}}]{Gray03}
{Gray}, R.~O., {Corbally}, C.~J., {Garrison}, R.~F., {McFadden}, M.~T., \&
  {Robinson}, P.~E. 2003, \aj, 126, 2048, \dodoi{10.1086/378365}

\bibitem[{{Gray} {et~al.}(2001){Gray}, {Napier}, \& {Winkler}}]{Gray_01}
{Gray}, R.~O., {Napier}, M.~G., \& {Winkler}, L.~I. 2001, \aj, 121, 2148,
  \dodoi{10.1086/319956}

\bibitem[{{Harada} {et~al.}(2024{\natexlab{a}}){Harada}, {Dressing}, {Kane}, \&
  {Ardestani}}]{Harada_24}
{Harada}, C.~K., {Dressing}, C.~D., {Kane}, S.~R., \& {Ardestani}, B.~A.
  2024{\natexlab{a}}, \apjs, 272, 30, \dodoi{10.3847/1538-4365/ad3e81}

\bibitem[{{Harada} {et~al.}(2024{\natexlab{b}}){Harada}, {Dressing}, {Kane},
  {Blunt}, {Dietrich}, {Hinkel}, {Li}, {Mamajek}, {Rice}, {Tuchow},
  {Turtelboom}, \& {Wittenmyer}}]{Harada_24b}
{Harada}, C.~K., {Dressing}, C.~D., {Kane}, S.~R., {et~al.} 2024{\natexlab{b}},
  arXiv e-prints, arXiv:2409.10679, \dodoi{10.48550/arXiv.2409.10679}

\bibitem[{{Hatzes} {et~al.}(2000){Hatzes}, {Cochran}, {McArthur}, {Baliunas},
  {Walker}, {Campbell}, {Irwin}, {Yang}, {K{\"u}rster}, {Endl}, {Els},
  {Butler}, \& {Marcy}}]{Hatzes_00}
{Hatzes}, A.~P., {Cochran}, W.~D., {McArthur}, B., {et~al.} 2000, \apjl, 544,
  L145, \dodoi{10.1086/317319}

\bibitem[{{Hatzes} {et~al.}(2022){Hatzes}, {Gandolfi}, {Korth}, {Rodler},
  {Sabotta}, {Esposito}, {Barrag{\'a}n}, {Van Eylen}, {Livingston}, {Serrano},
  {Luque}, {Smith}, {Redfield}, {Persson}, {P{\"a}tzold}, {Palle}, {Nowak},
  {Osborne}, {Narita}, {Mathur}, {Lam}, {Kab{\'a}th}, {Johnson}, {Guenther},
  {Grziwa}, {Goffo}, {Fridlund}, {Endl}, {Deeg}, {Csizmadia}, {Cochran},
  {Cuesta}, {Chaturvedi}, {Carleo}, {Cabrera}, {Beck}, \&
  {Albrecht}}]{Hatzes_22}
{Hatzes}, A.~P., {Gandolfi}, D., {Korth}, J., {et~al.} 2022, \aj, 163, 223,
  \dodoi{10.3847/1538-3881/ac5dcb}

\bibitem[{{Hauck} \& {Mermilliod}(1998)}]{Hauck_98}
{Hauck}, B., \& {Mermilliod}, M. 1998, \aaps, 129, 431,
  \dodoi{10.1051/aas:1998195}

\bibitem[{{Herschel}(1785)}]{Herschel_1785}
{Herschel}, W. 1785, Philosophical Transactions of the Royal Society of London
  Series I, 75, 40

\bibitem[{{Hinkley} {et~al.}(2023){Hinkley}, {Lacour}, {Marleau}, {Lagrange},
  {Wang}, {Kammerer}, {Cumming}, {Nowak}, {Rodet}, {Stolker}, {Balmer}, {Ray},
  {Bonnefoy}, {Molli{\`e}re}, {Lazzoni}, {Kennedy}, {Mordasini}, {Abuter},
  {Aigrain}, {Amorim}, {Asensio-Torres}, {Babusiaux}, {Benisty}, {Berger},
  {Beust}, {Blunt}, {Boccaletti}, {Bohn}, {Bonnet}, {Bourdarot}, {Brandner},
  {Cantalloube}, {Caselli}, {Charnay}, {Chauvin}, {Chomez}, {Choquet},
  {Christiaens}, {Cl{\'e}net}, {Coud{\'e} du Foresto}, {Cridland}, {Delorme},
  {Dembet}, {Drescher}, {Duvert}, {Eckart}, {Eisenhauer}, {Feuchtgruber},
  {Galland}, {Garcia}, {Garcia Lopez}, {Gardner}, {Gendron}, {Genzel},
  {Gillessen}, {Girard}, {Grandjean}, {Haubois}, {Hei{\ss}el}, {Henning},
  {Hippler}, {Horrobin}, {Houll{\'e}}, {Hubert}, {Jocou}, {Keppler},
  {Kervella}, {Kreidberg}, {Lapeyr{\`e}re}, {Le Bouquin}, {L{\'e}na}, {Lutz},
  {Maire}, {Mang}, {M{\'e}rand}, {Meunier}, {Monnier}, {Mouillet}, {Nasedkin},
  {Ott}, {Otten}, {Paladini}, {Paumard}, {Perraut}, {Perrin}, {Philipot},
  {Pfuhl}, {Pourr{\'e}}, {Pueyo}, {Rameau}, {Rickman}, {Rubini}, {Rustamkulov},
  {Samland}, {Shangguan}, {Shimizu}, {Sing}, {Straubmeier}, {Sturm}, {Tacconi},
  {van Dishoeck}, {Vigan}, {Vincent}, {Ward-Duong}, {Widmann}, {Wieprecht},
  {Wiezorrek}, {Woillez}, {Yazici}, {Young}, \& {Zicher}}]{Hinkley_23}
{Hinkley}, S., {Lacour}, S., {Marleau}, G.~D., {et~al.} 2023, \aap, 671, L5,
  \dodoi{10.1051/0004-6361/202244727}

\bibitem[{{Hirsch} {et~al.}(2021){Hirsch}, {Rosenthal}, {Fulton}, {Howard},
  {Ciardi}, {Marcy}, {Nielsen}, {Petigura}, {de Rosa}, {Isaacson}, {Weiss},
  {Sinukoff}, \& {Macintosh}}]{Hirsch_21}
{Hirsch}, L.~A., {Rosenthal}, L., {Fulton}, B.~J., {et~al.} 2021, \aj, 161,
  134, \dodoi{10.3847/1538-3881/abd639}

\bibitem[{{Huang} {et~al.}(2018){Huang}, {Burt}, {Vanderburg}, {G{\"u}nther},
  {Shporer}, {Dittmann}, {Winn}, {Wittenmyer}, {Sha}, {Kane}, {Ricker},
  {Vanderspek}, {Latham}, {Seager}, {Jenkins}, {Caldwell}, {Collins},
  {Guerrero}, {Smith}, {Quinn}, {Udry}, {Pepe}, {Bouchy}, {S{\'e}gransan},
  {Lovis}, {Ehrenreich}, {Marmier}, {Mayor}, {Wohler}, {Haworth}, {Morgan},
  {Fausnaugh}, {Ciardi}, {Christiansen}, {Charbonneau}, {Dragomir}, {Deming},
  {Glidden}, {Levine}, {McCullough}, {Yu}, {Narita}, {Nguyen}, {Morton},
  {Pepper}, {P{\'a}l}, {Rodriguez}, {Stassun}, {Torres}, {Sozzetti}, {Doty},
  {Christensen-Dalsgaard}, {Laughlin}, {Clampin}, {Bean}, {Buchhave}, {Bakos},
  {Sato}, {Ida}, {Kaltenegger}, {Palle}, {Sasselov}, {Butler}, {Lissauer},
  {Ge}, \& {Rinehart}}]{Huang_18}
{Huang}, C.~X., {Burt}, J., {Vanderburg}, A., {et~al.} 2018, \apjl, 868, L39,
  \dodoi{10.3847/2041-8213/aaef91}

\bibitem[{{Hurt} {et~al.}(2022){Hurt}, {Fulton}, {Isaacson}, {Rosenthal},
  {Howard}, {Weiss}, \& {Petigura}}]{Hurt_22}
{Hurt}, S.~A., {Fulton}, B., {Isaacson}, H., {et~al.} 2022, \aj, 163, 218,
  \dodoi{10.3847/1538-3881/ac5c47}

\bibitem[{{Iverson} \& {Nugent}(2015)}]{Iverson_15}
{Iverson}, E., \& {Nugent}, R. 2015, Journal of Double Star Observations, 11,
  91

\bibitem[{{Jaime} {et~al.}(2012){Jaime}, {Pichardo}, \& {Aguilar}}]{Jaime_12}
{Jaime}, L.~G., {Pichardo}, B., \& {Aguilar}, L. 2012, \mnras, 427, 2723,
  \dodoi{10.1111/j.1365-2966.2012.21839.x}

\bibitem[{{Johnson} {et~al.}(2016){Johnson}, {Endl}, {Cochran}, {Meschiari},
  {Robertson}, {MacQueen}, {Brugamyer}, {Caldwell}, {Hatzes}, {Ram{\'\i}rez},
  \& {Wittenmyer}}]{Johnson_16}
{Johnson}, M.~C., {Endl}, M., {Cochran}, W.~D., {et~al.} 2016, \apj, 821, 74,
  \dodoi{10.3847/0004-637X/821/2/74}

\bibitem[{{Jones} {et~al.}(2002){Jones}, {Paul Butler}, {Tinney}, {Marcy},
  {Penny}, {McCarthy}, {Carter}, \& {Pourbaix}}]{Jones_02}
{Jones}, H. R.~A., {Paul Butler}, R., {Tinney}, C.~G., {et~al.} 2002, \mnras,
  333, 871, \dodoi{10.1046/j.1365-8711.2002.05459.x}

\bibitem[{{Kaltenegger}(2017)}]{Kaltenegger_17}
{Kaltenegger}, L. 2017, \araa, 55, 433,
  \dodoi{10.1146/annurev-astro-082214-122238}

\bibitem[{{Kane} \& {Blunt}(2019)}]{Kane_19}
{Kane}, S.~R., \& {Blunt}, S. 2019, \aj, 158, 209,
  \dodoi{10.3847/1538-3881/ab4c3e}

\bibitem[{{Kane} {et~al.}(2024){Kane}, {Li}, {Turnbull}, {Dressing}, \&
  {Harada}}]{Kane_24}
{Kane}, S.~R., {Li}, Z., {Turnbull}, M.~C., {Dressing}, C.~D., \& {Harada},
  C.~K. 2024, \aj, 168, 195, \dodoi{10.3847/1538-3881/ad6a50}

\bibitem[{{Kasting} {et~al.}(1993){Kasting}, {Whitmire}, \&
  {Reynolds}}]{Kasting_93}
{Kasting}, J.~F., {Whitmire}, D.~P., \& {Reynolds}, R.~T. 1993, \icarus, 101,
  108, \dodoi{10.1006/icar.1993.1010}

\bibitem[{{Keenan} \& {McNeil}(1989)}]{Keenan_89}
{Keenan}, P.~C., \& {McNeil}, R.~C. 1989, \apjs, 71, 245,
  \dodoi{10.1086/191373}

\bibitem[{{Kepler} {et~al.}(2006){Kepler}, {Castanheira}, {Costa}, \&
  {Koester}}]{Kepler_06}
{Kepler}, S.~O., {Castanheira}, B.~G., {Costa}, A.~F.~M., \& {Koester}, D.
  2006, \mnras, 372, 1799, \dodoi{10.1111/j.1365-2966.2006.10992.x}

\bibitem[{{Kervella} {et~al.}(2019){Kervella}, {Arenou}, {Mignard}, \&
  {Th{\'e}venin}}]{Kervella_19}
{Kervella}, P., {Arenou}, F., {Mignard}, F., \& {Th{\'e}venin}, F. 2019, \aap,
  623, A72, \dodoi{10.1051/0004-6361/201834371}

\bibitem[{{Kervella} {et~al.}(2022){Kervella}, {Arenou}, \&
  {Th{\'e}venin}}]{Kervella_22}
{Kervella}, P., {Arenou}, F., \& {Th{\'e}venin}, F. 2022, \aap, 657, A7,
  \dodoi{10.1051/0004-6361/202142146}

\bibitem[{{Kharchenko}(2001)}]{Kharchenko_01}
{Kharchenko}, N.~V. 2001, Kinematika i Fizika Nebesnykh Tel, 17, 409

\bibitem[{{King} {et~al.}(2010){King}, {McCaughrean}, {Homeier}, {Allard},
  {Scholz}, \& {Lodieu}}]{King_10}
{King}, R.~R., {McCaughrean}, M.~J., {Homeier}, D., {et~al.} 2010, \aap, 510,
  A99, \dodoi{10.1051/0004-6361/200912981}

\bibitem[{{Kirkpatrick} {et~al.}(1991){Kirkpatrick}, {Henry}, \&
  {McCarthy}}]{Kirkpatrick91}
{Kirkpatrick}, J.~D., {Henry}, T.~J., \& {McCarthy}, Jr., D.~W. 1991, \apjs,
  77, 417, \dodoi{10.1086/191611}

\bibitem[{{Kirkpatrick} {et~al.}(2016){Kirkpatrick}, {Kellogg}, {Schneider},
  {Fajardo-Acosta}, {Cushing}, {Greco}, {Mace}, {Gelino}, {Wright},
  {Eisenhardt}, {Stern}, {Faherty}, {Sheppard}, {Lansbury}, {Logsdon},
  {Martin}, {McLean}, {Schurr}, {Cutri}, \& {Conrow}}]{Kirkpatrick_16}
{Kirkpatrick}, J.~D., {Kellogg}, K., {Schneider}, A.~C., {et~al.} 2016, \apjs,
  224, 36, \dodoi{10.3847/0067-0049/224/2/36}

\bibitem[{{Kopparapu} \& {Barnes}(2010)}]{Kopparapu_10}
{Kopparapu}, R.~K., \& {Barnes}, R. 2010, \apj, 716, 1336,
  \dodoi{10.1088/0004-637X/716/2/1336}

\bibitem[{{Kopparapu} {et~al.}(2014){Kopparapu}, {Ramirez}, {SchottelKotte},
  {Kasting}, {Domagal-Goldman}, \& {Eymet}}]{Kopparapu_14}
{Kopparapu}, R.~K., {Ramirez}, R.~M., {SchottelKotte}, J., {et~al.} 2014,
  \apjl, 787, L29, \dodoi{10.1088/2041-8205/787/2/L29}

\bibitem[{{Kopparapu} {et~al.}(2013){Kopparapu}, {Ramirez}, {Kasting}, {Eymet},
  {Robinson}, {Mahadevan}, {Terrien}, {Domagal-Goldman}, {Meadows}, \&
  {Deshpande}}]{Kopparapu13}
{Kopparapu}, R.~K., {Ramirez}, R., {Kasting}, J.~F., {et~al.} 2013, \apj, 765,
  131, \dodoi{10.1088/0004-637X/765/2/131}

\bibitem[{{Kraus} {et~al.}(2016){Kraus}, {Ireland}, {Huber}, {Mann}, \&
  {Dupuy}}]{Kraus_16}
{Kraus}, A.~L., {Ireland}, M.~J., {Huber}, D., {Mann}, A.~W., \& {Dupuy}, T.~J.
  2016, \aj, 152, 8, \dodoi{10.3847/0004-6256/152/1/8}

\bibitem[{{Lada}(2006)}]{Lada_06}
{Lada}, C.~J. 2006, \apjl, 640, L63, \dodoi{10.1086/503158}

\bibitem[{{Laliotis} {et~al.}(2023){Laliotis}, {Burt}, {Mamajek}, {Li},
  {Perdelwitz}, {Zhao}, {Butler}, {Holden}, {Rosenthal}, {Fulton}, {Feng},
  {Kane}, {Bailey}, {Carter}, {Crane}, {Furlan}, {Gnilka}, {Howell},
  {Laughlin}, {Shectman}, {Teske}, {Tinney}, {Vogt}, {Wang}, \&
  {Wittenmyer}}]{Laliotis_23}
{Laliotis}, K., {Burt}, J.~A., {Mamajek}, E.~E., {et~al.} 2023, \aj, 165, 176,
  \dodoi{10.3847/1538-3881/acc067}

\bibitem[{{Li} {et~al.}(2021){Li}, {Brandt}, {Brandt}, {Dupuy}, {Michalik},
  {Jensen-Clem}, {Zeng}, {Faherty}, \& {Mitra}}]{Li_21}
{Li}, Y., {Brandt}, T.~D., {Brandt}, G.~M., {et~al.} 2021, \aj, 162, 266,
  \dodoi{10.3847/1538-3881/ac27ab}

\bibitem[{{Llop-Sayson} {et~al.}(2021){Llop-Sayson}, {Wang}, {Ruffio}, {Mawet},
  {Blunt}, {Absil}, {Bond}, {Brinkman}, {Bowler}, {Bottom}, {Chontos}, {Dalba},
  {Fulton}, {Giacalone}, {Hill}, {Hirsch}, {Howard}, {Isaacson}, {Karlsson},
  {Lubin}, {Madurowicz}, {Matthews}, {Morris}, {Perrin}, {Ren}, {Rice},
  {Rosenthal}, {Ruane}, {Rubenzahl}, {Sun}, {Wallack}, {Xuan}, \&
  {Ygouf}}]{Llop-Sayson_21}
{Llop-Sayson}, J., {Wang}, J.~J., {Ruffio}, J.-B., {et~al.} 2021, \aj, 162,
  181, \dodoi{10.3847/1538-3881/ac134a}

\bibitem[{{Losse}(2010)}]{Losse_10}
{Losse}, F. 2010, Observations et Travaux, 75, 17

\bibitem[{{Lowrance} {et~al.}(2002){Lowrance}, {Kirkpatrick}, \&
  {Beichman}}]{Lowrance_02}
{Lowrance}, P.~J., {Kirkpatrick}, J.~D., \& {Beichman}, C.~A. 2002, \apjl, 572,
  L79, \dodoi{10.1086/341554}

\bibitem[{{Mamajek} \& {Stapelfeldt}(2023)}]{Mamajek_Stapelfeldt_23}
{Mamajek}, E., \& {Stapelfeldt}, K. 2023, {NASA ExEP Mission Star List for the
  Habitable Worlds Observatory}

\bibitem[{{Mann} {et~al.}(2019){Mann}, {Dupuy}, {Kraus}, {Gaidos}, {Ansdell},
  {Ireland}, {Rizzuto}, {Hung}, {Dittmann}, {Factor}, {Feiden}, {Martinez},
  {Ru{\'\i}z-Rodr{\'\i}guez}, \& {Thao}}]{Mann_19}
{Mann}, A.~W., {Dupuy}, T., {Kraus}, A.~L., {et~al.} 2019, \apj, 871, 63,
  \dodoi{10.3847/1538-4357/aaf3bc}

\bibitem[{{Marchal} \& {Bozis}(1982)}]{Marchal_82}
{Marchal}, C., \& {Bozis}, G. 1982, Celestial Mechanics, 26, 311,
  \dodoi{10.1007/BF01230725}

\bibitem[{{Marmier} {et~al.}(2013){Marmier}, {S{\'e}gransan}, {Udry}, {Mayor},
  {Pepe}, {Queloz}, {Lovis}, {Naef}, {Santos}, {Alonso}, {Alves}, {Berthet},
  {Chazelas}, {Demory}, {Dumusque}, {Eggenberger}, {Figueira}, {Gillon},
  {Hagelberg}, {Lendl}, {Mardling}, {M{\'e}gevand}, {Neveu}, {Sahlmann},
  {Sosnowska}, {Tewes}, \& {Triaud}}]{Marmier_13}
{Marmier}, M., {S{\'e}gransan}, D., {Udry}, S., {et~al.} 2013, \aap, 551, A90,
  \dodoi{10.1051/0004-6361/201219639}

\bibitem[{{Mason} {et~al.}(2017){Mason}, {Hartkopf}, \& {Miles}}]{Mason_17}
{Mason}, B.~D., {Hartkopf}, W.~I., \& {Miles}, K.~N. 2017, \aj, 154, 200,
  \dodoi{10.3847/1538-3881/aa803e}

\bibitem[{{Mason} {et~al.}(2021){Mason}, {Williams}, {Matson}, {Josties},
  {Eakens}, {Justice}, {Kilian}, \& {Warner}}]{Mason_21}
{Mason}, B.~D., {Williams}, S.~J., {Matson}, R.~A., {et~al.} 2021, \aj, 162,
  53, \dodoi{10.3847/1538-3881/abfaa2}

\bibitem[{{Mason} {et~al.}(2001){Mason}, {Wycoff}, {Hartkopf}, {Douglass}, \&
  {Worley}}]{Mason_01}
{Mason}, B.~D., {Wycoff}, G.~L., {Hartkopf}, W.~I., {Douglass}, G.~G., \&
  {Worley}, C.~E. 2001, \aj, 122, 3466, \dodoi{10.1086/323920}

\bibitem[{{Matthews} {et~al.}(2024){Matthews}, {Carter}, {Pathak}, \&
  {Morley}}]{Matthews_24}
{Matthews}, E.~S., {Carter}, Carter, A.~L., {Pathak}, P., \& {Morley}, C.~V.
  2024, Nature, \dodoi{10.1038/s41586-024-07837-8}

\bibitem[{{Mawet} {et~al.}(2019){Mawet}, {Hirsch}, {Lee}, {Ruffio}, {Bottom},
  {Fulton}, {Absil}, {Beichman}, {Bowler}, {Bryan}, {Choquet}, {Ciardi},
  {Christiaens}, {Defr{\`e}re}, {Gomez Gonzalez}, {Howard}, {Huby}, {Isaacson},
  {Jensen-Clem}, {Kosiarek}, {Marcy}, {Meshkat}, {Petigura}, {Reggiani},
  {Ruane}, {Serabyn}, {Sinukoff}, {Wang}, {Weiss}, \& {Ygouf}}]{Mawet_19}
{Mawet}, D., {Hirsch}, L., {Lee}, E.~J., {et~al.} 2019, \aj, 157, 33,
  \dodoi{10.3847/1538-3881/aaef8a}

\bibitem[{{Mayor} {et~al.}(2003){Mayor}, {Pepe}, {Queloz}, {Bouchy},
  {Rupprecht}, {Lo Curto}, {Avila}, {Benz}, {Bertaux}, {Bonfils}, {Dall},
  {Dekker}, {Delabre}, {Eckert}, {Fleury}, {Gilliotte}, {Gojak}, {Guzman},
  {Kohler}, {Lizon}, {Longinotti}, {Lovis}, {Megevand}, {Pasquini}, {Reyes},
  {Sivan}, {Sosnowska}, {Soto}, {Udry}, {van Kesteren}, {Weber}, \&
  {Weilenmann}}]{Mayor_03}
{Mayor}, M., {Pepe}, F., {Queloz}, D., {et~al.} 2003, The Messenger, 114, 20

\bibitem[{{McCaughrean} {et~al.}(2004){McCaughrean}, {Close}, {Scholz},
  {Lenzen}, {Biller}, {Brandner}, {Hartung}, \& {Lodieu}}]{McCaughrean_04}
{McCaughrean}, M.~J., {Close}, L.~M., {Scholz}, R.~D., {et~al.} 2004, \aap,
  413, 1029, \dodoi{10.1051/0004-6361:20034292}

\bibitem[{{Mermilliod}(1997)}]{Mermilliod_97}
{Mermilliod}, J.~C. 1997, {VizieR Online Data Catalog: Homogeneous Means in the
  UBV System (Mermilliod 1991)}, VizieR On-line Data Catalog: II/168.
  Originally published in: Institut d'Astronomie, Universite de Lausanne (1991)

\bibitem[{{Mesa} {et~al.}(2023){Mesa}, {Gratton}, {Kervella}, {Bonavita},
  {Desidera}, {D'Orazi}, {Marino}, {Zurlo}, \& {Rigliaco}}]{Mesa_23}
{Mesa}, D., {Gratton}, R., {Kervella}, P., {et~al.} 2023, \aap, 672, A93,
  \dodoi{10.1051/0004-6361/202345865}

\bibitem[{{Motalebi} {et~al.}(2015){Motalebi}, {Udry}, {Gillon}, {Lovis},
  {S{\'e}gransan}, {Buchhave}, {Demory}, {Malavolta}, {Dressing}, {Sasselov},
  {Rice}, {Charbonneau}, {Collier Cameron}, {Latham}, {Molinari}, {Pepe},
  {Affer}, {Bonomo}, {Cosentino}, {Dumusque}, {Figueira}, {Fiorenzano},
  {Gettel}, {Harutyunyan}, {Haywood}, {Johnson}, {Lopez}, {Lopez-Morales},
  {Mayor}, {Micela}, {Mortier}, {Nascimbeni}, {Philips}, {Piotto}, {Pollacco},
  {Queloz}, {Sozzetti}, {Vanderburg}, \& {Watson}}]{Motalebi_15}
{Motalebi}, F., {Udry}, S., {Gillon}, M., {et~al.} 2015, \aap, 584, A72,
  \dodoi{10.1051/0004-6361/201526822}

\bibitem[{{Naef} {et~al.}(2003){Naef}, {Mayor}, {Korzennik}, {Queloz}, {Udry},
  {Nisenson}, {Noyes}, {Brown}, {Beuzit}, {Perrier}, \& {Sivan}}]{Naef_03}
{Naef}, D., {Mayor}, M., {Korzennik}, S.~G., {et~al.} 2003, \aap, 410, 1051,
  \dodoi{10.1051/0004-6361:20031341}

\bibitem[{{NASA Exoplanet Archive}(2025)}]{NASA_Exo_Archive_PS}
{NASA Exoplanet Archive}. 2025, Planetary Systems, Version: 2025-03-31 11:50,
  NExScI-Caltech/IPAC, \dodoi{10.26133/NEA12}

\bibitem[{{National Academies of Sciences} \& Medicine(2021)}]{Astro2020}
{National Academies of Sciences}, E., \& Medicine. 2021, {Pathways to Discovery
  in Astronomy and Astrophysics for the 2020s}, \dodoi{10.17226/26141}

\bibitem[{{Newman} {et~al.}(2023){Newman}, {Plavchan}, {Burt}, {Teske},
  {Mamajek}, {Leifer}, {Gaudi}, {Blackwood}, \& {Morgan}}]{Newman_23}
{Newman}, P.~D., {Plavchan}, P., {Burt}, J.~A., {et~al.} 2023, \aj, 165, 151,
  \dodoi{10.3847/1538-3881/acad07}

\bibitem[{{Nielsen} {et~al.}(2019){Nielsen}, {De Rosa}, {Macintosh}, {Wang},
  {Ruffio}, {Chiang}, {Marley}, {Saumon}, {Savransky}, {Ammons}, {Bailey},
  {Barman}, {Blain}, {Bulger}, {Burrows}, {Chilcote}, {Cotten}, {Czekala},
  {Doyon}, {Duch{\^e}ne}, {Esposito}, {Fabrycky}, {Fitzgerald}, {Follette},
  {Fortney}, {Gerard}, {Goodsell}, {Graham}, {Greenbaum}, {Hibon}, {Hinkley},
  {Hirsch}, {Hom}, {Hung}, {Dawson}, {Ingraham}, {Kalas}, {Konopacky},
  {Larkin}, {Lee}, {Lin}, {Maire}, {Marchis}, {Marois}, {Metchev},
  {Millar-Blanchaer}, {Morzinski}, {Oppenheimer}, {Palmer}, {Patience},
  {Perrin}, {Poyneer}, {Pueyo}, {Rafikov}, {Rajan}, {Rameau}, {Rantakyr{\"o}},
  {Ren}, {Schneider}, {Sivaramakrishnan}, {Song}, {Soummer}, {Tallis},
  {Thomas}, {Ward-Duong}, \& {Wolff}}]{Nielsen_19}
{Nielsen}, E.~L., {De Rosa}, R.~J., {Macintosh}, B., {et~al.} 2019, \aj, 158,
  13, \dodoi{10.3847/1538-3881/ab16e9}

\bibitem[{{Paxton} {et~al.}(2011){Paxton}, {Bildsten}, {Dotter}, {Herwig},
  {Lesaffre}, \& {Timmes}}]{Paxton_11}
{Paxton}, B., {Bildsten}, L., {Dotter}, A., {et~al.} 2011, \apjs, 192, 3,
  \dodoi{10.1088/0067-0049/192/1/3}

\bibitem[{{Paxton} {et~al.}(2013){Paxton}, {Cantiello}, {Arras}, {Bildsten},
  {Brown}, {Dotter}, {Mankovich}, {Montgomery}, {Stello}, {Timmes}, \&
  {Townsend}}]{Paxton_13}
{Paxton}, B., {Cantiello}, M., {Arras}, P., {et~al.} 2013, \apjs, 208, 4,
  \dodoi{10.1088/0067-0049/208/1/4}

\bibitem[{{Paxton} {et~al.}(2015){Paxton}, {Marchant}, {Schwab}, {Bauer},
  {Bildsten}, {Cantiello}, {Dessart}, {Farmer}, {Hu}, {Langer}, {Townsend},
  {Townsley}, \& {Timmes}}]{Paxton_15}
{Paxton}, B., {Marchant}, P., {Schwab}, J., {et~al.} 2015, \apjs, 220, 15,
  \dodoi{10.1088/0067-0049/220/1/15}

\bibitem[{{Paxton} {et~al.}(2018){Paxton}, {Schwab}, {Bauer}, {Bildsten},
  {Blinnikov}, {Duffell}, {Farmer}, {Goldberg}, {Marchant}, {Sorokina},
  {Thoul}, {Townsend}, \& {Timmes}}]{Paxton_18}
{Paxton}, B., {Schwab}, J., {Bauer}, E.~B., {et~al.} 2018, \apjs, 234, 34,
  \dodoi{10.3847/1538-4365/aaa5a8}

\bibitem[{{Perryman} {et~al.}(1997){Perryman}, {Lindegren}, {Kovalevsky},
  {Hoeg}, {Bastian}, {Bernacca}, {Cr{\'e}z{\'e}}, {Donati}, {Grenon},
  {Grewing}, {van Leeuwen}, {van der Marel}, {Mignard}, {Murray}, {Le Poole},
  {Schrijver}, {Turon}, {Arenou}, {Froeschl{\'e}}, \& {Petersen}}]{Perryman_97}
{Perryman}, M.~A.~C., {Lindegren}, L., {Kovalevsky}, J., {et~al.} 1997, \aap,
  323, L49

\bibitem[{{Philipot} {et~al.}(2023){Philipot}, {Lagrange}, {Rubini}, {Kiefer},
  \& {Chomez}}]{Philipot_23}
{Philipot}, F., {Lagrange}, A.~M., {Rubini}, P., {Kiefer}, F., \& {Chomez}, A.
  2023, \aap, 670, A65, \dodoi{10.1051/0004-6361/202245396}

\bibitem[{{Quarles} {et~al.}(2020){Quarles}, {Li}, {Kostov}, \&
  {Haghighipour}}]{Quarles_20}
{Quarles}, B., {Li}, G., {Kostov}, V., \& {Haghighipour}, N. 2020, \aj, 159,
  80, \dodoi{10.3847/1538-3881/ab64fa}

\bibitem[{{Quarles} {et~al.}(2018){Quarles}, {Satyal}, {Kostov}, {Kaib}, \&
  {Haghighipour}}]{Quarles_18}
{Quarles}, B., {Satyal}, S., {Kostov}, V., {Kaib}, N., \& {Haghighipour}, N.
  2018, \apj, 856, 150, \dodoi{10.3847/1538-4357/aab264}

\bibitem[{{Raghavan} {et~al.}(2010){Raghavan}, {McAlister}, {Henry}, {Latham},
  {Marcy}, {Mason}, {Gies}, {White}, \& {ten Brummelaar}}]{Raghavan_10}
{Raghavan}, D., {McAlister}, H.~A., {Henry}, T.~J., {et~al.} 2010, \apjs, 190,
  1, \dodoi{10.1088/0067-0049/190/1/1}

\bibitem[{{Reyl{\'e}} {et~al.}(2022){Reyl{\'e}}, {Jardine}, {Fouqu{\'e}},
  {Caballero}, {Smart}, \& {Sozzetti}}]{Reyle_22}
{Reyl{\'e}}, C., {Jardine}, K., {Fouqu{\'e}}, P., {et~al.} 2022, in The 21st
  Cambridge Workshop on Cool Stars, Stellar Systems, and the Sun, Cambridge
  Workshop on Cool Stars, Stellar Systems, and the Sun, 218,
  \dodoi{10.5281/zenodo.7669746}

\bibitem[{{Rickman} {et~al.}(2022){Rickman}, {Matthews}, {Ceva},
  {S{\'e}gransan}, {Brandt}, {Zhang}, {Brandt}, {Forveille}, {Hagelberg}, \&
  {Udry}}]{Rickman_22}
{Rickman}, E.~L., {Matthews}, E., {Ceva}, W., {et~al.} 2022, \aap, 668, A140,
  \dodoi{10.1051/0004-6361/202244633}

\bibitem[{{Rickman} {et~al.}(2024){Rickman}, {Ceva}, {Matthews},
  {S{\'e}gransan}, {Bowler}, {Forveille}, {Franson}, {Hagelberg}, {Udry}, \&
  {Vigan}}]{Rickman_24}
{Rickman}, E.~L., {Ceva}, W., {Matthews}, E.~C., {et~al.} 2024, arXiv e-prints,
  arXiv:2401.10058, \dodoi{10.48550/arXiv.2401.10058}

\bibitem[{{Rodriguez} {et~al.}(2015){Rodriguez}, {Duch{\^e}ne}, {Tom},
  {Kennedy}, {Matthews}, {Greaves}, \& {Butner}}]{Rodriguez_15}
{Rodriguez}, D.~R., {Duch{\^e}ne}, G., {Tom}, H., {et~al.} 2015, \mnras, 449,
  3160, \dodoi{10.1093/mnras/stv483}

\bibitem[{{Rosenthal} {et~al.}(2021){Rosenthal}, {Fulton}, {Hirsch},
  {Isaacson}, {Howard}, {Dedrick}, {Sherstyuk}, {Blunt}, {Petigura}, {Knutson},
  {Behmard}, {Chontos}, {Crepp}, {Crossfield}, {Dalba}, {Fischer}, {Henry},
  {Kane}, {Kosiarek}, {Marcy}, {Rubenzahl}, {Weiss}, \&
  {Wright}}]{Rosenthal_21}
{Rosenthal}, L.~J., {Fulton}, B.~J., {Hirsch}, L.~A., {et~al.} 2021, \apjs,
  255, 8, \dodoi{10.3847/1538-4365/abe23c}

\bibitem[{{Scholz} {et~al.}(2003){Scholz}, {McCaughrean}, {Lodieu}, \&
  {Kuhlbrodt}}]{Scholz_03}
{Scholz}, R.~D., {McCaughrean}, M.~J., {Lodieu}, N., \& {Kuhlbrodt}, B. 2003,
  \aap, 398, L29, \dodoi{10.1051/0004-6361:20021847}

\bibitem[{{Solomon} \& {Head}(1991)}]{Solomon_91}
{Solomon}, S.~C., \& {Head}, J.~W. 1991, Science, 252, 252,
  \dodoi{10.1126/science.252.5003.252}

\bibitem[{{Takeda} {et~al.}(2007){Takeda}, {Ford}, {Sills}, {Rasio}, {Fischer},
  \& {Valenti}}]{Takeda_07}
{Takeda}, G., {Ford}, E.~B., {Sills}, A., {et~al.} 2007, \apjs, 168, 297,
  \dodoi{10.1086/509763}

\bibitem[{{Tokovinin}(2021)}]{Tokovinin_21}
{Tokovinin}, A. 2021, \aj, 161, 144, \dodoi{10.3847/1538-3881/abda42}

\bibitem[{{Torres} {et~al.}(2006){Torres}, {Quast}, {da Silva}, {de La Reza},
  {Melo}, \& {Sterzik}}]{Torres_06}
{Torres}, C.~A.~O., {Quast}, G.~R., {da Silva}, L., {et~al.} 2006, \aap, 460,
  695, \dodoi{10.1051/0004-6361:20065602}

\bibitem[{{Tuchow} {et~al.}(2024){Tuchow}, {Stark}, \& {Mamajek}}]{Tuchow_24}
{Tuchow}, N.~W., {Stark}, C.~C., \& {Mamajek}, E. 2024, \aj, 167, 139,
  \dodoi{10.3847/1538-3881/ad25ec}

\bibitem[{{van Leeuwen}(2007)}]{vanLeeuwen_07}
{van Leeuwen}, F. 2007, \aap, 474, 653, \dodoi{10.1051/0004-6361:20078357}

\bibitem[{{Van Zandt} \& {Petigura}(2024)}]{VanZandt_24}
{Van Zandt}, J., \& {Petigura}, E.~A. 2024, \aj, 167, 250,
  \dodoi{10.3847/1538-3881/ad390b}

\bibitem[{{Venner} {et~al.}(2021){Venner}, {Vanderburg}, \&
  {Pearce}}]{Venner_21}
{Venner}, A., {Vanderburg}, A., \& {Pearce}, L.~A. 2021, \aj, 162, 12,
  \dodoi{10.3847/1538-3881/abf932}

\bibitem[{{Vogt} {et~al.}(2005){Vogt}, {Butler}, {Marcy}, {Fischer}, {Henry},
  {Laughlin}, {Wright}, \& {Johnson}}]{Vogt_05}
{Vogt}, S.~S., {Butler}, R.~P., {Marcy}, G.~W., {et~al.} 2005, \apj, 632, 638,
  \dodoi{10.1086/432901}

\bibitem[{{Vogt} {et~al.}(1994){Vogt}, {Allen}, {Bigelow}, {Bresee}, {Brown},
  {Cantrall}, {Conrad}, {Couture}, {Delaney}, {Epps}, {Hilyard}, {Hilyard},
  {Horn}, {Jern}, {Kanto}, {Keane}, {Kibrick}, {Lewis}, {Osborne},
  {Pardeilhan}, {Pfister}, {Ricketts}, {Robinson}, {Stover}, {Tucker}, {Ward},
  \& {Wei}}]{Vogt_94}
{Vogt}, S.~S., {Allen}, S.~L., {Bigelow}, B.~C., {et~al.} 1994, in Society of
  Photo-Optical Instrumentation Engineers (SPIE) Conference Series, Vol. 2198,
  Instrumentation in Astronomy VIII, ed. D.~L. {Crawford} \& E.~R. {Craine},
  362, \dodoi{10.1117/12.176725}

\bibitem[{{Vogt} {et~al.}(2015){Vogt}, {Burt}, {Meschiari}, {Butler}, {Henry},
  {Wang}, {Holden}, {Gapp}, {Hanson}, {Arriagada}, {Keiser}, {Teske}, \&
  {Laughlin}}]{Vogt_15}
{Vogt}, S.~S., {Burt}, J., {Meschiari}, S., {et~al.} 2015, \apj, 814, 12,
  \dodoi{10.1088/0004-637X/814/1/12}

\bibitem[{{Volk} {et~al.}(2003){Volk}, {Blum}, {Walker}, \& {Puxley}}]{Volk_03}
{Volk}, K., {Blum}, R., {Walker}, G., \& {Puxley}, P. 2003, \iaucirc, 8188, 2

\bibitem[{{Xuan} \& {Wyatt}(2020)}]{Xuan_20}
{Xuan}, J.~W., \& {Wyatt}, M.~C. 2020, \mnras, 497, 2096,
  \dodoi{10.1093/mnras/staa2033}

\bibitem[{{Zhao} {et~al.}(2011){Zhao}, {Oswalt}, {Rudkin}, {Zhao}, \&
  {Chen}}]{Zhao_11}
{Zhao}, J.~K., {Oswalt}, T.~D., {Rudkin}, M., {Zhao}, G., \& {Chen}, Y.~Q.
  2011, \aj, 141, 107, \dodoi{10.1088/0004-6256/141/4/107}

\end{thebibliography}

\end{document}